\documentclass[fleqn,usenatbib]{mnras}

\usepackage{newtxtext,newtxmath}

\usepackage[T1]{fontenc}

\DeclareRobustCommand{\VAN}[3]{#2}
\let\VANthebibliography\thebibliography
\def\thebibliography{\DeclareRobustCommand{\VAN}[3]{##3}\VANthebibliography}

\usepackage{graphicx}	
\usepackage{amsmath}	
\usepackage{comment}
\usepackage{xspace} 
\usepackage{mathtools}
\usepackage{hyperref}
\usepackage{float}
\usepackage{bm}
\usepackage{pdflscape}
\usepackage{pdfpages}
\DeclareFontFamily{U}{mathx}{}
\DeclareFontShape{U}{mathx}{m}{n}{<-> mathx10}{}
\DeclareSymbolFont{mathx}{U}{mathx}{m}{n}
\DeclareMathAccent{\widecheck}{0}{mathx}{"71}

\usepackage{multirow} \usepackage{threeparttable}
\usepackage{booktabs} \usepackage{stfloats}

\newcommand{\PACO}{{\texttt{PACO}}\xspace} 
 
\newcommand{\dPACO}{{\texttt{deep PACO}}\xspace} 

\newcommand*{\V}[1]{\boldsymbol{#1}}   
\newcommand*{\M}[1]{\mathbf{#1}}       
\newcommand*{\TransposeLetter}{\hspace*{-.25ex}\top\hspace*{-.25ex}}
\newcommand*{\T}{^{\TransposeLetter}} 
\DeclareFontFamily{U}{mathx}{\hyphenchar\font45}
\DeclareFontShape{U}{mathx}{m}{n}{<-> mathx10}{}
\DeclareSymbolFont{mathx}{U}{mathx}{m}{n}

\DeclarePairedDelimiterX{\Paren}[1]{(}{)}{#1}
\DeclarePairedDelimiterX{\Brace}[1]{\{}{\}}{#1}
\DeclarePairedDelimiterX{\Brack}[1]{[}{]}{#1}
\DeclarePairedDelimiterX{\Abs}[1]{\rvert}{\lvert}{#1}
\DeclarePairedDelimiterX{\Norm}[1]{\lVert}{\rVert}{#1}
\DeclarePairedDelimiterX{\Avg}[1]{\langle}{\rangle}{#1}
\DeclarePairedDelimiterX{\Round}[1]{\lfloor}{\rceil}{#1}
\DeclarePairedDelimiterX{\Floor}[1]{\lfloor}{\rfloor}{#1}
\DeclarePairedDelimiterX{\Ceil}[1]{\lceil}{\rceil}{#1}
\DeclarePairedDelimiterX{\Inner}[2]{\langle}{\rangle}{#1,#2}

\DeclareMathOperator{\Trace}{tr}

\DeclarePairedDelimiterXPP{\Expect}[1]{\mathbb{E}}(){}{#1}

\newcommand*{\estim}[1]{\widehat{#1}}
\newcommand{\Sest}{\widehat{\M{S}}}

\usepackage[squaren,Gray]{SIunits}

\usepackage{numprint}

\makeatletter
\def\widebreve{\mathpalette\wide@breve}
\def\wide@breve#1#2{\sbox\z@{$#1#2$}%
     \mathop{\vbox{\m@th\ialign{##\crcr
\kern0.08em\brevefill#1{0.8\wd\z@}\crcr\noalign{\nointerlineskip}%
                    $\hss#1#2\hss$\crcr}}}\limits}
\def\brevefill#1#2{$\m@th\sbox\tw@{$#1($}%
  \hss\resizebox{#2}{\wd\tw@}{\rotatebox[origin=c]{90}{\upshape(}}\hss$}
\makeatletter

\title[deep PACO: deep learning meets PAtch COvariances]{deep PACO: Combining statistical models with deep learning for exoplanet detection and characterization in direct imaging at high contrast}

\author[O. Flasseur et al.]{
Olivier Flasseur$^{1,2,5}$\thanks{O. Flasseur is currently with CRAL (5), and was previously with LESIA (1) and Inria (2) where most of this work was performed. E-mail: olivier.flasseur@univ-lyon1.fr}, Théo Bodrito$^{2}$, Julien Mairal$^{3}$, \newauthor Jean Ponce$^{2,4}$, Maud Langlois$^{5}$, and Anne-Marie Lagrange$^{1,6}$
\\
$^{1}$Laboratoire d'{\'E}tudes Spatiales et d'Instrumentation en Astrophysique, Observatoire de Paris, Univ. PSL, Sorbonne Univ., Univ. Paris Diderot, France\\
$^{2}$Département d'Informatique de l'{\'E}cole Normale Supérieure (ENS-PSL, CNRS, Inria), France\\
$^{3}$Univ. Grenoble Alpes, Inria, CNRS, Grenoble INP, LJK, 38000 Grenoble, France\\
$^{4}$Courant Institute of Mathematical Sciences, Center for Data Science, New York Univ., USA\\
$^{5}$Centre de Recherche Astrophysique de Lyon, CNRS, Univ. de Lyon, Univ. Claude Bernard Lyon 1, ENS de Lyon, France\\
$^{6}$Univ. Grenoble Alpes, Institut de Planétologie et d'Astrophysique de Grenoble, France
}

\pubyear{2023}

\begin{document}
\label{firstpage}
\pagerange{\pageref{firstpage}--\pageref{lastpage}}
\maketitle

\begin{abstract}
Direct imaging is an active research topic in astronomy for the detection and the characterization of young sub-stellar objects. The very high contrast between the host star and its companions makes the observations particularly challenging. In this context, post-processing methods combining several images recorded with the pupil tracking mode of telescope are needed. In previous works, we have presented a data-driven algorithm, \PACO, capturing locally the spatial correlations of the data with a multi-variate Gaussian model. \PACO delivers better detection sensitivity and confidence than the standard post-processing methods of the field. However, there is room for improvement due to the approximate fidelity of the \PACO statistical model to the time evolving observations. In this paper, we propose to combine the statistical model of \PACO with supervised deep learning. The data are first pre-processed with the \PACO framework to improve the stationarity and the contrast. A convolutional neural network (CNN) is then trained in a supervised fashion to detect the residual signature of synthetic sources. Finally, the trained network delivers a detection map. The photometry of detected sources is estimated by a second CNN. We apply the proposed approach to several datasets from the VLT/SPHERE instrument. Our results show that its detection stage performs significantly better than baseline methods (cADI, PCA), and leads to a contrast improvement up to half a magnitude compared to \PACO. The characterization stage of the proposed method performs on average on par with or better than the comparative algorithms (PCA, \PACO) for angular separation above 0.5’’.
\end{abstract}




\section{Introduction}
\label{sec:introduction}

High-contrast imaging is an observational method used to study the close environment of stars \citep{traub2010direct,bowler2016imaging,pueyo2018direct}. It is particularly adapted to detect young, massive and hot exoplanets (see e.g. \cite{chauvin2004giant,chauvin2005companion,schneider2011defining,nielsen2019gemini}), thus complementing well indirect exoplanet detection methods such as transit photometry or Doppler spectroscopy \citep{santos2008extra}. Direct imaging offers other appealing characteristics like the detection of candidate companions from a few hours of observations and the ability to characterize them in terms of age, surface gravity, effective temperature and composition \citep{allard2003model,allard2007k}, or to predict their evolution \citep{burrows1997nongray,chabrier2000evolutionary}. Despite these promises, only a few dozens exoplanets have been unveiled and characterized since the emergence of direct imaging in the early 2000s \citep{marois2008direct,lagrange2009probable,nielsen2012gemini,macintosh2015discovery,chauvin2017discovery,keppler2018discovery}. This is mainly due to (i) the relatively low occurrence of giant exoplanets, (ii) the very high contrast between the host star and the exoplanets (typically, higher than $10^5$ in the infrared), and (iii) the required high angular resolution (typically, better than a tenth of arcsec). 

In this context, cutting-edge ground-based facilities like VLT/SPHERE \citep{beuzit2019sphere}, Gemini/GPI \citep{macintosh2014first}, Keck/NIRC2 \citep{castella2016commissioning}, Magellan/MagAO \citep{morzinski2014magao} or SUBARU/SCExAO \citep{jovanovic2015subaru} are equipped with an (extreme) adaptive optics system and a coronagraph to attenuate as much as possible the glare of the star. Currently, the non-blocked residual starlight contamination and its temporal fluctuations remain the main limitation. It takes the form of spatially-correlated \textit{speckles} that resemble the expected signature of a point-like source (e.g., an exoplanet, a brown dwarf, a background star). The observations are also impacted by additional sources of noise (i.e., thermal background flux, detector readout, photon noise). Together, speckles and noise form a spatially and non-stationary \textit{nuisance component} that corrupts the signals of the sought objects. Off-axis objects can either take the form of point-like sources or that of spatially extended features like circumstellar disks. In this paper, we focus on the detection of point-like sources,  leaving the problem of disk reconstruction for future work.

In order to unmix the sought objects from the nuisance component, high-contrast observations are performed with dedicated strategies like angular differential imaging (ADI; \cite{marois2006angular}), that we consider in this paper. ADI consists in tracking the observed target over time, with the telescope derotator tuned to keep the telescope pupil stable while the field of view rotates around the host star. Consequently, in the resulting 3-D datasets (2-D + time), the objects of interest follow an apparent motion along a deterministic circular trajectory centered on the star while the telescope pupil remains static. With ADI, speckles due to residual starlight aberrations are \textit{quasi-static}, i.e., they are strongly correlated across exposures.
ADI also allows to extract the astrometry and the photometry of detected sources. These estimates can be used to characterize the physical properties of the detected sources by fitting orbital and atmospheric models \citep{vigan2010photometric, cheetham2019spectral, mesa2019determining}. In this paper, we address two tasks: (i) the detection of point-like sources, and (ii) the estimation of their photometry.


The last cornerstone of high-contrast imaging is data \textit{reduction}, i.e., the combination of the recorded images by dedicated post-processing algorithms. This critical step brings the additional gain in contrast (between one and three orders of magnitudes for existing methods) needed to detect the faint signals coming from thermal self-emission of giant exoplanets. The classical principle is to estimate a reference image (so-called \textit{on-axis PSF}) of the nuisance component, that can be subtracted from the data in order to unveil the sought objects. A simple practical implementation of this general strategy consists in subtracting the temporal mean or median of the dataset from each frame of the ADI stack. The residual images are then co-aligned to the true-North so that the signals of the sought objects are superimposed and can be combined by temporal stacking. This is the principle of the cADI method designed to process the first direct imaging observations \citep{marois2006angular,lagrange2009probable}. In the last decade, several more advanced methods have been developed, see, e.g., \cite{pueyo2018direct} for a review. In particular, TLOCI \citep{marois2013tloci,marois2014gpi} (or its variants such as LOCI \citep{lafreniere2007new}, ALOCI \citep{currie2012direct1,currie2012direct2}, MLOCI \citep{wahhaj2015improving}), and KLIP/PCA \citep{soummer2012detection,amara2012pynpoint} are currently implemented in most reduction pipelines \citep{amara2012pynpoint,gonzalez2017vip,galicher2018astrometric}. LOCI-based and PCA-based algorithms are considered as standards to process high-contrast observations. With the (A, M, T)-LOCI algorithm, the on-axis PSF is estimated by combining images selected in a library. The combined images are selected and weighted to minimize the residual noise and maximize the throughput of point-like sources simultaneously. The PCA algorithm performs a principal component analysis of the data, and a low-rank estimate of the on-axis PSF is formed by keeping the first principal components of the decomposition. In the same vein, the LLSG algorithm \citep{gonzalez2016low} decomposes the dataset into low-rank, sparse and Gaussian components. Because few sources are expected in the field of view, their signatures are mainly recovered in the sparse component. The RSM algorithm \citep{dahlqvist2020regime, dahlqvist2021improving, dahlqvist2021auto} combines residual images obtained with different post-processing algorithms (e.g., cADI and PCA)  to leverage their specific benefits and mitigate their respective drawbacks. Since all of these algorithms are based on the estimation and subtraction of an off-axis PSF, they are facing a common pitfall: they fail in deriving a statistically grounded detection map, especially at short angular separations. As a consequence, the identification of candidate sources partly relies on visual inspection of the detection map.

To circumvent this issue, several works have considered alternative ways to reduce the data in order to produce more quantitative outputs. The derivation of a custom signal-to-noise ratio (S/N) through a $t$-test empirically corrected for the varying number of samples as a function of the angular separation \citep{mawet2014fundamental} is a pioneering work in this direction. \cite{jensen2017new} recommend the adoption of metrics combining the achievable contrast with the fraction of detected sources. Instead of normalizing the combined residual images by the empirical standard-deviation of the noise (roughly) approximated on annuli, \cite{pairet2019stim} propose to build a detection map directly from the set of residual images by comparing the variance of samples located on the expected trajectory of putative sources. Other methods reformulate the detection task as an inverse problem. Among them, ANDROMEDA \citep{mugnier2009optimal,cantalloube2015direct} and FMMF \citep{ruffio2017improving} build a model of the residual off-axis signal after subtraction of the estimated on-axis PSF. The \PACO algorithm \citep{flasseur2018unsupervised,flasseur2018exoplanet,flasseur2018SPIE,flasseur2020robust,flasseur2020pacoasdi} builds a more consistent statistical model, self-calibrated on the data, that accounts for the spatial correlations of the nuisance component at the scale of small image patches of a few tens of pixels, see Sect. \ref{subsubsec:statistical_modeling}.

Given the success of data-driven approaches in solving various high-level imaging tasks (e.g., detection, segmentation, classification) in very diverse fields (e.g., photography, microscopy, bio-medical imaging, remote sensing), machine learning and deep learning approaches have also been investigated by the direct imaging community. \cite{fergus2014s4} report one of the first steps in this direction with a discriminative approach exploiting the specific structure of high-contrast data. Based on support vector machines, the underlying model is trained from two-classes samples generated by resorting to massive injections of fake companions. \cite{gonzalez2018supervised} formalize the detection problem as a binary classification task and propose a fully supervised deep learning approach also trained in a supervised fashion. They use collections of patches pre-processed by PCA for different numbers of principal components as input of a random forest or of a convolutional neural network that decides in favor of the presence or on the absence of a point-like source in each patch. While demonstrating powerful detection capabilities, this algorithm 
showed to be prone to a high level of false alarms in some cases \citep{cantalloube2020exoplanet}. Besides, the tuning of hyper-parameters remains a critical point making the operating point difficult to reach. Generative adversarial networks (GANs, \cite{goodfellow2014generative}) have been used to produce multiple realizations of \textit{pure} nuisance component as an alternative way to generate a large basis of labeled samples used to train a deep discriminative model \citep{yip2019pushing}. Recent works \citep{samland2021trap, gebhard2022half} recast the unmixing problem as a regularized regression task. The underlying linear model is in charge of explaining the evolution of the nuisance component (and possibly, the evolution of the source signals) in a time series extracted at a given pixel location from temporal series of reference selected to be signal free and causally independent from the putative source signals.


Intensive testing of \PACO, both on public \citep{cantalloube2020exoplanet} as well as on consortia data challenges, and more recently on a sub-sample of about 75 observations from the SPHERE's SHINE survey \citep{chomez2022survey} shows that \PACO is one of the algorithms of choice to process high-contrast observations. In particular, the latter work \citep{chomez2022survey} demonstrates the expected gain in terms of achievable contrast (up to $10^{-7}$ at a few arcsec), and in terms of the underlying exoplanet population (with a mass up to $5\,\text{M}_{\text{Jup}}$ at 5 AU for stars at about 60 parsec away). 
Thanks to its unique data-driven modeling of the nuisance component, accounting for its non-stationary spatial correlations, \PACO is especially well suited to process observations in which the typical spatial extent of speckles lies in a patch of a few tens of pixels. However, the statistical model of \PACO is approximate in case of spatial correlations spread over a patch of a few tens of pixels (e.g., for background-limited observations and/or in case of unstable observing conditions). This motivates the path we follow in this work: we propose a new detection algorithm, named \dPACO, that combines the statistical model of \PACO with a supervised deep learning framework. The statistical model of \PACO is used to improve the stationarity and the contrast of the data in a pre-processing step, while deep learning is in charge of correcting for the (putative) approximate fidelity of the statistical model of \PACO to the reality of the observations. To do so, the data are centered and whitened locally using the \PACO framework, and a CNN is trained in a supervised fashion to detect the residual signature of synthetic sources from pre-processed science data. The network is trained from scratch using full frame samples generated with a custom data augmentation strategy allowing to build a large training set from a single ADI dataset. Finally, the underlying discriminative model is applied to the pre-processed observations and delivers a detection map. Detected sources are then photometrically characterized by a second deep neural network, also trained from scratch using patch samples generated with a dedicated data augmentation step. On this latter part, while the recent astronomy literature reports several works for photometry estimation through deep learning models (see e.g., \cite{boucaud2020photometry, cabayol2021pau, huertas2022dawes} for galaxie's photometry or red-shift estimation), to the best of your knowledge this is the first time that deep learning is employed to estimate the flux of detected sources in direct imaging at high-contrast.

This paper\footnote{\samepage A preliminary version of this work was presented in the form of a conference contribution in \cite{flasseur2022exoplanet}. The present manuscript contains a significant amount of additional methodological developments, technical improvements, and experiments.} is organized as follow. Section \ref{sec:algorithm_detection} presents the main ingredients of the detection part of the proposed algorithm. Section \ref{sec:algorithm_characterization} focuses on the characterization stage of the proposed method. Section \ref{sec:results} evaluates the detection and characterization performance on several high-contrast observations from the IRDIS imager \citep{dohlen2008prototyping} of the VLT/SPHERE instrument \citep{beuzit2019sphere}. Finally, Sect. \ref{sec:conclusion} draws the paper conclusions and gives future research prospects.

\medskip

\begin{figure*}
	\begin{center}
		\includegraphics[width=0.85\textwidth]{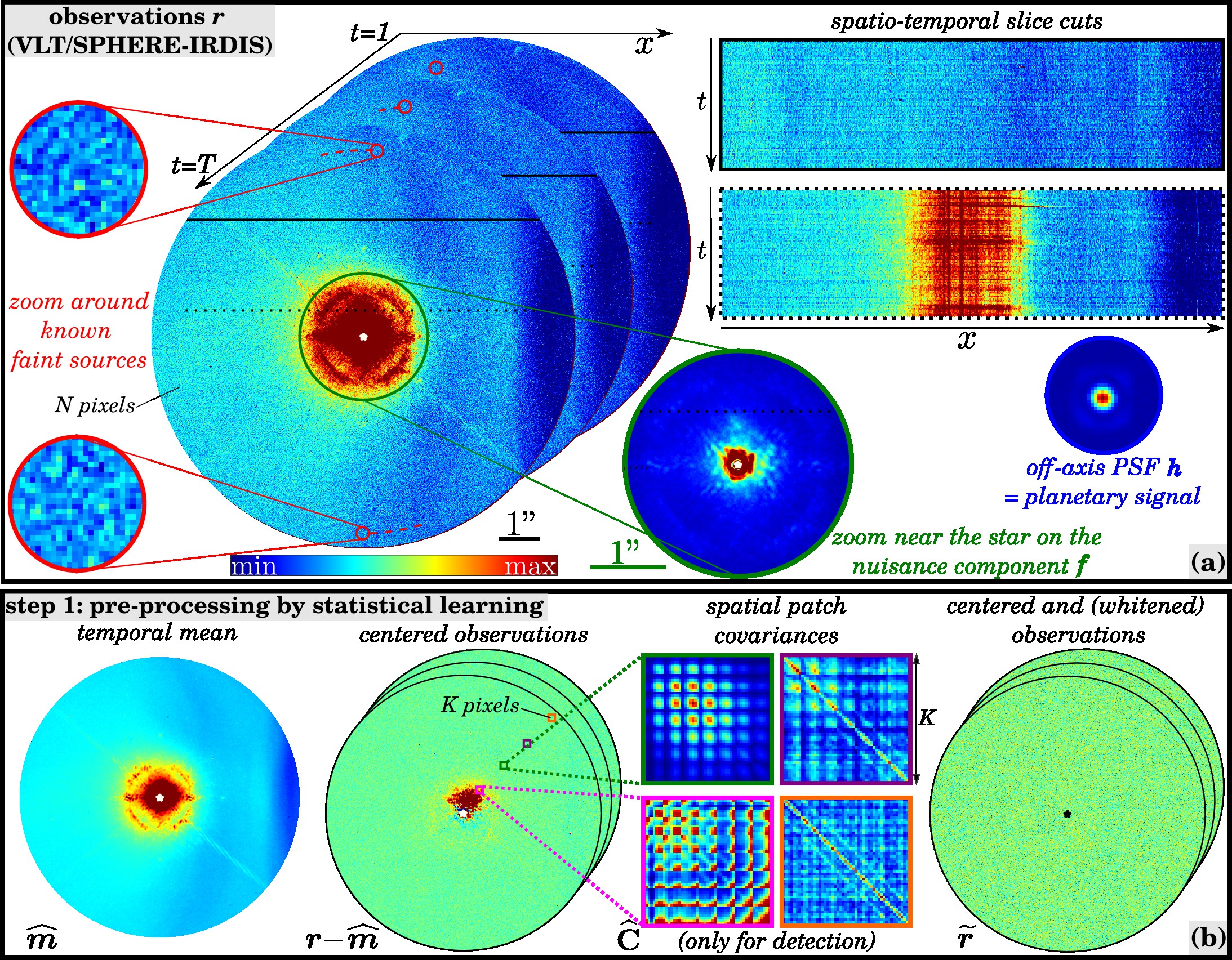}
		\caption{(a) Typical observations $\V r$ from the VLT/SPHERE-IRDIS instrument conducted in pupil tracking mode (i.e., with the ADI technique). Zooms around two known background faint sources are displayed in the red circles, a zoom on the nuisance component $\V f$ near the star is highlighted in the green circle, and a view of the sought exoplanetary signal, taking the form of the off-axis PSF $\V h$, is shown in the blue circle. Two spatio-temporal slice cuts along the black solid and dashed lines are shown on the right, as an illustration of the spatial non-stationarity of the correlations of the nuisance. (b) Illustration of the main operations performed during step 1 of the proposed approach, namely the pre-processing of the observations by statistical learning. The estimation of the mean $\widehat{m}$ and of the covariance matrices $\widehat{\M C}$ are based on the \PACO model of the nuisance component. Examples of estimated spatial covariance matrices are displayed in squares for four regions of interest pickled at about $0.5$ (pink), 1.0 (green), 1.5 (purple), and 2.0 (orange) arcsec. It results from this pre-processing step centered and whitened observations from which our detection and characterization models are built by supervised deep learning, see Fig. \ref{fig:data_pipeline_part2}. Dataset: HIP 72192 (2015-05-10), see Sect. \ref{sec:results} for the recording logs.}
		\label{fig:data_pipeline_part1}
	\end{center}
\end{figure*}

\begin{figure*}
	\begin{center}
		\includegraphics[width=\textwidth]{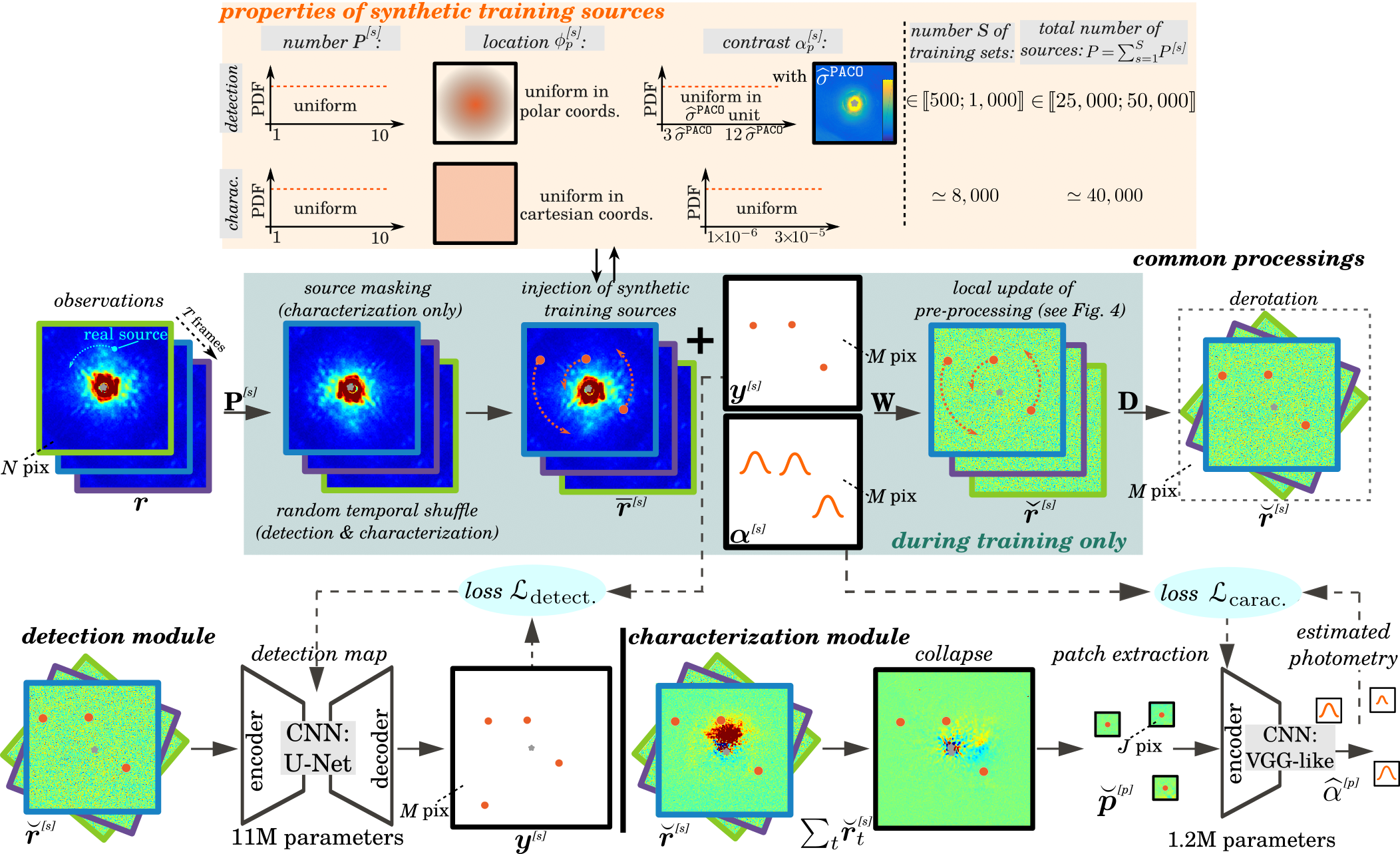}
		\caption{Schematic representation of the main operations performed during the detection and characterization steps of the proposed algorithm by supervised deep learning. The first line displays a view of the main parameters defining synthetic sources injected into the pre-processed observations (see Fig. \ref{fig:data_pipeline_part1}) at training time. The second line shows common operations performed for both the detection and the characterization steps. The left (respectively, the right) part of the third line is for operations applied solely during the detection (respectively, the characterization) step. Throughout this paper, synthetic training sources injected to build our models are highlighted in orange while (possibly unknown) real and synthetic sources that we aim to detect and to characterize at inference time are displayed in light blue in the schematic representations. Dataset: HIP 72192 (2015-05-10), see Sect. \ref{sec:results} for the recording logs.}
		\label{fig:data_pipeline_part2}
	\end{center}
\end{figure*}

\noindent Throughout the text, the reader can refer to Table \ref{tab:notations}, Figs. \ref{fig:data_pipeline_part1}-\ref{fig:data_pipeline_part2}, and Table \ref{tab:summary_settings}, summarizing respectively the main notations, the processing pipeline of the proposed approach, and the main setting of the different parameters.

\begin{table}
    \centering
    \caption{Summary of the main notations.}
    \label{tab:notations}
    \begin{tabular}{@{}c@{}c@{\hspace*{1ex}}l@{}}
        \toprule     
        \textbf{Not.}\; & \textbf{Range}\; & \textbf{Definition} \\
        \midrule
        \multicolumn{3}{c}{$\blacktriangleright$ \textit{constants}}\\ 
        \midrule
        $N$ & $\mathbb{N}$ & number of pixels in a frame\\
        $M$ & $\mathbb{N}$ & number of pixels in a detection map\\
        $T$ & $\mathbb{N}$ & number of temporal frames\\
        $K$ & $\mathbb{N}$ & number of pixels in a patch (pre-processing)\\
        $J$ & $\mathbb{N}$ & number of pixels in a patch (characterization)\\
        $Q$ & $\mathbb{N}$ & number of samples involved in shrinkage\\
        $P$ & $\mathbb{N}$ & total number of training sources\\
        $S$ & $\mathbb{N}$ & total number of training sets\\
        \midrule
        \multicolumn{3}{c}{$\blacktriangleright$ \textit{indexes}}\\ 
        \midrule
        $n$ & $\llbracket 1; N \rrbracket$ & pixel index\\
        $t$ & $\llbracket 1; T \rrbracket$ & temporal index\\
        $p$ & $\llbracket 1; P \rrbracket$ & source index\\
        $s$ & $\llbracket 1; S \rrbracket$ & training set index\\
        $\phi$ & $\mathbb{R}_+^2$ & 2-D (sub-pixel) angular location of a source\\
        \midrule
        \multicolumn{3}{c}{$\blacktriangleright$ \textit{data quantities}}\\ 
        \midrule
        $\V r$ & $\mathbb{R}^{N \times T}$ & observations\\
        $\V f$ & $\mathbb{R}^{N \times T}$ & nuisance component\\
        $\V h$ & $\mathbb{R}^{N}$ & off-axis PSF\\
        $\widetilde{\V r}$ & $\mathbb{R}^{N \times T}$ & pre-processed observations without injections\\
        $\widebar{\V r}$ & $\mathbb{R}^{N \times T}$ & observations with injections\\
        $\widecheck{\V r}$ & $\mathbb{R}^{N \times T}$ & pre-processed observations with injections\\
        $\widebreve{\V r}$ & $\mathbb{R}^{N \times T}$ & input (images) of the CNN (detection)\\
        $\widebreve{\V p}$ & $\mathbb{R}^{J}$ & input (patch) of the CNN (characterization)\\
        \midrule
        \multicolumn{3}{c}{$\blacktriangleright$ \textit{operators}}\\
        \midrule
        $\M E$ & $\mathbb{R}^{N \times \,.} \mapsto \mathbb{R}^{K \times \,.}$ & patch extractor (pre-processing)\\
        $\M W$ & $\mathbb{R}^{K \times \,.} \mapsto \mathbb{R}^{K \times \,.}$ & centering and whitening (pre-processing)\\
       	$\M D$ & $\mathbb{R}^{\,. \times T} \mapsto \mathbb{R}^{\,. \times T}$ & frame derotator by parallactic angles\\
        \midrule
        \multicolumn{3}{c}{$\blacktriangleright$ \textit{losses and metrics}}\\
        \midrule
        $\mathcal{L}_{\text{detect.}}$ & $\left[ 0; 1 \right]^{2\times M} \mapsto \mathbb{R}_+$ & Dice2 score (detection loss)\\
        $\mathcal{L}_{\text{carac.}}$ & $\mathbb{R}_+^{2} \mapsto \mathbb{R}_+$ & absolute relative error  (characterization loss)\\
        TPR & $\left[ 0; 1 \right]$ & true positive rate (detection metric)\\ 
        FDR & $\left[ 0; 1 \right]$ & false discovery rate (detection metric)\\ 
        F1R & $\left[ 0; 1 \right]$ & harmonic mean of TPR and FDR (detection metric)\\ 
        \midrule
        \multicolumn{3}{c}{$\blacktriangleright$ \textit{estimated quantities}}\\ 
        \midrule
        $\widehat{\V y}$ & $\left[ 0; 1 \right]^M$ & detection map\\
        $\widehat{\alpha}$ & $\mathbb{R}_+$ & photometry (in source to star contrast)\\
        $\widehat{\V m}$ & $\mathbb{R}^{N}$ & temporal mean\\
        $\widehat{\rho}$ & $\left[ 0; 1 \right]$ & shrinkage factor\\
       	$\widehat{\M S}$ & $\mathbb{R}^{K \times K}$ & empirical spatial covariance\\
       	$\widehat{\M C}$ & $\mathbb{R}^{K \times K}$ & shrunk spatial covariance\\
	    $\widehat{\M {L}}$ & $\mathbb{R}^{K \times K}$ & Cholesky's factorization of $\widehat{\M C}$\\
        \bottomrule
    \end{tabular}
\end{table}

\section{Source detection algorithm}
\label{sec:algorithm_detection}

\subsection{\samepage Stastical model of the non-stationary patch covariances}
\label{subsec:algorithm_step1}

\noindent Section \ref{subsubsec:statistical_modeling} recalls for completeness the main ingredients of the statistical model embedded in the \PACO algorithm for ADI observations \citep{flasseur2018exoplanet,flasseur2020robust}. Section \ref{subsubsec:preprocessing_centering_whitening} describes how to use this model to attenuate the strong and spatially non-stationary correlations of the data as well as to improve the contrast. Resulting residuals from this pre-processing step are centered and whitened observations from which our detection and characterization models are built by supervised deep learning, see Sect. \ref{subsec:algorithm_step2}. While obtained through the statistical model of \PACO, these custom pre-processed observations are not directly produced by the \PACO algorithm.

Ablation studies have shown that this pre-processing step is of primary importance: the training step fails to converge in its absence due to the high non-stationarity of the nuisance component and the large fluctuations it displays near the star, see Fig. \ref{fig:data_pipeline_part1}. This effect is due to the fact that standard deep learning architectures assume some degree of invariance, in particular that the data are normalized in a specific range and are corrupted by a stationary noise, see also Sect. \ref{subsec:detection_results_known_real_sources}.

\subsubsection{Statistical model of the nuisance component}
\label{subsubsec:statistical_modeling}

\noindent A dataset $\V r$ in $\mathbb{R}^{N\times T}$ recorded with the ADI technique is formed by $N$-pixel images captured at different times $t$ in $\llbracket 1 ; T \rrbracket$. The direct model for the observed intensity is:
\begin{equation}
	\V r = \V f + \sum\limits_{p=1}^P \alpha_p \, \V h(\phi_p)\,,
	\label{eq:image_formation}
\end{equation} 
where $\V f$ in $\mathbb{R}^{N \times T}$ is the nuisance component, and $\V h \left(\phi_p \right)$ in $\mathbb{R}^{N \times T}$ stands for the contribution of a point source $p \in \llbracket 1 ; P \rrbracket$ with a contrast $\alpha_p$ in $\mathbb{R}_+$ that is assumed constant during the few hours of the total observations. The contribution of a source $p$ takes the form of the off-axis PSF centered at location $\mathcal{F}_t(\phi_p)$ in the $t$-th image,  where $\phi_p$ is its initial location on an image at a reference time $t_{\text{ref}}$ (e.g., $t_{\text{ref}} = t_1$). The function $\mathcal{F}_t$ is a geometric transform (typically in ADI, a circular translation with respect to the star located at the center of the images) modeling the apparent motion of the field of view between the observation configurations at time $t_{\text{ref}}$ and time $t$. The mapping  $\mathcal{F}_t$ is deterministic since it depends solely on the measured parallactic angles. Given that very few sources are expected in the field of view, we assume that the measured intensity is the superimposition of the nuisance component and at most one unresolved point-like source $p$ at each pixel location $n$, i.e., multiple sources do not overlap.

\medskip

\noindent In previous works on the \PACO algorithm \citep{flasseur2018unsupervised, flasseur2018exoplanet,flasseur2018SPIE}, we have proposed to describe the random fluctuations of the nuisance component $\V f$ by a statistical model whose parameters are estimated in a data-driven fashion. We recall hereafter the main ingredients of this statistical model. 

Given the spatial non-stationarity of the nuisance component, the model is built locally at the scale of a patch with an area of a few tens of pixels. It models the distribution of $T$ patches\footnote{For the convenience, we use in the equations a vectorized version of 2-D patches.} $\V f_n = \lbrace \M E_{n,t} \, \V f \rbrace_{t=1:T}$ in $\mathbb{R}^{K \times T}$ extracted around pixel $n$ ($\M E_{n,t}$ denotes the $K$-pixel patch extraction operator at location $n$ and time $t$) with a multi-variate Gaussian $\mathcal{N}(\V{m}_{n},\M{C}_{n})$. The covariance matrix $\M C_n$ is non-diagonal, i.e., it accounts for the local correlations of $\V f$. The sample estimators $\lbrace \widehat{\V m}_n ; \widehat{\M S}_n \rbrace$ of the local mean and covariances coming from the maximum likelihood are the following:
\begin{equation}
	\begin{cases}
		\widehat{\V m}_n = \frac{1}{T} \sum\limits_{t=1}^T \M E_{n,t} \, \V r \, \in \mathbb{R}^K\,, \vspace{0.5mm}\\
		\widehat{\M S}_n = \frac{1}{T} \sum\limits_{t=1}^T (\M E_{n,t}\, \V r - \widehat{\V m}_n) (\M E_{n,t}\, \V r - \widehat{\V m}_n)\T \, \in \mathbb{R}^{K \times K}\,.
	\end{cases}
	\label{eq:sample_estimators}
\end{equation}
Since the number of samples available, i.e. the number $T$ of temporal frames, is typically equivalent or smaller than the number $K$ of pixels in a patch\footnote{The number of pixels in a patch is determined in a data-driven fashion, as described in \cite{flasseur2018exoplanet} for the \PACO algorithm. It corresponds to twice the full width at half maximum (FWHM) of the measured off-axis PSF at the given wavelength. In practice, this empirical rule typically leads to $K$ in $\llbracket 7^2 ; 12^2 \rrbracket$ pixels for the VLT/SPHERE instrument operating at a wavelength $\lambda \in \left[ 0.9 ; 2.2 \right]\, \micro\meter$.}, the sample covariance $\widehat{\M S}_n$ is very noisy and may even be rank deficient. A form of regularization must be enforced to stabilize the estimate and allow the inversion of the covariance matrix involved in the data whitening step (see Sect. \ref{subsubsec:preprocessing_centering_whitening}). We use a \textit{shrinkage} estimator \citep{ledoit2004well,chen2010shrinkage} formed by a convex combination between the low bias/high variance estimator $\widehat{\M S}_n$ and a high bias/low variance estimator $\widehat{\M F}_n$: 
\begin{equation}
	\estim{\M C}_n = (1 - \widehat{\rho}_n)\,\widehat{\M S}_n + \widehat{\rho}_n\,\widehat{\M F}_n\,,
\end{equation}
where $\widehat{\M F}_n$ is a diagonal matrix encoding the sample variances:
\begin{equation}
	\left[ \widehat{\M F}_n \right]_{kk'} =
	\begin{cases}
		\left[ \widehat{\M S}_n \right]_{kk'} ~~~~~ &\text{if} ~~~~~ k = k' \\
		0 ~~~~~ &\text{if} ~~~~~ k \neq k'\,.
	\end{cases}
\end{equation}
The hyper-parameter $\widehat{\rho}_n$ plays a central role since it governs a bias-variance trade-off. In our previous works \citep{flasseur2018exoplanet,flasseur2021rexpaco}, we have derived its closed-form expression, which is an extension of the results of \cite{chen2010shrinkage} in the case of a non-constant shrinkage matrix $\widehat{\M F}_n$:
\begin{equation}
  \estim\rho_n= \frac{
    \Trace\Paren[\big]{\Sest_{n}^2}
    + \Trace^2\Paren[\big]{\Sest_{n}}
    - 2\sum_{k=1}^K\Brack[\big]{\Sest_{n}}_{kk}^2
  }{
    (Q+1)\,\Paren[\Big]{
      \Trace\Paren[\big]{\Sest_{n}^2}
      - \sum_{k=1}^K\Brack[\big]{\Sest_{n}}_{kk}^2
    }
  } \,,
   \label{eq:shrinkfactor}
\end{equation}
where $Q$ is the number of non-null patches involved in the computation of $\widehat{\M S}_n$. Here, $Q$ is equal to $T$ everywhere.

In Appendix \ref{subapp:refinement_statistical_model}, we discuss a refinement of this statistical model to account for the temporal fluctuations of the observations. It leads to a slight improvement  in terms of detection performance at the cost of an increase of the computational burden by a factor 10 to 30. For these reasons, it is not applied by default in the following. We recommend to use it, in a second step, to refine the analysis of ambiguous candidate detections found by the proposed method embedding a multi-variate Gaussian model.

\subsubsection{Centering and local whitening of the observations}
\label{subsubsec:preprocessing_centering_whitening}

\begin{figure}
	\begin{center}
		\includegraphics[width=0.48\textwidth]{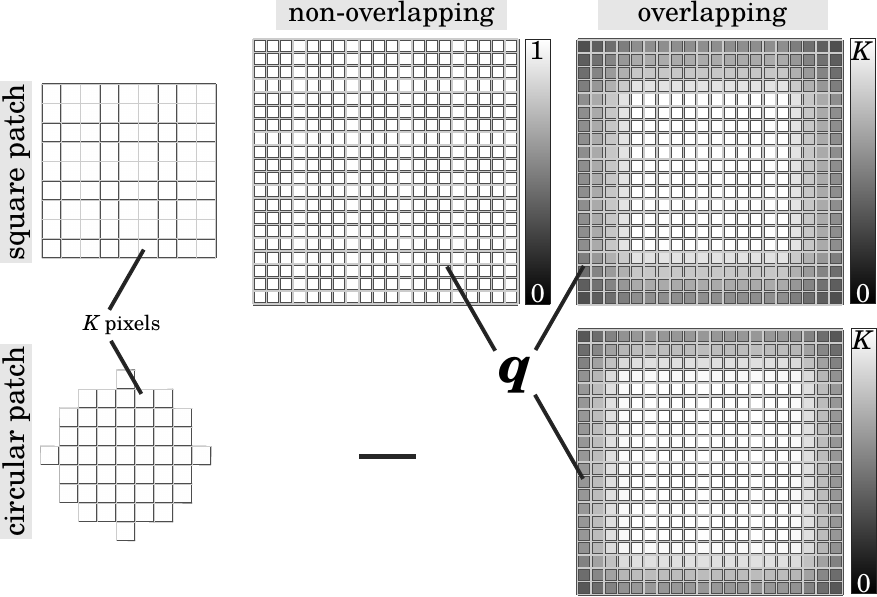}
		\caption{Schematic view of the quantity $\V q$, representing the number of patches contributing to the computation of the pre-processed observations $\widetilde{\V r}$ at each pixel of the field of view, as a function of the patch shape and of the tessellation of the field of view. Non-overlapping circular patches are not considered in this work since they do not allow a complete paving of the field of view. The spatial scale is not respected for the purpose of illustration.}
		\label{fig:patches_qn}
	\end{center}
\end{figure}

\noindent We consider a set of locations $\mathbb{P}$ where the statistics of the nuisance component should be computed. The cardinal of $\mathbb{P}$ depends solely on the patch shape and on the \textit{patch stride}\footnote{We define the patch stride as the distance (in pixels) between the centers of two adjacent patches, both in the $x$ and $y$ directions.} used to cover the whole field of view. 
 For a given patch stride, we first define the number of (centered and whitened) patches averaged at each location $n'$ of the field of view:
\begin{equation}
	q_{n'} = \sum\limits_{n \in \mathbb{P}} \delta_{\M 1_{n} \, \M 1_{n'}\T \, \neq \, 0}\,, \forall n' \in \llbracket 1 ; N \rrbracket \,,
	\label{eq:nb_patches_general_case}
\end{equation}
where $\delta$ is the indicator function (i.e., $\delta_{x=y}$ is either equal to 1 if the condition $x=y$ is met, and equal to 0 otherwise), and $\M 1_n \in \mathbb{R}^{K}$ (respectively, $\M 1_{n'} \in \mathbb{R}^{K}$) is the vectorization of a 1-valued patch centered at location $n$ (respectively, $n'$). Figure \ref{fig:patches_qn} gives a view of the quantity $\V q \in \mathbb{N}^N$ in the general case, i.e. for either square and circular patches as well as non-overlapping and overlapping patches. By default, we consider square patches of $K$ pixels area. 
The pre-processed images $\widetilde{\V r}$ in $\mathbb{R}^{N \times T}$, after centering and whitening, are obtained by:
\begin{align}
	\widetilde{r}_{n'} &= \frac{\left[ \sum\limits_{n \in \mathbb{P}} \M E_n\T \, \M W_n \, \M E_n \, \V r \right]_{n'}}{q_{n'}} \,, \nonumber \\&= \frac{\left[ \sum\limits_{n \in \mathbb{P}} \M E_n\T \, \widehat{\M{L}}_n\T \, (\V r_n - \widehat{\V m}_n) \right]_{n'}}{q_{n'}}\,, \forall n' \in \llbracket 1 ; N \rrbracket \,,
	\label{eq:whitening_general_case}
\end{align}
where $\M W_n$ is an operator performing centering and whitening of the collection of patches $\V r_n$  $\mathbb{R}^{K \times T}$ at location $n$, such that $\widehat{\M{L}}_n$ is the Cholesky's factorization of $\widehat{\M C}_n^{-1}$ (i.e., $\widehat{\M{L}}_n \, \widehat{\M{L}}_n\T = \widehat{\M C}_n^{-1}$). In the specific case (considered by default in this paper) of non-overlapping square patches of $K$ pixels, $\text{card}(\mathbb{P}) = \lfloor N / K \rceil$, and Eqs. (\ref{eq:nb_patches_general_case})-(\ref{eq:whitening_general_case}) simplify as:
\begin{equation}
	\begin{cases}
		q_n' = 1 \,, \forall n' \in \llbracket 1 ; N \rrbracket\,,\\
		\widetilde{\V r}_n = \M W_n \V r_n = \widehat{\M{L}}_n\T \, \left( \V r_n - \widehat{\V m}_n \right) \,, \forall n \in \mathbb{P} \,.
	\end{cases}
	\label{eq:whitening_non_overlapping_patches}
\end{equation}

In Appendix \ref{subapp:refinement_whitening_observations}, we discuss a refinement of the intermediate quantity $\widetilde{\V r}$ produced by the pre-processing step under the same statistical model as described in Sect. \ref{subsubsec:statistical_modeling} or in Appendix \ref{subapp:refinement_statistical_model}. It can be noted that overlapping patches should be used with this refinement, and that the patch shape can be either squared or circular. This alternative approach leads to a better stability and robustness of the method for datasets recorded under bad observing conditions\footnote{Since the notion of \textit{observing conditions} is relative and can be characterized by several metrics (e.g., Strehl ratio, coherence time, air mass, etc.), we did not find quantitative values for these measures indicating the strict use of the variant version described in Appendix \ref{subapp:refinement_whitening_observations}. This large effort would require the processing of hundreds of datasets, which is left for future work. Qualitatively, we observed that datasets impacted by a bright wind-driven halo, displaying the same apparent motion than the objects of interest, are more subject to lead to an increased false alarm rate without the pre-processing described in Appendix \ref{subapp:refinement_whitening_observations}.}. However, the computational burden of this variant is increased by a factor $2 \times K$, typically lying in $\llbracket 100 ; 250 \rrbracket$ for the VLT/SPHERE instrument. For these reasons, this variant is not applied by default in the following. We recommend to apply it, in a second step, when the training, validation, or inference results are clearly impacted by a significant number of false alarms, much higher than expected, which is an unambiguous sign of the limited fidelity to the observations of the pre-processing procedure used by default. 

\subsection{Exoplanet detection by supervised deep learning}
\label{subsec:algorithm_step2}

\noindent We formalize the detection problem as a supervised pixel-wise classification task: 
starting from a temporal series of pre-processed images including synthetic sources, the goal is to infer a detection map $\widehat{\V y}$ in $\left[0 ; 1 \right]^{M}$, where each pixel value represents a score between 0 and 1 such that a high (respectively, a low) score values the presence (respectively, the absence) of a source centered at that location. Interpreting this score as a true probability of presence of a source requires a control of the uncertainties with dedicated methods (see e.g., \cite{gawlikowski2021survey,hullermeier2021aleatoric} for recent review papers) that is left for future work. For this reason, in the following, we refer to this score as a \textit{pseudo-probability}. Besides, the $M$-pixel detection map is larger than a $N$-pixel single temporal frame of the pre-processed data. Due to the apparent rotation induced by ADI, it is theoretically possible to detect a source lying in the sensor field of view at a single date, see Figs. \ref{fig:data_pipeline_part2}(c) and \ref{fig:HD95086_20150505_snr_prediction_maps_comparisons_natural_outputs_full_1}. By derotating each individual temporal frame with the corresponding parallactic angle and combining the resulting transformed measurements, an area larger than $N$ pixels can be scrutinized. Its resulting spatial extent depends solely on the total amount of parallactic rotation between the first and the last frame.

Section \ref{subsubsec:generation_training_samples} details the construction process of the training set, Sect. \ref{subsubsec:model_architecture} describes the selected model architecture, and Sect. \ref{subsubsec:loss_metrics} discusses the metrics we consider to evaluate the performance of the proposed method.

\subsubsection{Construction of the training basis}
\label{subsubsec:generation_training_samples}

\noindent In high-contrast imaging, obtaining real ground-truth data is a twofold challenge.

First, the overall number of positive samples is limited as relatively few point-like sources have been confirmed to date. Second, negative samples are hard to define since some undiscovered sources might be present in the observed data. To overcome these difficulties, we adopt the following training strategy: the training set consists of $S$ pairs $\lbrace {\widebreve{\V r}^{[s]} ; \V y^{[s]}} \rbrace_{s=1:S}$ of samples resulting from the massive injection of synthetic point-like sources. 
In this framework, $\widebreve{\V r}^{[s]} \in \mathbb{R}^{N \times T}$ represents observations, with injected synthetic sources, that have been pre-processed. The quantity $\V y^{[s]} \in \llbracket 0, 1 \rrbracket^M$ is the ground-truth map pointing the injection locations of any synthetic training source falling within the field of view at least in one temporal frame. The ground truth map is built for a given (and arbitrary) orientation of the field of view, e.g. aligned with the true North.
The implemented simulation process is quite realistic since the injected source signature corresponds to the off-axis PSF of the target star usually measured just before or just after the main sequence of observations by decentering the coronagraph. 

Second, the nuisance component varies drastically from one observation to the other, as it is highly dependent on the observing conditions, the magnitude of the star, and the instrument settings. As a consequence, we follow an observation-dependent approach, and train a different model on each observation. It means that the model parameters (except algorithmic and optimization hyper-parameters, see Sect. \ref{subsubsec:implementation_details_step2}) are optimized from scratch for each dataset.	

This setup implies the design of a custom data-augmentation strategy (i) to prevent over-fitting of the model that is trained from a unique temporal series of images, and (ii) to account for our lack of knowledge about real sources \textit{--unknown at training time but that we aim to detect at test time--}. To circumvent these issues, we apply a random permutation of the $T$ images forming the observations $\V r$ for each new training sample $s \in \llbracket 1 ; S \rrbracket$. This operation allows us (i) to create artificially different training sets, and (ii) to break the temporal consistency of (known and unknown) real sources. Synthetic sources are then injected in the temporally permuted data using the parallactic angles and the best fit of the off-axis PSF by a Gaussian and an Airy pattern. Besides, the off-axis PSF is assumed to be time-invariant. We have checked numerically that this assumption is reasonable for our classification task. Similarly, assuming a slightly different pattern for the off-axis PSF (e.g., measured \textit{versus} fitted model) between the data generation process and the training step does not lead to a significant drop in the detection performance. At this intermediate stage, each training sample $\widebar{\V r}^{\left[ s \right]}$ is obtained by:
\begin{equation} 
	\widebar{\V r}^{\left[ s \right]} = \M P^{\left[ s \right]} \, \V r + \sum\limits_{p=1}^{P^{\left[ s \right]}} \alpha_p^{\left[ s \right]} \, \V h \left(\phi_p^{\left[ s \right]}\right)\,,
\end{equation}
where $\M P$ is an operator performing the random temporal permutation of the images of $\V r$ and $\V h \left( \phi_p \right)$ in $\mathbb{R}^{N\times T}$ represents the spatio-temporal contribution of a synthetic source centered at location $\phi_p$ on a reference image at time $t_{\text{ref}}$, see Sect. \ref{eq:image_formation}. The number of sources $P^{\left[ s \right]}$, their contrasts $\lbrace \alpha_p^{\left[ s \right]} \rbrace_{p=1:P^{\left[ s \right]}}$ and their initial locations $\lbrace \phi_p^{\left[ s \right]} \rbrace_{p=1:P^{\left[ s \right]}}$ are free parameters. In practice, the number $P^{\left[ s \right]}$ of injected sources in each training sample is drawn uniformly in $\llbracket 1 ; 10\rrbracket$. This setting represents a realistic scenario since we expect a few faint point-like sources in the field of view. We denote by $P$ the total number of injected sources over the $S$ training sets, i.e. $P = \sum_{s=1}^S P^{\left[ s \right]}$. The initial locations $\lbrace \phi_p \rbrace_{p=1:P}$ of the injected sources are drawn uniformly 
in polar coordinates (i.e., the center of the field of view is more sampled than farther away). Note that we have compared this setting with a uniform sampling in cartesian coordinates (i.e., uniform density over the field of view), and we found very similar detection performance for both settings. The selected one (i.e., uniform in polar coordinates) slightly speeds up the training procedure, likely because the pre-processed observations fluctuate slightly more near the star than farther away, thus requiring more training samples to discriminate a source from the nuisance. The range of injected flux is also a critical choice. For instance, if the lower bound is too low, class overlap can occur and the model is not able to discriminate between sources and the nuisance component leading to a high level of false alarms. In the opposite case, if the upper bound is too low, evident bright sources will not be detected since there are no similar examples in the training set. In practice, we set the injection range in an unsupervised fashion. We train our model on sources which are challenging to detect with other methods: the contrast $\lbrace \alpha_p \rbrace_{p=1:P}$ of the injected sources is drawn uniformly in $\left[ 3\widehat{\sigma}_{\phi_p}^{\texttt{PACO}} ; 12\widehat{\sigma}_{\phi_p}^{\texttt{PACO}} \right]$ where $\widehat{\sigma}_{\phi_p}^{\texttt{PACO}}$ is the 1-sigma contrast reached by \PACO at location $\phi_p$. This setting covers both sources that are detectable above the standard $5\sigma$ detection confidence and sources whose detection confidence remains below the $5\sigma$ detection limit reached by \PACO. In practice, we found that this setting is suitable to detect both faint sources and bright sources without generating large number of false alarms.  

\begin{figure}
	\begin{center}
		\includegraphics[width=0.5\textwidth]{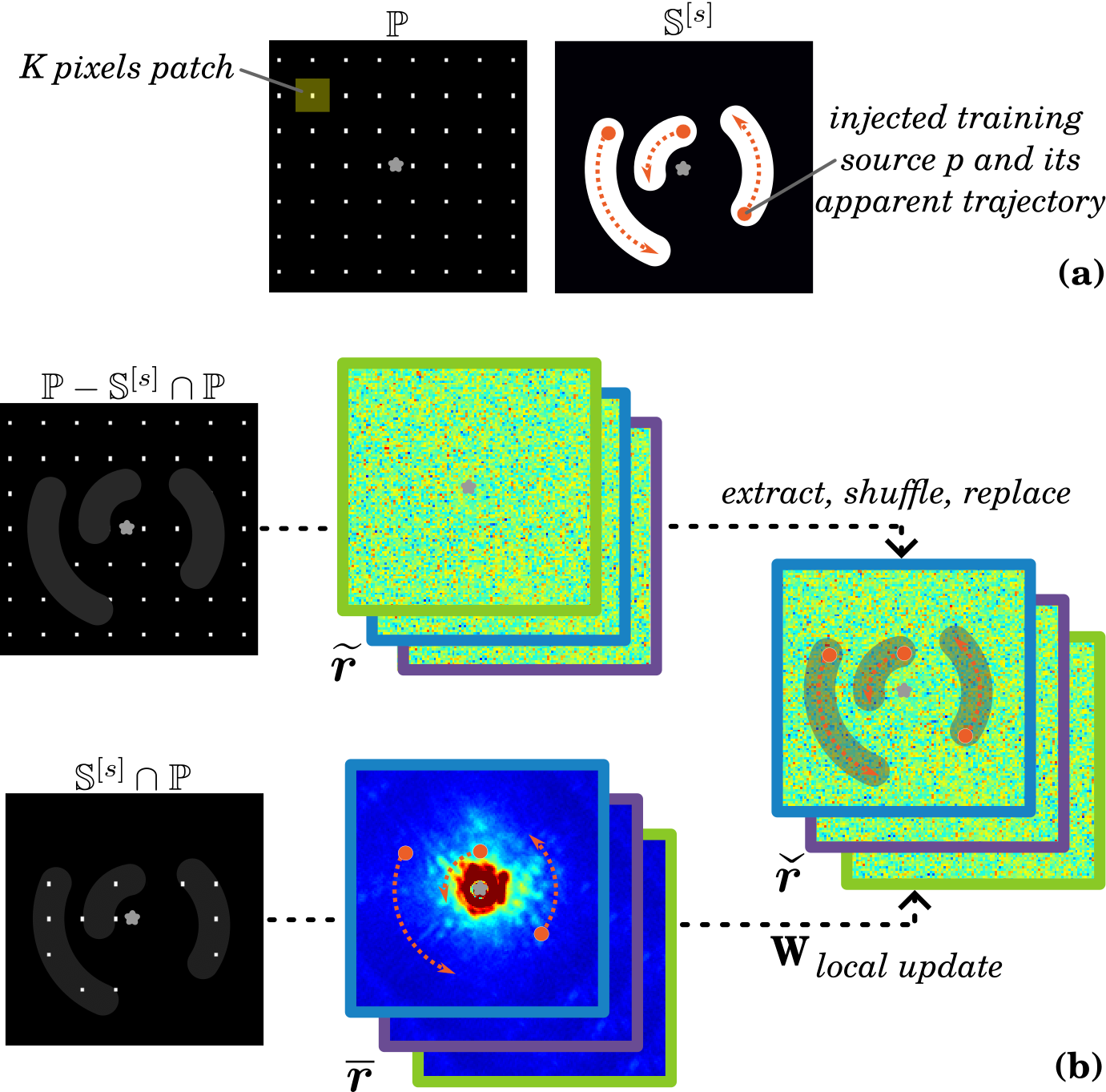}
		\caption{Schematic illustration of the local update of the pre-processing embedded in the training step of the proposed detection algorithm. (a) Illustration of the sets $\mathbb{P}$ and $\mathbb{S}^{[s]}$ for a given training set $s$ with three injected training sources displayed in orange. (b) Illustration of the computation of the pre-processed observations $\widecheck{\V r}$ in the presence of injected training sources from pre-computed (i.e., before injections) centered and whitened observations $\widetilde{\V r}$. The spatial scale is not respected for the purpose of illustration.}
		\label{fig:update_preprocessing}
	\end{center}
\end{figure}

As the pre-processing is an expensive procedure and becomes the bottleneck during online data generation, we adopt a local update strategy to reduce its computational cost. Prior to the injection of synthetic sources, the whole dataset is pre-processed, i.e. centered and spatially whitened, see Fig. \ref{fig:data_pipeline_part1} (b). We denote by $\widetilde{\V r}$ the pre-computed cube. After each batch $s$ of injections, the set $\mathbb{S}^{[s]}$ of locations impacted by the signal of the $P^{[s]}$ sources is determined. Outside $\mathbb{S}^{[s]}$, the pre-processed images are obtained from the temporal permutation of $\widetilde{\V r}$. Inside $\mathbb{S}^{[s]}$, the statistics of the nuisance component are updated given the contamination of the $P^{[s]}$ injected sources, and the pre-processed images are updated with these refined statistics.
At this intermediate stage, each training sample $\widecheck{\V r}^{[s]}$ is obtained by:
\begin{equation}
	\widecheck{\V r}_n^{[s]} = \begin{cases}
  \M W_n \, \widebar{\V r}_n^{[s]} , & \text{for } n \in \mathbb{S}^{[s]}\, \cap\, \mathbb{P}\,, \\
	\M P^{\left[ s \right]} \, \widetilde{\V r}_n, & \text{for } n \in \mathbb{P} - \mathbb{S}^{[s]}\, \cap\, \mathbb{P}\,.
\end{cases}
\end{equation}
This dual strategy, illustrated by Fig. \ref{fig:update_preprocessing}, is applied to prevent any detection bias (i.e., an over-estimation of the actual detection performance of the proposed algorithm) since we have shown in previous work on the \PACO algorithm \citep{flasseur2018exoplanet} that the statistics of the nuisance component can suffer from a (slight) bias, in particular at short angular separations and/or when the total amount of parallactic rotation is low. This slight potential bias is due to the contamination of a source whose signal is partly encoded both in the mean and in the spatial covariances of the nuisance component. 

Finally, the intermediate images of each training sample are derotated with the opposite of the parallactic angles so that signal of the synthetic sources are spatially co-aligned along the temporal axis:
\begin{equation}
	\widebreve{\V r}^{[s]} = \M D \, \widecheck{\V r}^{[s]}\,,
\end{equation} 
where $\M D$ is a derotation operator. 
This derotation step is mandatory to perform a semantic segmentation with the CNN we consider (see Sect. \ref{subsubsec:model_architecture}) given the limited spatial extent of its receptive field.

The binary ground-truth segmentation map $\V y^{[s]}$ is obtained by setting to \textit{1} circular areas centered at the locations $\lbrace \phi_p \rbrace_{p=1:P^{[s]}}$ of the $P^{[s]}$ injected sources. Other regions of $\V y^{[s]}$ are set to \textit{0}. The radius of the circles is set to the full width at half maximum of the off-axis PSF, which corresponds to the expected spatial extent of an exoplanetary signature in the data.

A schematic summary of the construction of the training set is given in Fig. \ref{fig:data_pipeline_part2}. 

\subsubsection{Model and architecture}
\label{subsubsec:model_architecture}

\noindent Deep convolutional neural networks reach state-of-art performance in solving pixel-wise classification tasks for various imaging fields including microscopy, astronomy, medical imaging or remote sensing. A large variety of model architecture has been studied in the literature (e.g., auto-encoder \citep{badrinarayanan2017segnet}, VGG \citep{simonyan15}, ResNet \citep{he2016deep}) and their performance often rely on an intricate trade-off between model complexity, the amount of data available for training, and their fidelity with the data used at test time. A  common feature of some classical deep architectures is to encode the diversity of the training samples in a low dimensional subspace by transforming the network input with a cascade of convolution and downsampling operations. Starting from this latent representation, the initial image size is retrieved by a decoder performing reverse transformations with a cascade of deconvolution and upsampling operations. 

We also based our model on the above-mentioned category of architectures. We chose a U-Net \citep{ronneberger2015u}  with a ResNet18 \citep{he2016deep} as encoder backbone ($\simeq$11 millions of free parameters), which is an architecture widely used for image segmentation. Its residual connections preserve of the input's spatial information along the cascade of convolution and downsampling operations thanks to a direct mapping of the output of each layer of the compression arm into the corresponding layer of the decompression arm. We use the architecture implemented in the \texttt{SMP} package\footnote{The SMP package containing the network architecture used in this paper is available at \href{https://github.com/qubvel/segmentation\_models.pytorch}{https://github.com/qubvel/segmentation\_models.pytorch}.}. The encoder and decoder parts are formed by four blocks, each one is composed by a series of convolution layers, batch normalization layers, rectified linear unit (ReLU) activation, and max pooling layers. The final layer of the network has a sigmoid activation function to produce a detection map $\widehat{\V y} \in \left[0 ; 1\right]^M$. The detailed description of the architecture, the number of parameters per layer, and the input / output shapes of each layer can be found at the above mentioned link. 

The network weights are trained from scratch with the samples generated with the procedure described in Sect. \ref{subsubsec:generation_training_samples}. Initial weights are drawn uniformly through a He-Kaiming distribution \citep{he2015delving}. The \texttt{SMP} package also provides pre-trained weights. Pre-training is performed with the ImageNet dataset (RGB conventional images) either in a supervised, semi-supervised, or weakly-supervised learning fashion \citep{yalniz2019billion}. In case of pre-training, the weights of the first convolutional layer are replicated in order to match the $T$-depth of our inputs. We compared all of these strategies against a supervised learning from scratch with our custom training set (see Sect. \ref{subsubsec:generation_training_samples}). We found similar performance with all approaches, and opted for an end-to-end learning.

\subsubsection{Loss function and accuracy metrics}
\label{subsubsec:loss_metrics}

\noindent The design of loss function used for optimizing the network weights at training time is driven by three criteria: (i) handling with the strong class imbalance (the number of background pixels being much larger than the number of pixels from the sources), (ii) being computationally efficient, (iii) matching the astrophysical goals (i.e., having a measure close to a detection accuracy score). We compare losses classically used for semantic segmentation, such as the binary cross-entropy (BCE), $\ell_1$ and $\ell_2$ norms, mean square error and Hinge loss. We have also compared losses based on a similarity measure such as the Dice score \citep{milletari2016v} and hybrid losses combining at least two individual loss measurements (e.g., BCE with Dice score). Our experiments have consistently shown better performance with Dice-based scores. We selected the Dice2 loss (the \textit{2} means for \textit{two} classes), first introduced for biomedical imaging segmentation with very unbalanced classes \citep{sudre2017generalised,wang2020improved}. Given a training set of ground-truth and predicted detection maps $\lbrace  \V y^{[s]} ; \widehat{\V y}^{[s]} \rbrace$, the Dice2 score is defined by:
\begin{align}
	\mathcal{L}_{\text{detect.}}\left(\V y^{[s]}, \widehat{\V y}^{[s]}\right) \, = \, &1 - \underbrace{\frac{\sum\limits_{m=1}^M (1-\V y_m^{[s]})(1-\widehat{\V y}_m^{[s]} + \epsilon)}{\sum\limits_{m=1}^M 2 - \V y_m^{[s]} - \widehat{\V y}_m^{[s]} + \epsilon}}_{\text{background error}}\, \nonumber \vspace{1mm}\\ &- \,\underbrace{\frac{\sum\limits_{m=1}^M \V y_m^{[s]} \, \widehat{\V y}_m^{[s]} + \epsilon}{\sum\limits_{m=1}^M \V y_m^{[s]} + \widehat{\V y}_m^{[s]} + \epsilon}}_{\text{source error}}\,,
	\label{eq:loss}
\end{align}
where $\epsilon$ is a minimum-value smoothing and stability parameter added to avoid division by zero. Targeted loss property (i) is satisfied since errors in the source and background areas are penalized equally regardless the relative occurrence of these two classes in $\V y^{[s]}$. Property (ii) is also satisfied, and we illustrate numerically in the two following paragraphs that property (iii) is also reached.

\begin{figure*}
	\begin{center}
		\includegraphics[width=\textwidth]{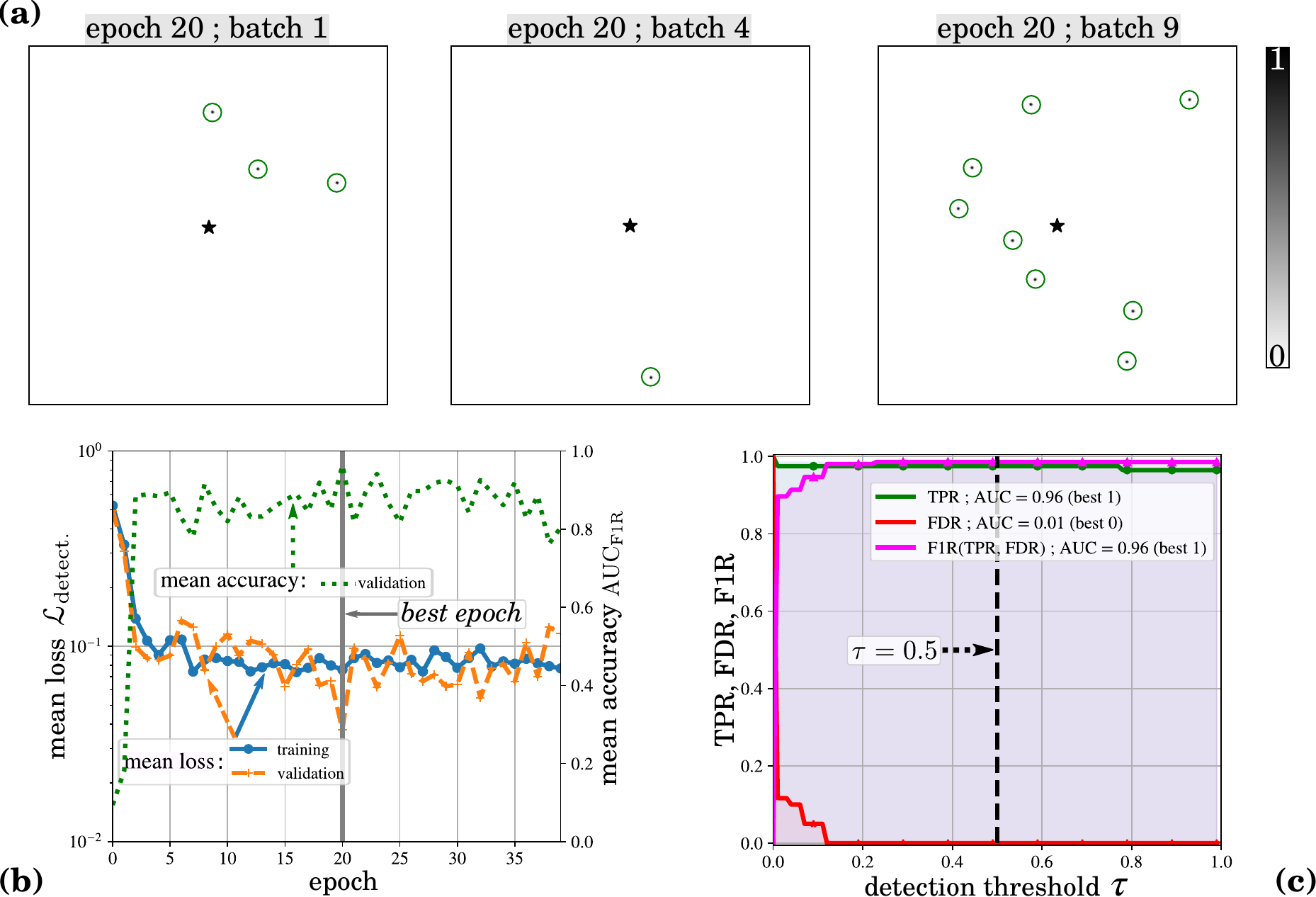}
		\caption{Example of training and validation results. (a) Examples of detection maps obtained at validation time for the best epoch (number 20). (b) Evolution of the loss function at training and validation time, as well as the evolution of the F1R accuracy metric at validation time. (c)  ROCs representing the TPR, FDR and F1R as a function of the prescribed detection threshold $\tau$ for the best epoch (number 20, symbolized by the gray vertical  line in panel (b)). Dataset: HD 95086 (2015-05-05), see Sect. \ref{sec:results} for the recording logs.}
		\label{fig:training}
	\end{center}
\end{figure*}

At validation time, we evaluate the ability of the model to detect point-like sources while simultaneously avoiding false alarms at best as possible. In other words, we aim to obtain a model obeying a precision-recall trade-off. For a predicted detection map $\widehat{\V y}^{[s]}$ in $\left[ 0 ; 1 \right]^M$ thresholded at $\tau$ in $\left[0 ; 1\right]$, we count the number of true positives (TP, i.e., true detections), false positives (FP, i.e., false alarms) and false negatives (FN, i.e., missed detections). Following standard practice in direct imaging (see e.g., \cite{flasseur2018exoplanet,gonzalez2018supervised,cantalloube2020exoplanet}), detections are treated as blobs of one resolution element radius which corresponds to the expected spatial extent of an exoplanetary signature in the data. From TP, FP, and FN, we derive the true positive rate (TPR, i.e., the recall), the false discovery rate (FDR, i.e., the precision), and the F1R score, which is the harmonic mean between TPR and FDR, that we use as an overall measure of the precision-recall trade-off:
\begin{equation}
	\begin{cases}
		\text{TPR} = \frac{\text{TP}}{\text{TP} + \text{FN}} \in [0;1]\vspace{1mm}\,,\\ \text{FDR} = \frac{\text{FP}}{\text{FP} + \text{TP}}\in [0;1]\vspace{1mm}\,,\\ \text{F1R} = \frac{2}{\frac{1}{\text{TPR}} + \frac{1}{\text{FDR}}} = \frac{2\text{TP}}{2\text{TP} + \text{FN}+\text{FP}}\in [0;1]\,.
	\end{cases}	
	\label{eq:detection_metrics}
\end{equation}
From TPR, FDR, and F1R, receiver operating curves (ROCs; \cite{kay1993fundamentals}) are built. ROCs are obtained by evaluating the figures of merit defined in Eq.~(\ref{eq:detection_metrics}) as a function of the detection threshold $\tau$. Finally, the area under the curve (AUC) for the F1R score is computed as an aggregate score of the model performance (best when close to 1). \cite{gonzalez2016low,gonzalez2018supervised,flasseur2018exoplanet,dahlqvist2020regime,cantalloube2020exoplanet,daglayan2022likelihood} exemplified the relevance of ROCs in high-contrast imaging to derive an aggregate measurement of the targeted precision-recall trade-off.

Figure \ref{fig:training}(a) displays some examples of detection maps obtained at validation time for the best validation epoch. These maps illustrate qualitatively the ability of our model to detect synthetic sources while simultaneously avoiding false alarms. Figure \ref{fig:training}(b) shows the evolution of the empirical risk (see Eq. (\ref{eq:loss})) at training and validation time as well as the evolution of the F1R accuracy metric (see Eq. (\ref{eq:detection_metrics})) at validation time. The loss function does not exhibit significant discrepancy between training and validation steps and the convergence is reached in a few epochs\footnote{\samepage In machine or deep learning, an epoch refers to a group of multiple training sets from which the network weights are optimized sequentially by stochastic gradient descent. Multiple epochs, formed by a random selection and ordering of some training sets taken from the training base, are generally needed to reach convergence of the network weights. Learning and optimization hyper-parameters can also be tuned between two consecutive epochs according to a pre-defined scheduling, see Sect. \ref{subsubsec:implementation_details_step2}.}. Besides, the accuracy score is high and well anti-correlated with the loss. This latter observation illustrates that the loss function is a satisfactory estimate of the overall accuracy metric (see targeted property (iii) at the beginning of Sect. \ref{subsubsec:loss_metrics}). Finally, Fig. \ref{fig:training}(c) gives an illustration of validation ROC obtained for the best epoch (symbolized by a gray vertical line in Fig. \ref{fig:training}(b)). The validation ROC confirms the good ability of the trained model to discriminate (synthetic) point-like sources from the nuisance component.  

\subsubsection{Implementation details}
\label{subsubsec:implementation_details_step2}

\noindent Pairs of samples $\lbrace \widebreve{\V r}^{[s]} ; \V y^{[s]} \rbrace$ are generated on the fly at training and evaluation time following the procedure described in Sect. \ref{subsubsec:generation_training_samples}. To avoid over-fitting, each realization $s$ is unique with no repetition for the different epochs. The notion of \textit{epoch} is used only as a way to evaluate regularly the performance of the model with the validation procedure described in Sect. \ref{subsubsec:loss_metrics}, and also to adapt the learning rate through a pre-defined scheduling. The optimization of the network weights is performed with an iterative stochastic gradient-descent strategy on mini-batches of samples formed (possibly) by the concatenation of multiple training sets. It means that, at each iteration, the model weights are updated in the opposite direction to the gradient of the loss. The loss is evaluated from the current mini-batch of samples with respect to the model weights. Since we work on series of $T$ images, with $T$ typically lying between 50 and 100 images, the batch-size (i.e., the number of training sets comprised within a mini-batch) is fixed at 1 given memory constraints. This setting does not affect the overall performance of the method and only requires to perform more iterations to reach convergence since the cost function is quite noisy, see Fig. \ref{fig:training}(a). Even under this setting, the convergence is typically reached in a few epochs, see Sect. \ref{subsubsec:loss_metrics} and Fig. \ref{fig:training}(b). In practice, for each training epoch, $S=100$ pairs of samples $\lbrace \widebreve{\V r}^{[s]} ; \V y^{[s]} \rbrace$ are generated and fed sequentially as input of the network. For each validation epoch, $S$ is fixed at 10 given the computational burden required to build ROCs representing the F1R score as a function of the detection threshold $\tau$. The training process stops when the accuracy metric $\text{AUC}_\text{F1R}$ (i.e., AUC under ROC representing the F1R score as a function of the detection threshold $\tau$ at validation time) evolves in less than 2\% during the ten previous epochs. The model optimization is performed with the adaptive gradient algorithm AMSGrad \citep{reddi2019convergence} which is a variant of the Adam \citep{kingma2014adam} optimizer with a longer-term memory of past gradients. The parameters of the optimizer and of the scheduler have been fine tuned on two datasets and are kept constant for all our experiments. In practice, we observed that the optimized values are quite robust with respect to the dataset diversity. The weight decay\footnote {In machine or deep learning, the weight decay refers to a regularization technique reducing the complexity of a model to prevent over-fitting.} is fixed at $10^{-5}$ and the initial learning rate is set to $10^{-3}$ with a regular decrease by 10\% every 10 epochs. The optimization of the network weights is performed with the high-performance deep learning library PyTorch \citep{NEURIPS2019_9015} on GPUs server with NVIDIA system equipped with either Tesla V100 or GTX 1080 Ti cards. The pre-processing step is highly parallelized and has a double implementation so that it can performed either on CPUs or on GPUs depending on the number of available CPUs cores and on the server specifications.

\section{Source characterization algorithm}
\label{sec:algorithm_characterization}

Once sources have been detected, they can be characterized in terms of astrometry and of photometry. In this section, we present a new method based on supervised deep learning to estimate the photometry of detected point-like sources. 
The sub-pixel estimation of the astrometry is not addressed in this paper because it requires a sub-pixel estimation of the detection criterion as well as statistical guarantees on its significance. These specific developments are left for future work, and the proposed characterization algorithm can be use to estimate the photometry at the pixel barycenter of (candidate) sources revealed by the detection stage presented in Sect. \ref{sec:algorithm_detection}. As in Sect. \ref{sec:algorithm_detection}, we successively discuss the pre-processing stage, the formalization of the problem as a regression task, the model and the underlying architecture, the metric we use for training and evaluation, and some implementation details.

\subsection{Pre-processing aspects}
\label{subsec:preprocessing_characterization}

\begin{figure}
	\begin{center}
		\includegraphics[width=0.49\textwidth]{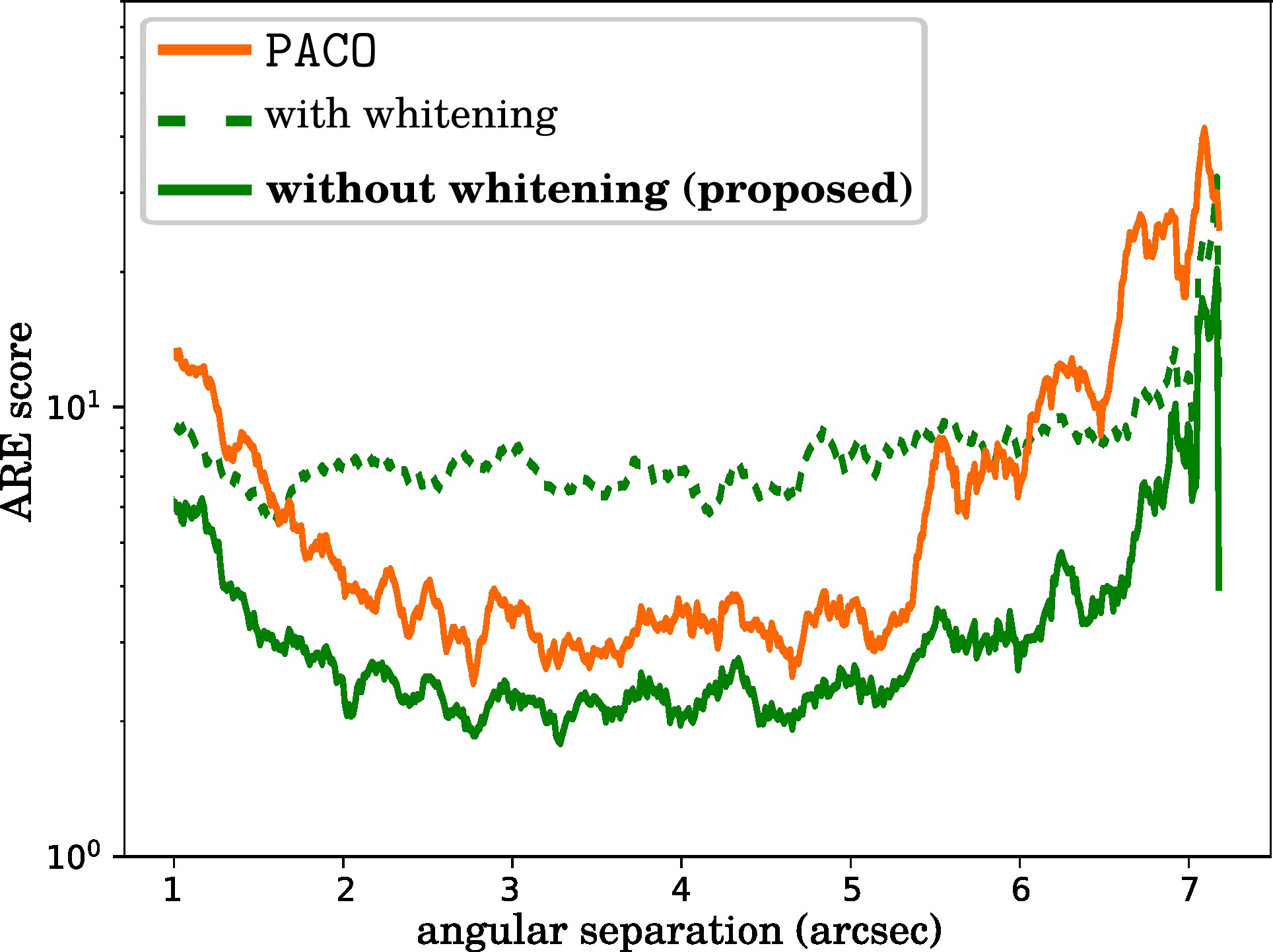}
		\caption{Influence of the whitening of the spatial correlations during the pre-processing step of the characterization algorithm. ARE (see Sect. \ref{ref:subsubsec:loss_metrics_characterization} for the definition of the metric) on the estimated photometry is reported as a function of the angular separation, with and without whitening of the spatial correlations. The performance of \PACO are displayed as a purpose of comparison. The results are averaged azimuthally for 40,000 sources of flux drawn uniformly between $1 \times 10^{-6}$ and  $3 \times 10^{-5}$. The known real sources were masked out and were not considered. Dataset: HD 95086 (2015-05-05), see Sect. \ref{sec:results} for the recording logs.}
		\label{fig:whitening}
	\end{center}
\end{figure}

We adopt a simple patch-based approach, in which we predict the flux of a putative source from a unique (reduced) patch centered on the approximate location of the source. We propose to parameterize the mapping between the input patch and the flux with a CNN trained in a supervised fashion, from the dataset of interest (i.e., the underlying model is data-dependent, as for the detection stage of this paper). 

The dataset is first reduced to a single frame, from which the input patches are extracted. This pre-processing consists in three steps. First, the temporal mean is computed and subtracted pixel-wise in order to remove most of the quasi-static speckles. Second, the dataset is derotated by the opposite of the parallactic angles to co-align the signal of the sources. Third, the dataset is averaged temporally, resulting in a single averaged frame. 
This last step allows to obtain an efficient training procedure as it reduces the size of the input data by a factor $T$. Besides, we observed empirically that this step is beneficial to improve the estimation accuracy as it acts as a simple denoiser: the source signal is constant along time, while residual speckles are not quasi-static after cube derotation, thus canceling out. Keeping the notation introduced in Sect. \ref{sec:algorithm_detection}, these operations transform a given (intermediate) training dataset $\widebar{\V r}$ in $\mathbb{R}^{N \times T}$ with injected synthetic sources as follows:
\begin{equation}
	\begin{cases}
		\widecheck{\V r}_n = \widebar{\V r}_n - \widehat{\V m}_n \,, \forall n \in \mathbb{P}\,, \text{(step 1)} \,,\\
		\widebreve{\V r} = \frac{1}{T}\sum\limits_{t=1}^T \left[ \M D \, \widecheck{\V r} \right]_t\,, \text{(steps 2 and 3)} \,.
	\end{cases}
\end{equation}
In this framework, and unlike the detection stage, we do not apply a whitening of the spatial correlations at step 1. 
Indeed, we observed empirically that whitening the dataset between steps 1 and 2-3 degrades the performance of our model, as illustrated by Fig. \ref{fig:whitening}. More quantitatively, the absolute error of estimation is increased, whatever the angular separation, by a factor between three and five. Besides, keeping a whitening step for photometry estimation does not allow to obtain better results than \PACO for most of the field of view. 
This is expected as the whitening  distorts the shape and the norm of the source signal, thus hampering the recovery of its flux.
The detection algorithm is not subject to this constraint as its task is to determine whether a source is present or not, regardless of its flux. This fact illustrates that deriving a quantitative result (as a flux estimate) is a more complex task than providing a qualitative result (as related to the presence or to the absence of a source) with our algorithmic setting. We can expect that building the model from several datasets of observations (instead of a single one in this work) would relax these constraints. These specific developments are left for future work, see also Sect. \ref{sec:conclusion} for a discussion. 
After pre-processing, square patches $\widebreve{\V p}^{[p]} \in \mathbb{R}^J$ are finally extracted around the location of each injected synthetic source $p$ during the training and validation steps, or around each (candidate) real point-like source $p$ at inference time. The patch size $J$ is an hyper-parameter whose setting is discussed in more details in Sect. \ref{ref:subsubsec:implementation_details_characterization}.  

\begin{figure}
	\begin{center}
		\includegraphics[width=0.48\textwidth]{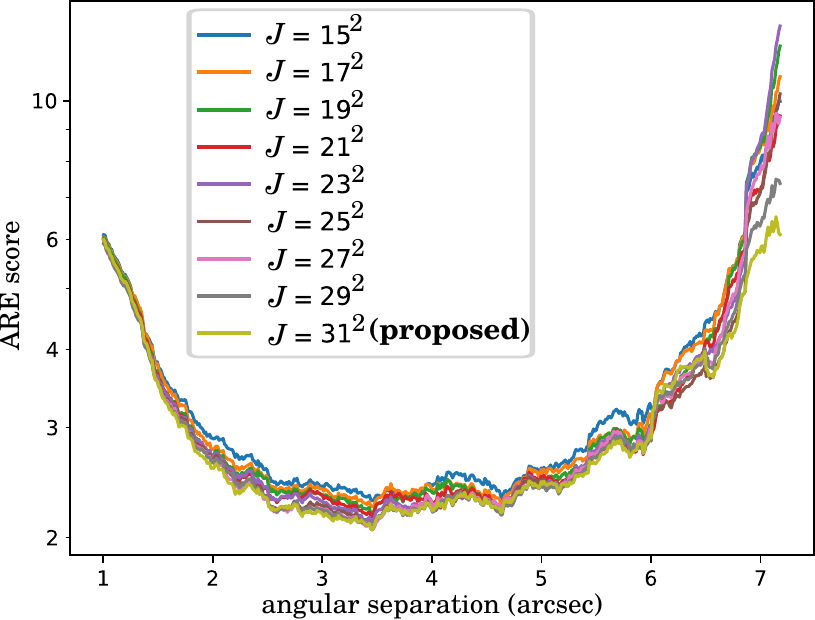}
		\caption{Influence of the patch size $J$ in the characterization algorithm. ARE (see Sect. \ref{ref:subsubsec:loss_metrics_characterization} for the definition of the metric) on the estimated photometry is reported as a function of the angular separation, for patches between $J=15^2$ and $J=31^2$ pixels area. The results are averaged azimuthally for 40,000 sources of flux drawn uniformly between $1 \times 10^{-6}$ and $3 \times 10^{-5}$. The known real sources were masked out and were not considered. Dataset: HD 95086 (2015-05-05), see Sect. \ref{sec:results} for the recording logs.}
		\label{fig:patch_size}
	\end{center}
\end{figure}

\subsection{Regression by supervised deep learning}

\subsubsection{Construction of the training basis}
\label{subsubsec:construction_training_basis_characterization}

As with the detection procedure of Sect. \ref{sec:conclusion}, we resort to massive injections of synthetic sources with various fluxes to build our training basis. 

Prior to the injections, the first step consists in masking any real and/or synthetic detected sources that we aim to estimate the photometry at inference time. This operation prevents, as best as possible, data leakages between training and test sets so that training patches do not contain any pixel from patches considered at inference time. In practice, source masking is performed by replacing, for each temporal frame, the local area impacted by the signal of the sources of interest by their pixel-wise temporal mean. Like for the generation of the training basis of the detection stage, we also apply, as a data-augmentation strategy, a random permutation of the temporal frames prior to the construction of a training set $s$ from which training samples with injected sources are extracted. 

We then build each training set $s$ by injecting a dozen of synthetic sources, with a relative flux (i.e., exoplanet to star contrast) ranging from $1 \times 10^{-6}$ to $3 \times 10^{-5}$ with respect to the flux of the host star. Given the current instrumental and processing performance in direct imaging, this setting corresponds to sources with relatively low flux, for which the estimation is usually the most flawed. This operation range can be easily modified in the algorithm to characterize (expected) sources with a lower or higher contrast level, if needed. The contrast of synthetic sources is drawn uniformly in the above-mentioned range, regardless of the angular separation. After injection, each training set is pre-processed and the input patches are extracted following the method described in Sect. \ref{subsec:preprocessing_characterization}. This procedure is repeated to get a prescribed number $P$ of training patches $\lbrace \widecheck{\V p}^{[p]} \in \mathbb{R}^{J} \rbrace_{p=1:P}$. The number of training patches, hence the total number $P$ of injected synthetic sources, is an additional hyper-parameter whose setting is discussed in more details in Sect. \ref{ref:subsubsec:implementation_details_characterization}.

\subsubsection{Model and architecture}
\label{subsub:model_architecture_characterization}

\begin{table}
	\caption{Architecture of the proposed CNN for source characterization. The shapes of the layers are indicated for a unit batch size.}
	\label{tab:cnn_architecture_characterization}
	\centering
	\begin{tabular}{cc}
		\toprule
		\textbf{layer} & \textbf{shape} \\
		\midrule
		Input & $1 \times 31 \times 31$ \\
		\hline	
		Conv2D + ReLU & $128 \times 25 \times 25$ \\
		Conv2D + ReLU & $128 \times 21 \times 21$ \\
		Conv2D + ReLU & $256 \times 17 \times 17$ \\
		\hline	
		MaxPooling & $256 \times 1 \times 1$ \\
		\hline	
		DenseLayer + ReLU & $256 \times 1 \times 1$ \\
		DenseLayer & $1 \times 1 \times 1$ \\
		\bottomrule
	\end{tabular}
\end{table} 

We built a custom network based on VGG, an architecture initially proposed for image classification \citep{simonyan15}. The underlying model has 1.2 million of free parameters, and its detailed architecture is described in Table \ref{tab:cnn_architecture_characterization}. We use a stride\footnote{In deep learning, the convolutional  stride (in pixels) set how far the convolutional filters move from one node of the image grid to the next one.} of one for all convolutional layers. We have also tested alternative models, both with a deeper and shallower architecture, all leading to worst estimation performance  than the selected one. In particular, we experienced a significant degradation of the performance at short angular separations with deeper architectures. The later are the more subject to over-fitting (due to the increase in terms of model complexity), especially in the absence of a whitening procedure preventing memorization of the nuisance structures by the network.

From an input patch $\widebreve{\V p}^{[p]} \in \mathbb{R}^{J}$, the network produces a single scalar $\widehat{\alpha}^{[p]} \in \mathbb{R}_+$, representing the estimated source's photometry. 

\subsubsection{Loss function and accuracy metric}
\label{ref:subsubsec:loss_metrics_characterization}

Our choice of the loss function $\mathcal{L}_{\text{carac.}}$ is driven by the two following criteria: (i) being computationally efficient, and (ii) matching the astrophysical goals. We chose the absolute relative error (ARE) between the ground-truth and the predicted photometry, which is a classical loss for regression problems:
\begin{equation}
	\mathcal{L}_{\text{carac.}}\left(\alpha^{[p]}, \widehat{\alpha}^{[p]}\right) \,\,\,(\%) = 100 \times \frac{\big| \alpha^{[p]} - \widehat{\alpha}^{[p]} \big|}{\alpha^{[p]}}\,.
	\label{eq:loss_characterization}
\end{equation}
The ARE has the advantage of giving the same contribution to each individual source $p$ regardless of its flux, when it is averaged over multiples ones. Since this metric is computationally very efficient, we also use it to evaluate the overall performance of the method at validation time.

\subsubsection{Implementation details}
\label{ref:subsubsec:implementation_details_characterization}

\begin{figure}
	\begin{center}
		\includegraphics[width=0.48\textwidth]{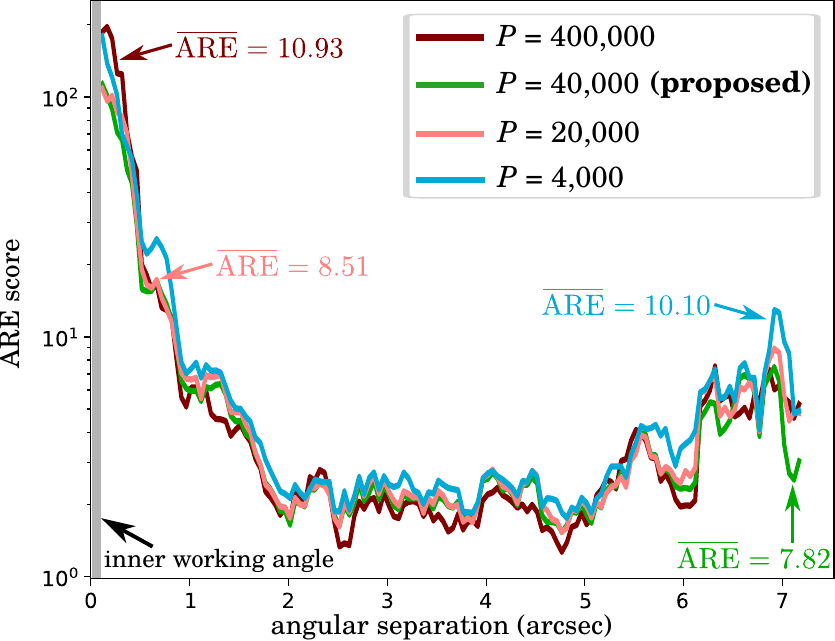}
		\caption{Influence of the number of training sources $P$ in the characterization algorithm. ARE (see Sect. \ref{ref:subsubsec:loss_metrics_characterization} for the definition of the metric) on the estimated photometry is reported as a function of the angular separation, for $P \in \{$400,000; 40,000; 20,000; 4,000$\}$ sources. The results are averaged azimuthally for sources of flux drawn uniformly between $1 \times 10^{-6}$ and $3 \times 10^{-5}$. The mean ARE (denoted by $\overline{\text{ARE}}$) averaged over the angular separation is also reported as an overall measure of the performance. The known real sources were masked out and were not considered. Datasets: the eleven SPHERE-IRDIS datasets considered in this work, see Sect. \ref{sec:results} for the recording logs.}
		\label{fig:training_set_size_characterization}
	\end{center}
\end{figure}

\noindent In this section, we successively discuss the setting of the patch size, the number of training sources, the sampling strategy of injected synthetic sources, and some optimization aspects.

\medskip 

\noindent Due to the pre-processing described in Sect. \ref{subsec:preprocessing_characterization}, part of the signal of the sources can be encoded in the temporal mean that is subtracted to the full frames in order to attenuate the quasi-static speckles. As a result, a (negative) contribution, taking the form of an arc, can spread out along the trajectory of the sources in the reduced frame. This well-known phenomenon in direct imaging is usually referred as \textit{self-subtraction}. As a consequence, we can expect that the performance of the predictor would increase with the patch size $J$, to be able to capture the (extended) signature of sources induced by self-subtraction. Besides, increasing the patch size increases the context (i.e., local realizations of the nuisance component) that could be beneficial to unmix the different contributions. As shown by Fig. \ref{fig:patch_size}, increasing the patch size is indeed beneficial as it reduced the mean relative error of estimation. The gain is more pronounced as the angular separation increases, which could be due to the larger extent of the self-subtraction signature (as the apparent motion of sources induced by ADI increases with the angular separation). However, large patches are not convenient in the case of adjacent sources, as both signals will be contained in both input patches. As a trade-off, we chose a patch size of $J=31^2$ pixels, as it encompasses the core of the signal of the source without being impractical when multiple sources are present in the field of view.

The number $P$ of synthetic sources (which also corresponds to the number of patches) used at training time is an additional hyper-parameter obeying a trade-off.
On the one hand, it should be large enough to be representative of the variety of real sources in terms of flux and locations. On the other hand, it should be sufficiently small to avoid over-fitting on the training set.
These expected behaviors are confirmed by numerical experiments presented in Fig. \ref{fig:training_set_size_characterization}. The error at small angular separations increases with $P$ since this is the area of the field of view the more subject to data leakages when generating multiple samples from a few tens of pixels only. The overall performance also degrades when $P$ is too small. Based on this study, we include $P=40,000$ patches in our training set since it leads to the smallest ARE averaged over the whole of view. 

Concerning the source's sampling strategy, we are interested at evaluation time in assessing the performance of the model per angular separation. As such, it is natural to sample test sources uniformly in the polar coordinates system. However, polar sampling is detrimental during the training phase, as pixels at short angular separations would be over-represented in the training set, leading to an over-fitting of the model in this area. It can be noted that this effect does not occur in the detection stage of the proposed approach given that the training sets consist of full frames of pre-processed observations; each pixel of the field of view being equally represented in the training base. We experimentally observed that sampling training sources uniformly in the Cartesian coordinates system reduces significantly over-fitting at short angular separations, without degrading performance in the rest of the field of view. As a result, we opt during training for a uniform sampling of the coordinates of synthetic sources in the Cartesian system.

Concerning the optimization process, pairs of samples $\lbrace \widebreve{\V p}^{[p]} ;  \alpha^{[p]} \rbrace$ are pre-generated before training and evaluation.
As for the detection stage, each realization $p$ is unique to limit over-fitting as best as possible. Unlike the detection stage, the characterization stage operates on a single patch (instead of the $T$ temporal images), thus allowing to chose a batch size higher than one to improve the stability and to reduce the computation time of the optimization process. In practice, the batch size is fixed at 1024. The number of epochs is fixed at 300, which shown to be sufficient to reach convergence of the network weights for all the considered observations. For each training epoch, the full set of training samples is fed as input of the network in a random order. 
The model optimization is performed with Adam \citep{kingma2014adam} with a learning rate of $10^{-3}$. As for the detection stage, the optimization of the network weights is done in PyTorch \citep{NEURIPS2019_9015} on either Tesla V100 or GTX 1080 Ti cards. The pre-processing step is highly parallelized, and it is run on CPUs.

\section{Results}
\label{sec:results}

\subsection{Datasets description and reduction strategies}
\label{subsec:datasets}

\begin{table*}
		\caption{Observing parameters of ADI sequences from the VLT/SPHERE-IRDIS instrument considered in this paper. Columns are: target name, ESO survey ID, observation date, spectral filter $\lambda$, number $T$ of available temporal frames, total apparent rotation $\Delta_{\text{par}}$ of the field of view, number NDIT of sub-integration exposures, individual exposure time DIT, average coherence time $\tau_0$, average seeing, and the first paper reporting analysis of the same data. All the observations are performed with the apodized Lyot coronagraph \citep{carbillet2011apodized} of the VLT/SPHERE instrument. $^{\text{(a)}}$Since the EIDC aimed to perform a \textit{blind} benchmark, information that would allow to identify the datasets are not known. $^{\text{(b)}}$This dataset was recorded with the star-hopping technique recently available for the VLT/SPHERE instrument \citep{wahhaj2021search} and its analysis is not reported yet. We do not exploit the dataset associated to the observation of the reference star. The dataset associated to the target star (HD 95086) is processed like all other datasets considered in this paper. $^{\text{(c)}}$This dataset is only used for additional experiments conducted in Appendix \ref{subapp:refinement_whitening_observations}.}
		\label{tab:dataset_logs}
		\centering
		\begin{tabular}{cccccccccccc} 
				\toprule 
				Target & ESO ID & Obs. date & $\lambda$ ($\micro\meter$) & $T$ & $\Delta_{\text{par}}$ (°) & NDIT & DIT (s) & $\tau_0$ (ms) & Seeing ('') & Related paper\\
				\midrule
				\multicolumn{11}{c}{\textit{VLT/SPHERE-IRDIS observations from the EIDC challenge}}\\
				\midrule
				IRDIS 1$^{\text{(a)}}$ & $-^{\text{(a)}}$ & $-^{\text{(a)}}$ & 1.625 & 252 & 40.3 & $-^{\text{(a)}}$ & $-^{\text{(a)}}$ & $-^{\text{(a)}}$ & $-^{\text{(a)}}$ & \cite{cantalloube2020exoplanet}\\			
				IRDIS 2$^{\text{(a)}}$ & $-^{\text{(a)}}$ & $-^{\text{(a)}}$ & 1.593 & 80 & 31.5 & $-^{\text{(a)}}$ & $-^{\text{(a)}}$ & $-^{\text{(a)}}$ & $-^{\text{(a)}}$ & \cite{cantalloube2020exoplanet}\\	
				IRDIS 3$^{\text{(a)}}$ & $-^{\text{(a)}}$ & $-^{\text{(a)}}$ & 1.593 & 228 & 80.5 & $-^{\text{(a)}}$ & $-^{\text{(a)}}$ & $-^{\text{(a)}}$ & $-^{\text{(a)}}$ & \cite{cantalloube2020exoplanet}\\
				\midrule
				\multicolumn{11}{c}{\textit{VLT/SPHERE-IRDIS observations}}\\
				\midrule
				HD 95086 & 095.C-0298(A) & 2015-05-05 & 2.110 & 52 & 18.2 & 4 & 64 & 2.3 & 0.89 & \cite{chauvin2018investigating}\\ 
				HD 95086 & 1100.C-0481(E) & 2018-01-05 & 2.110 & 70 & 41.0 & 10 & 96 & 7.8 & 0.32 & \cite{desgrange2022depth}\\ 
				HD 95086 & 106.21VL.001 & 2021-03-11 & 2.110 & 104 & 41.4 & 2 & 32 & 7.0 & 0.77 & $-^{\text{(b)}}$\\ 
				HIP 88399 & 095.C-0298(A) & 2015-05-10 & 1.593 & 46 & 34.3 & 4 & 64 & 1.2 & 1.05 & \cite{langlois2021sphere}\\ 
				HIP 88399 & 097.C-0865(A) & 2016-04-16 & 1.593 & 54 & 37.3 & 5 & 64 & 2.0 & 1.45 & \cite{langlois2021sphere}\\ 					
				HIP 88399 & 1100.C-0481(F) & 2018-04-11 & 1.593 & 40 & 31.9 & 10 & 96 & 5.5 & 0.74 & \cite{langlois2021sphere}\\ 
				HD 131399 & 095.C-0389(A) & 2015-06-12 & 2.110 & 92 & 36.7 & 6 & 16 & 1.9 & 0.90 & \cite{wagner2016direct}\\ 	
				HD 131399 & 296.C-5036(A) & 2016-05-07 & 2.110 & 56 & 39.5 & 7 & 32 & 3.6 & 0.98 & \cite{wagner2016direct}\\ 	
				HIP 65426 & 198.C-0209(E) & 2017-02-09 & 2.110 & 55 & 49.1 & 4 & 64 & 4.4 & 0.82 & \cite{chauvin2017discovery}\\ 					
				HIP 65426 & 1100.C-0481(G) & 2018-05-13 & 2.110 & 40 & 31.7 & 10 & 96 & 4.3 & 0.81 & \cite{cheetham2019spectral}\\
				HIP 72192 & 095.C-0389(A) & 2015-06-11 & 2.110 & 96 & 17.3 & 6 & 16 & 1.9 & 1.03 & \cite{flasseur2018exoplanet}\\
				HR 8799$^{\text{(c)}}$ & 095.C-0298(C) & 2015-07-04 & 2.110 & 112 & 17.9 & 8 & 32 & 2.3 & 0.94 & \cite{langlois2021sphere}\\				
				\bottomrule
		\end{tabular} 
\end{table*}
		
\noindent For our comparative analysis, we have selected 15 datasets recorded with the SPHERE-IRDIS instrument.

Three of the 15 SPHERE-IRDIS datasets were extracted from the \textit{exoplanet imaging data challenge} (EIDC) initially designed to ground the detection performance of existing post-processing algorithms for high-contrast imaging \citep{cantalloube2020exoplanet}. These datasets are used as a sanity check to assess the ability of the proposed algorithm to detect injected sources at moderate levels of contrast. 

To study in more details the performance of the proposed method, we selected 12 additional datasets, mostly part from the SHINE survey of the SPHERE-IRDIS instrument \citep{desidera2021sphere, langlois2021sphere, vigan2021sphere}. 
They were obtained by the observation of the following stars:\\
-- HD 95086, a A8III type star of the Carina constellation, hosting an exoplanet discovered by direct imaging with the SPHERE instrument \citep{rameau2013discovery,rameau2013confirmation}. Ten known background point sources are also within the SPHERE-IRDIS field of view. In addition, based on the analysis of \PACO and \dPACO detection maps, we have identified two additional (unbounded) candidate point-like sources. Given their unknown status, we exclude these sources from our general analysis (i.e., there are not considered as true detections or false alarms). We briefly discuss there status in Sect. \ref{subsubsec:confirmation_bckgd_sources}.\\
-- HIP 88399, a F6V type star of the Vela constellation, without known bounded exoplanet. However, six faint background sources fall within the SPHERE-IRDIS field of view.\\
-- HD 131399 A, a A1V type star of a triple system located in the Centaurus constellation, with a known faint point source (HD 131399 Ab) discovered by direct imaging \citep{wagner2016direct}. While first supposed to be an exoplanet, follow-up analysis of its astro-photometry show that HD 131399 Ab is more likely a background brown dwarf \citep{nielsen2017evidence}. Besides, a bright background star falls within the SPHERE-IRDIS field of view.\\
-- HIP 65426, a A8III type star of the Carina constellation, hosting an exoplanet discovered by direct imaging with the SPHERE instrument \citep{chauvin2017discovery}.\\
-- HIP 72192, a A0V type star of the Lupus constellation, without known bounded exoplanet. However, two faint background sources fall within the SPHERE-IRDIS field of view.\\
\indent Considered SPHERE-IRDIS observations were obtained in \textit{H2}-\textit{H3} (i.e., $\lambda_{\textit{H2}} = 1.593 \,\micro\meter$, $\lambda_{\textit{H3}} = 1.667 \,\micro\meter$) or in \textit{K1}-\textit{K2} (i.e., $\lambda_{\textit{K1}} = 2.110 \,\micro\meter$, $\lambda_{\textit{K2}} = 2.251 \,\micro\meter$) dual spectral bands. 
The main parameters of each observation are summarized in Table \ref{tab:dataset_logs}. The diversity in the experienced observing conditions is quite representative of the SPHERE observations.

\medskip

The raw observations were pre-reduced\footnote{In this paper, we use the term \textit{pre-reduction} to stand for the extraction, mapping, and correction of the raw data. We use the term \textit{pre-processing} to stand for the centering and whitening of the pre-reduced observations with the statistical model detailed in Sect. \ref{subsec:algorithm_step1}, that serve as inputs of the supervised deep learning stage detailed in Sect. \ref{subsec:algorithm_step2}.} with the DRH pipeline \citep{pavlov2008advanced} of the SPHERE instrument, which performs thermal background subtraction, flat-field correction, anamorphism correction, compensation for spectral transmission, flux normalization, bad pixels identification and interpolation, frame centering, true-North alignment, wavelength calibration, astrometric calibration, and frame selection. These operations are complemented by custom routines implemented in the SPHERE data center \citep{delorme2017sphere}, in particular to improve bad pixels correction. Finally, the SPHERE data center combines the pre-reduced observations and delivers the calibrated ADI datasets we consider in this work. ADI reduction is performed by considering the first spectral channel (i.e., either at $\lambda_{\textit{H2}} = 1.593 \,\micro\meter$ or at $\lambda_{\textit{K1}} = 2.110 \,\micro\meter$) given that the contrast is significantly more favorable in this channel than in the second one.

\medskip

To ground the performance of the proposed algorithm, we resort to massive injections of synthetic sources, as done to train the deep model of the proposed algorithm, see Sect. \ref{subsubsec:generation_training_samples}. Given the simplicity of the signature of the sought objects (i.e., taking the form of a blob spatially correlated over only a few pixel width), we did not find significant bias in using the same injection procedure for the training and evaluation steps. Injections of synthetic sources in the EIDC benchmark were performed by there authors with the VIP pipeline. 

The detection performance of the proposed method are compared in Sect. \ref{subsec:detection_results} with the cADI, PCA, and \PACO algorithms, (see Sect. \ref{sec:introduction} for their respective principle). For cADI, we have re-implemented the original method \citep{marois2006angular} based on a full-frame estimation of the off-axis PSF and of the S/N map, i.e., without angular-specific processing. We have also used the refined implementation of cADI available in the VIP package \citep{gonzalez2017vip}, which includes a protection angle strategy accounting for a minimal field rotation between successive images when building the off-axis PSF in order to limit the self-subtraction effect. After computation of the off-axis PSF, a S/N map is derived by accounting for an annular-based estimation of the noise in the residual images. We also applied the VIP implementation of the PCA-based algorithm combined with the same protection angle strategy and the annular-based computation of the S/N. For PCA reductions, the number of modes has been optimized in annuli by maximizing the S/N of synthetic sources with similar ranges of contrast than the ones we consider for our comparisons. The other parameters of the VIP implementation of cADI and PCA are less critical and are fixed at pre-set values \citep{gonzalez2017vip}. For \PACO, we performed the data reduction with our fully unsupervised processing pipeline \citep{flasseur2018unsupervised}. 

Concerning the photometry estimation, we compare in Sect. \ref{subsec:characterization_results} the performance of the proposed method with \PACO and the VIP implementation of the PCA. \PACO parameters are estimated automatically in a data-driven fashion. In a nutshell, the characterization of a point-like source $p$ is performed by a joint estimation of (i) the statistics (i.e., mean $\V m_n$ and covariances $\M C_n$) of the nuisance component $\V f$, and (ii) of the photometry (i.e., flux $\alpha_p$) of the given source, see Eq. \eqref{eq:image_formation}. For the PCA, we resort to a similar procedure than for the detection step to set the parameters. In particular, the number of modes is optimized for each injected source to be characterized by maximizing its S/N. Once the setting fixed, the flux of a given source is estimated by minimizing the residuals through the injection of negative fake companions \citep{wertz2017vlt}. This is performed with a two steps procedure, as recommended in the VIP package: (i) a first guess estimate is obtained by performing a grid-search, (ii) a local optimization is performed with a Nelder-Mead simplex algorithm \citep{nocedal1999numerical}. Given computational constraints (largely dominated by the PCA), the exact (known) sub-pixel location $\phi_p$ of each injected source $p$ is provided to the different algorithms (i.e., it is not optimized), and the photometry is estimated at this ground-truth position. When estimating the photometry of real sources, both the astrometry and the photometry are optimized by the different algorithms.

\subsection{Detection results}
\label{subsec:detection_results}

\subsubsection{Detection of known real sources}
\label{subsec:detection_results_known_real_sources}

\begin{figure*}
	\begin{center}
		\includegraphics[width=0.90\textwidth]{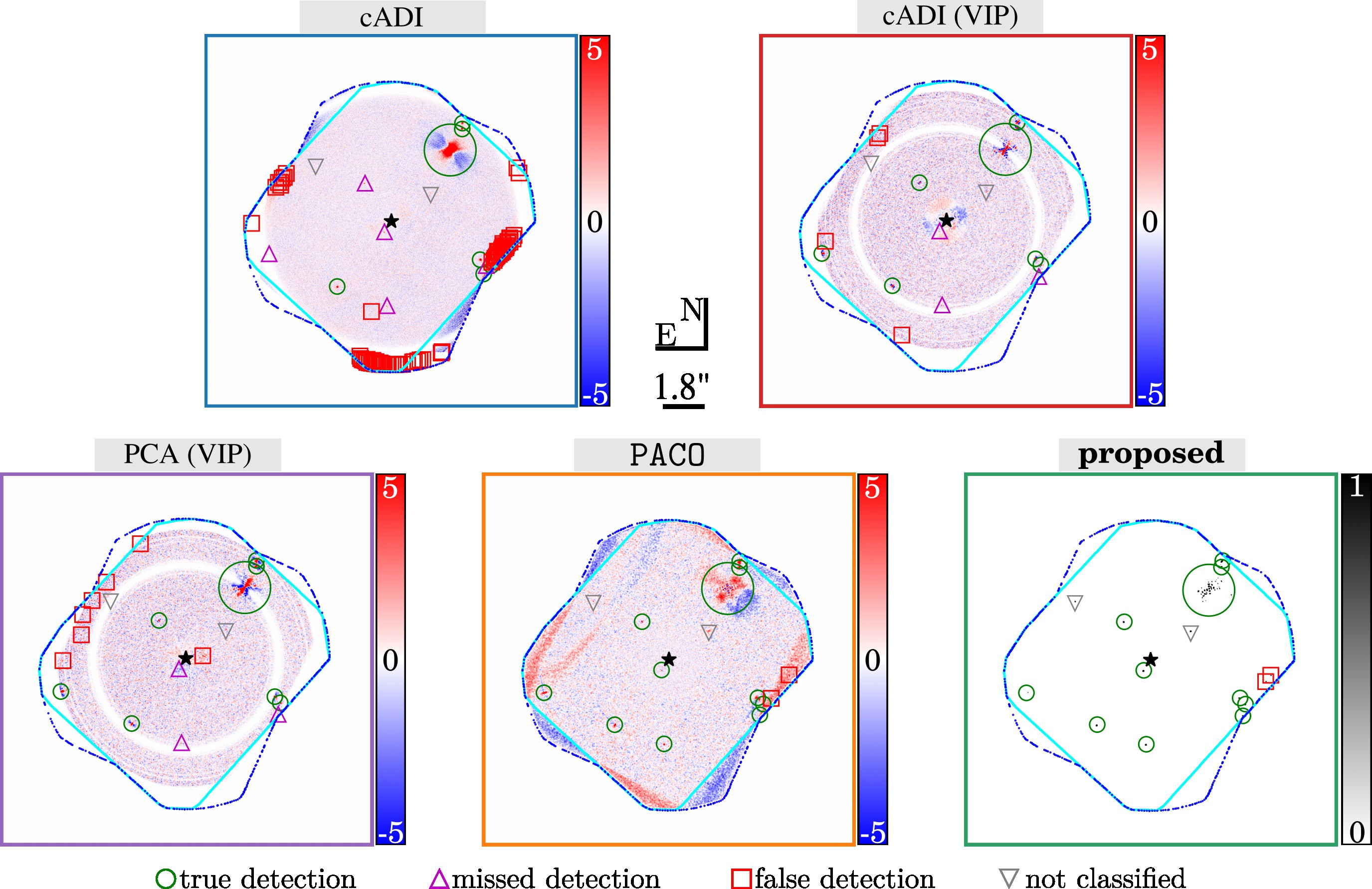}
		\caption{Detection maps obtained with the selected algorithms (see Sect. \ref{subsec:datasets}). Sources are  classified as true, missed and false detections. The two additional candidate point-like sources whose identification is discussed in Sect. \ref{subsubsec:confirmation_bckgd_sources} are not classified. 
        The detection threshold is set to $\tau=5$ for cADI, cADI (VIP), PCA (VIP) and \PACO. It is set to $\tau=0.5$ for the proposed algorithm. The light blue line represents the sensor field of view (encompassed withing a $N$-pixels square support) while the dashed blue line represents the extended field of view (encompassed withing a $M$-pixels square support) on which the detection can be performed due to the apparent rotation of the field induced by ADI. Dataset: HD 95086 (2015-05-05), see Sect. \ref{sec:results} for the observation logs.}
		\label{fig:HD95086_20150505_snr_prediction_maps_comparisons_natural_outputs_full_1}
	\end{center}
	
	\bigskip
	\vspace{-5mm}
	
	\begin{center}
		\includegraphics[width=0.90\textwidth]{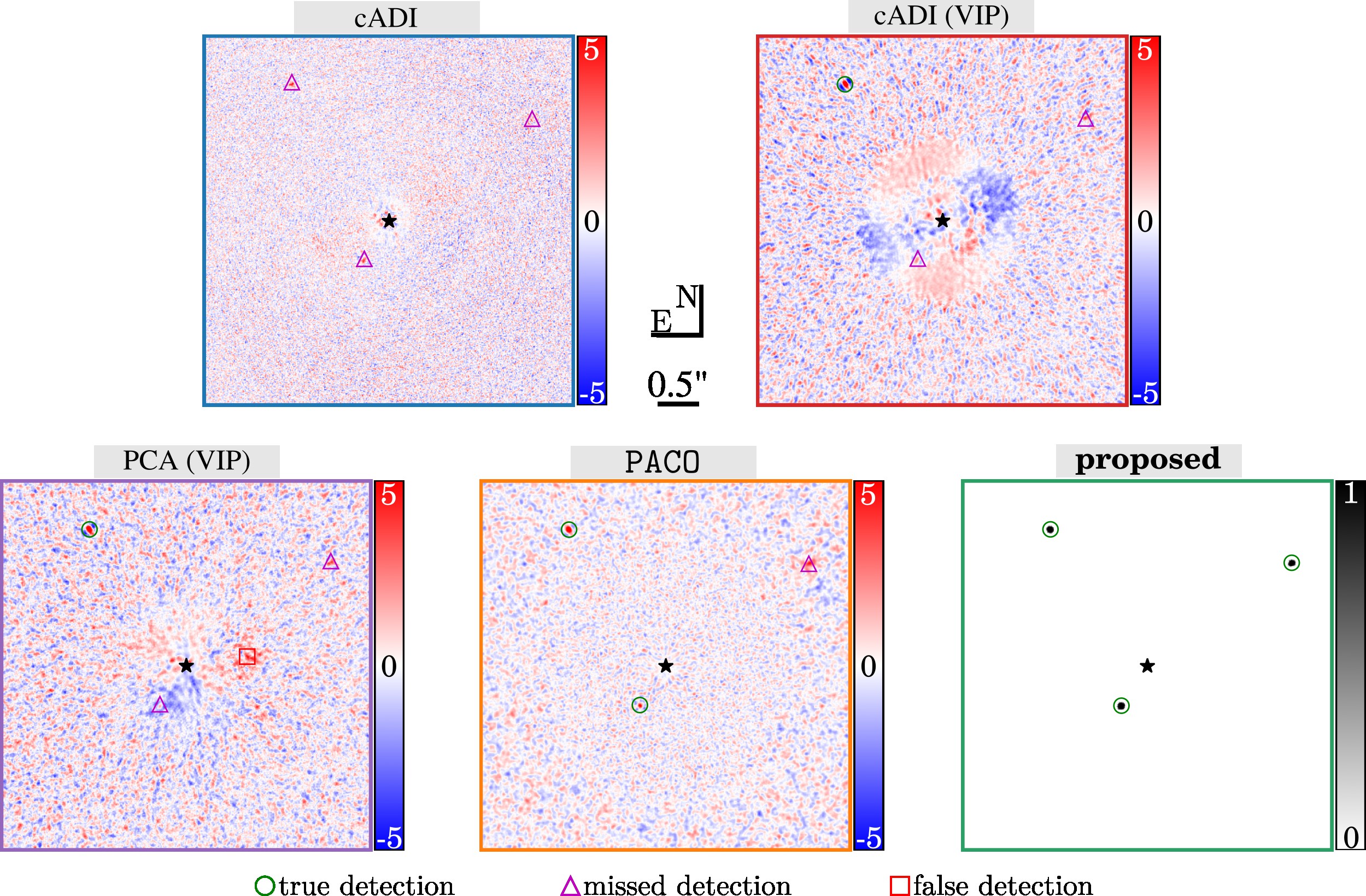}	
		\caption{Same caption than Fig. \ref{fig:HD95086_20150505_snr_prediction_maps_comparisons_natural_outputs_full_1}. Zoom near the host star. Dataset: HD 95086 (2015-05-05), see Sect. \ref{sec:results} for the observation logs.}
		\label{fig:HD95086_20150505_snr_prediction_maps_comparisons_natural_outputs_zoom_1}
	\end{center}
\end{figure*}

\begin{figure}
	\begin{center}
		\includegraphics[width=0.48\textwidth]{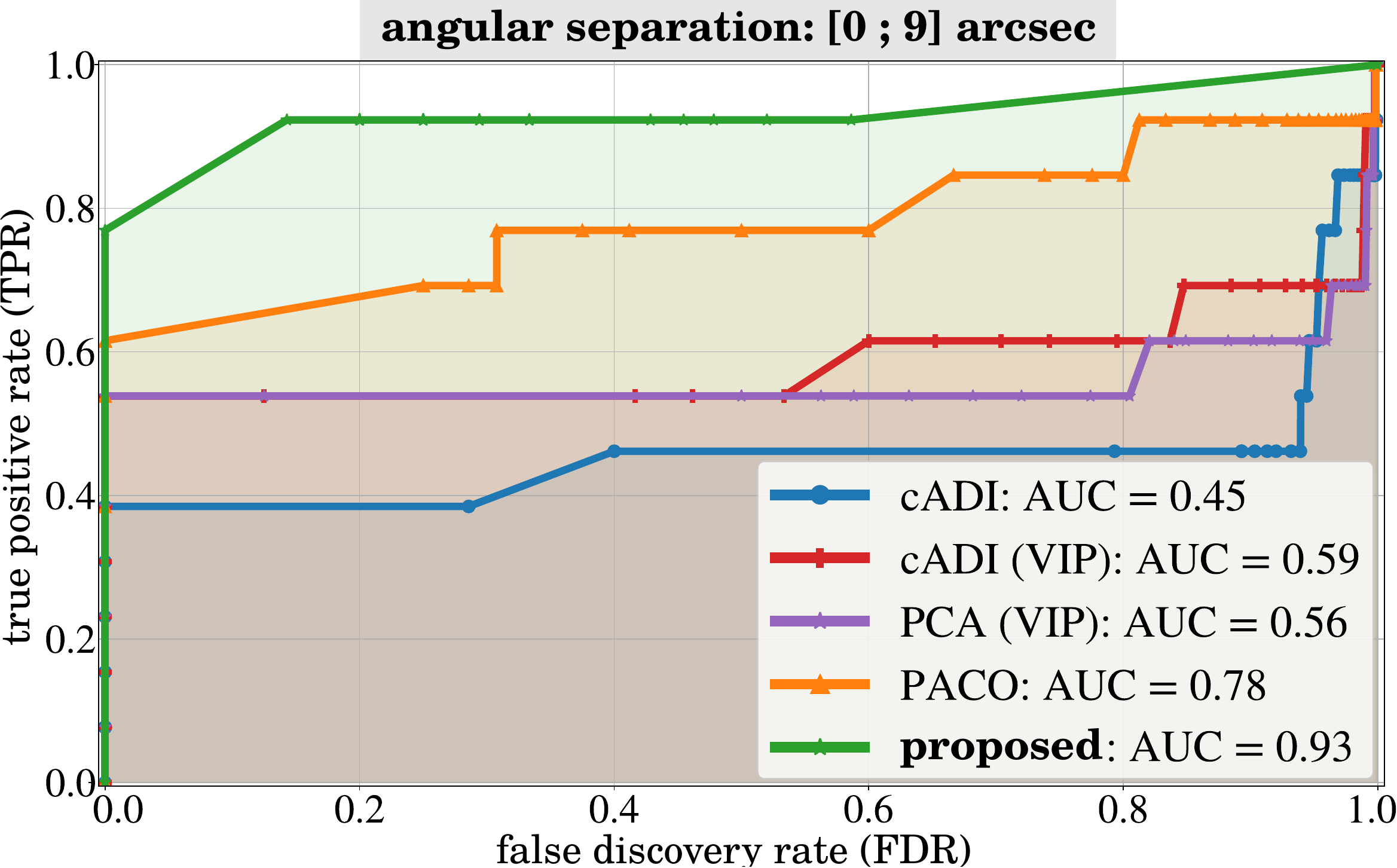}
		\caption{ROCs showing the TPR as a function of the FDR for real sources. Dataset: HD 95086 (2015-05-05), see Sect. \ref{sec:results} for the observation logs.}
	\label{fig:HD95086_20150505_IRDIS_ADI_TPR_f(FDR)_cube_index_0_real_sources_only_ang_sep_[0_;_9]}
	\end{center}
\end{figure}

\begin{table}
	\caption{Mean results of AUC for ROCs giving the TPR as a function of the FDR. The scores are averaged over the eleven SPHERE-IRDIS datasets considered in this paper, see Sect. \ref{sec:results} for the observation logs. Only the 59 known real sources present in these datasets were considered. Figure 7 of \citep{flasseur2022exoplanet} display the corresponding ROC from which these mean results were aggregated.}
	\label{tab:roc_mean_results_real_sources}
	\centering
	\begin{tabular}{cccccc}
		\toprule
		\textbf{sep. ('')} & \textbf{cADI} & \textbf{cADI (VIP)} & \textbf{PCA (VIP)} & \textbf{PACO} & \textbf{proposed}\\
		\midrule
		$\left[ 0 ; 9 \right]$ & 0.38 & 0.72 & 0.66 & 0.88 & \textbf{0.95}\\
		\bottomrule
	\end{tabular}
\end{table} 

\noindent A first classical sanity test to evaluate the detection performance of a post-processing algorithm is to study qualitatively its ability to re-detect real known sources initially detected with different algorithms, and possibly from different datasets. We present detailed results for one dataset (HD 95086, 2015-05-05, see Table \ref{tab:dataset_logs}) selected among the eleven SPHERE-IRDIS observations we consider because HD 95086 is the star having the larger number of known real sources in the SPHERE-IRDIS field of view. Results obtained for the ten other datasets are given in supplementary material. Figures \ref{fig:HD95086_20150505_snr_prediction_maps_comparisons_natural_outputs_full_1} and \ref{fig:HD95086_20150505_snr_prediction_maps_comparisons_natural_outputs_zoom_1} give detection maps produced with the five tested algorithms. The detection threshold is set to $\tau=5$ for the algorithms producing a S/N map (i.e., cADI, cADI (VIP), PCA (VIP), \PACO), and to $\tau=0.5$ for the proposed method producing a pseudo-probability map. Due to the binary pixel-wise classification task we consider for the training step of the proposed method (see Sect. \ref{subsec:algorithm_step2}), its detection map is almost binary (i.e., each pixel value is close either to 0 or 1) so that the setting of the threshold $\tau$ is quite flexible. Based on the analysis of the detection maps, \PACO and the proposed method lead to the best qualitative results since there are the only algorithms able to detect all real known sources without any false alarm in most of the field of view. With the proposed method, only two false alarms occur very near the borders of the field of view due to the limited number (much lower than $T$) of temporal samples available in this area to build a consistent model of the nuisance. This claim is also supported by the \PACO detection map that displays a few false alarms in the same area.

Figure \ref{fig:HD95086_20150505_IRDIS_ADI_TPR_f(FDR)_cube_index_0_real_sources_only_ang_sep_[0_;_9]} gives a more quantitative analysis of the previous results through ROCs representing the TPR as a function of the FDR (see Sect. \ref{subsubsec:loss_metrics} and Eq. (\ref{eq:detection_metrics})) for the same dataset of HD 95086 (2015-05-05). This type of representation gives a comparison of the precision-recall trade-off reached by each method, regardless the detection quantity (S/N or pseudo-probability) they produce. These curves are obtained by counting the number of true positives (TPs), and false alarms (FAs) for the full range of possible detection thresholds, i.e. $\tau \in \left[ 0 ; 1 \right]$ for the proposed method, and $\tau \in \left[\text{min}(\widehat{\V y}) ;\text{max}(\widehat{\V y})\right]$ for cADI, PCA, and \PACO. Table \ref{tab:roc_mean_results_real_sources} presents averaged results over the eleven SPHERE-IRDIS datasets we consider in this study, and the detailed scores for each dataset are given in Table 1 of the supplementary material. These results illustrate the benefits of the proposed method in terms of precision-recall trade-off: the AUC under ROC is improved by at least 7$\%$ with respect to the comparative algorithms.

\begin{figure}
	\begin{center}
		\includegraphics[width=0.48\textwidth]{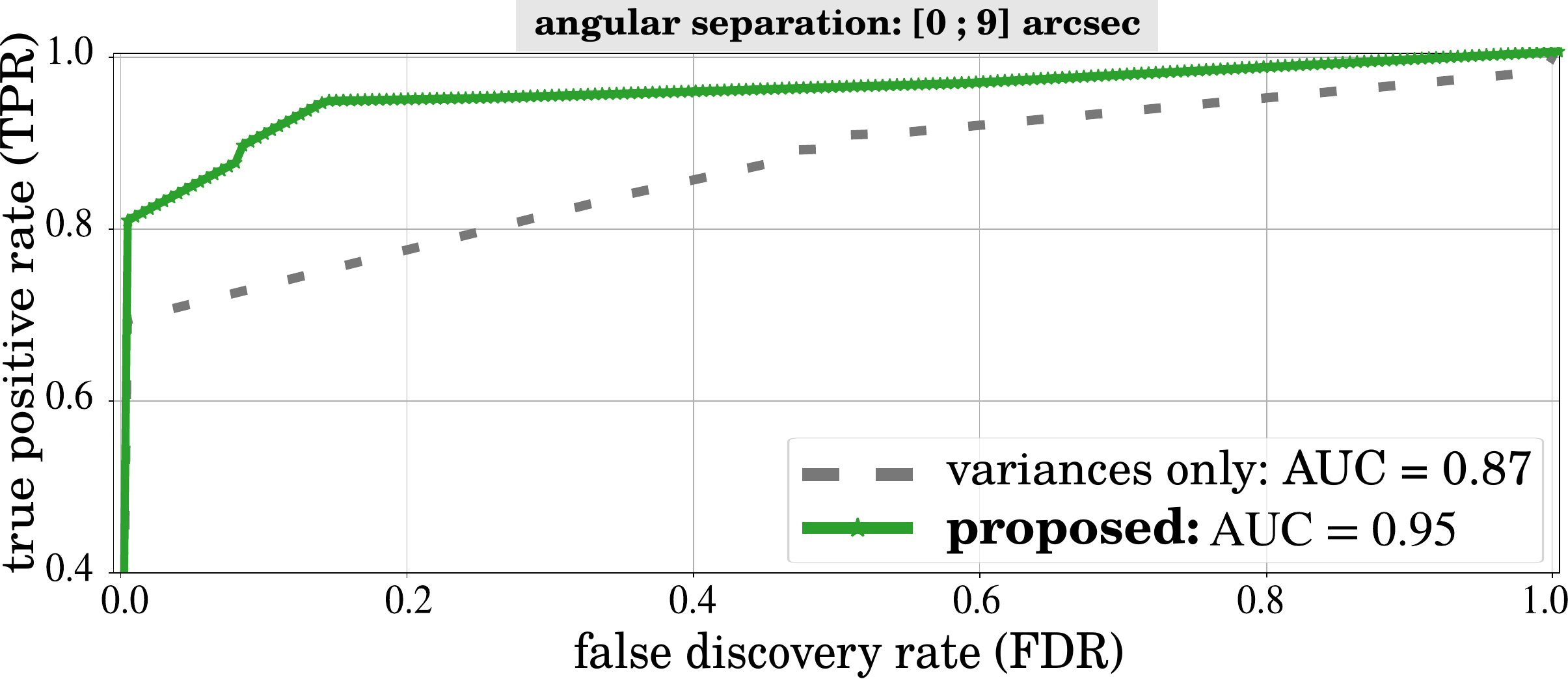}
		\caption{Ablation study on the influence of the whitening procedure of the pre-preprocessing step. The ROCs show the TPR as a function of the FDR built from the 11 SPHERE-IRDIS datasets we consider in this work (containing 59 known real sources), see Sect. \ref{sec:results} for the observation logs.}
	\label{fig:variances_vs_covariances}
	\end{center}
\end{figure}

As a final study based on the detection of known real sources, we evaluate the importance of our pre-processing step by resorting to model ablation. Removing the whitening procedure and keeping only the temporal centering in the pre-processing step does not allow to reach convergence of the network weights at training time. This is due to the high dynamics and to the high spatial non-stationarity of the residual images. We also test to account only for the pixel variances in the whitening procedure, i.e. we neglect the spatial covariances so that  matrices $\widehat{\M S}_n$ in Eq. \eqref{eq:sample_estimators} are considered diagonal. Figure \ref{fig:variances_vs_covariances} compares the precision-recall trade-off of this downgraded model to the model of the proposed approach (accounting locally for the spatial covariances). When neglecting covariances, the overall precision-recall trade-off of the detector is decreased by 8\% in average and the sensitivity of detection is especially lowered for low false discovery rates (which is, in practice, the most useful regime in high-contrast imaging). While several whitening methods could be used in conjunction with our method, this study confirms the importance of our custom whitening procedure accounting for the spatial covariances of the nuisance in order to reach the best performance of a detector built by supervised deep learning. This observation is also in agreement with studies performed in other works through alternative whitening procedures. For instance, the SODINN algorithm (see Sect. 1) that also trains a CNN to perform a detection task by supervised deep learning is sometimes prone to a large false alarm rate \citep{cantalloube2020exoplanet}. Likely, this side effect is (at least in part) due to the embedded pre-processing step that builds an empirical model of the nuisance component through PCA; an approach that does not explicitly model the spatial covariances of the nuisance, thus leading to spatially non-stationary residual images used at training time.

\subsubsection{Detection of synthetic sources}
\label{subsec:detection_synthetic_sources}

\begin{figure}
	\begin{center}
		\includegraphics[width=0.48\textwidth]{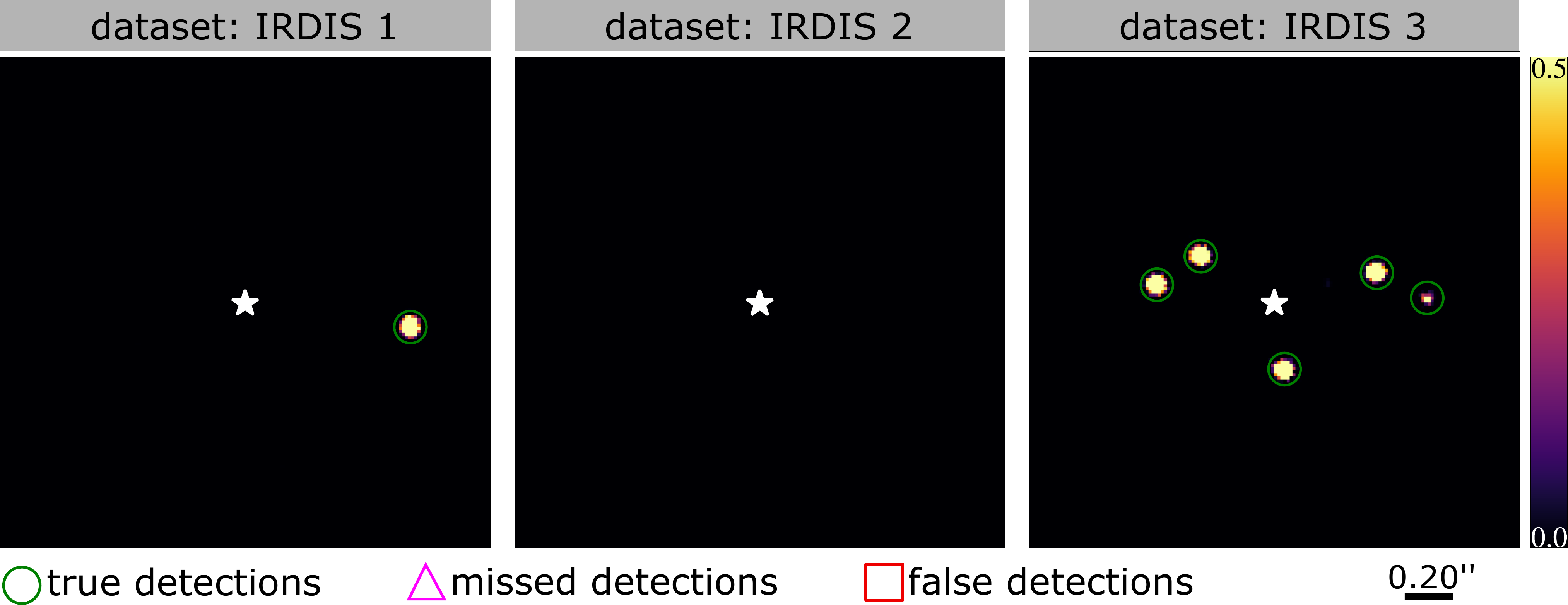}
		\caption{Detection maps obtained with the proposed algorithm on the three SPHERE-IRDIS datasets from the EIDC data challenge, see Sect. \ref{sec:results} for the observation logs. Sources are classified as true, missed and false detections. The detection threshold is set to $\tau=0.5$.}
		\label{fig:eidc_detection_maps}
	\end{center}
\end{figure}

\begin{table*}
		\caption{Detection scores: F1R at detection threshold $\tau=0.5$ (the higher, the better), AUC under the ROC representing the TPR as a function of $\tau$ (the higher, the better), and AUC under the ROC representing the FDR as a function of $\tau$ (the lower, the better). Scores reported for \PACO are extracted from the EIDC data challenge \citep{cantalloube2020exoplanet}, and \dPACO scores are computed with a similar procedure for the three SPHERE-IRDIS datasets from the EIDC. --$^\text{(a)}$Metrics can not be computed since there is no injected source in this dataset.}
		\label{tab:eidc_scores}
		 \begin{tabular}{cccccc|ccccc}
			\toprule
			& \textbf{IRDIS 1} & \textbf{IRDIS 2} & \textbf{IRDIS 3} & \textbf{mean} & \textbf{rank} & \textbf{IRDIS 1} & \textbf{IRDIS 2} & \textbf{IRDIS 3} & \textbf{mean} & \textbf{rank}\\
			\hline
			 & \multicolumn{5}{c}{\textit{PACO}} ~~~ & \multicolumn{5}{c}{\textit{deep PACO}}\\
			 \hline
			 \textbf{F1R}& 1.00 & --$^\text{(a)}$ & 1.00 & 1.00 & 1\textsuperscript{st}/22 \textit{(on par)} & 1.00 & --$^\text{(a)}$ & 1.00 & 1.00 & 1\textsuperscript{st}/23 \textit{(on par)}\\
			 $\textbf{AUC}_{\textbf{TPR}}$ & 1.00 & --$^\text{(a)}$ & 0.93 & 0.97 & 1\textsuperscript{st}/22 \textit{(on par)} & 1.00 & --$^\text{(a)}$ & 0.96 & 0.98 & 1\textsuperscript{st}/23\\
			 $\textbf{AUC}_{\textbf{FDR}}$ & 0.39 & --$^\text{(a)}$ & 0.32 & 0.36 & 6/22 & 0.01 & --$^\text{(a)}$ & 0.01 & 0.01 & 1\textsuperscript{st}/23\\			 
			\bottomrule
		\end{tabular}
\end{table*}

In this section, we evaluate quantitatively the detection performance of the proposed method with ROCs and contrast curves built from massively injected synthetic sources.

As a first step, we apply the proposed detection algorithm on the three SPHERE-IRDIS datasets from the public EIDC data challenge \citep{cantalloube2020exoplanet}. The resulting detection maps are given in Fig. \ref{fig:eidc_detection_maps}. Using the same procedure than \cite{cantalloube2020exoplanet} to benchmark 22 post-processing algorithms (including cADI, PCA, and \PACO), we compute the F1R score at the set detection threshold ($\tau=0.5$), and ROCs for the TPR and the FDR scores, as defined in Eqs. (\ref{eq:detection_metrics}), by varying the detection threshold. The AUC is then computed from ROCs. Table \ref{tab:eidc_scores} summarizes the results we obtained with the proposed algorithm. As a purpose of comparison, we also report the \PACO results that have been published in \cite{cantalloube2020exoplanet}. It emphasizes the interesting precision-recall of the proposed approach that performs, on these datasets, on par with or better than the 22 post-processing algorithms considered in the EIDC data challenge. However, these results should be taken with caution since they are based on a few datasets, with only six injected sources at relatively bright levels of contrast. Besides, several algorithms (including \PACO) are also able to detect the injected sources without any false alarm in the field of view for a sufficiently large detection threshold. At this stage, the better performance of \dPACO in terms of false alarms rejection (quantified by the $\text{AUC}_{\text{FDR}}$ metric) are mostly explained by the fact that the proposed algorithm produces detection maps with almost binary values. It can be noted that this effect is also encountered for all the algorithms of the EIDC data challenge producing the same type of outputs like RSM \citep{dahlqvist2020regime} or SODINN \citep{gonzalez2018supervised}. 

\begin{figure*}
	\begin{center}
		\includegraphics[width=0.8\textwidth]{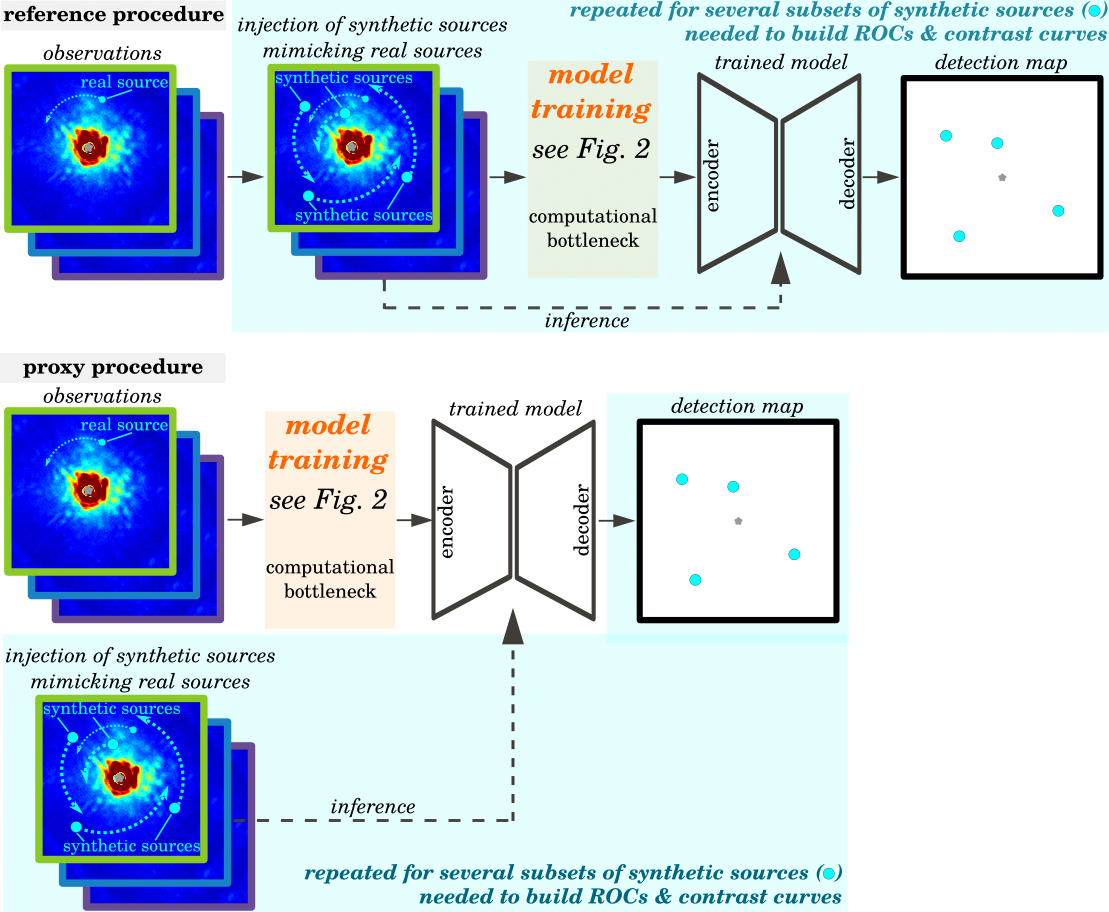}
		\caption{Schematic representation of the reference and proxy procedures used to evaluate the performance of the proposed approach through massive injections of synthetic sources mimicking the behavior (i.e., same apparent motion) of real sources.}
		\label{fig:reference_proxy_procedures}
	\end{center}
\end{figure*}

\begin{figure}
	\begin{center}
		\includegraphics[width=0.48\textwidth]{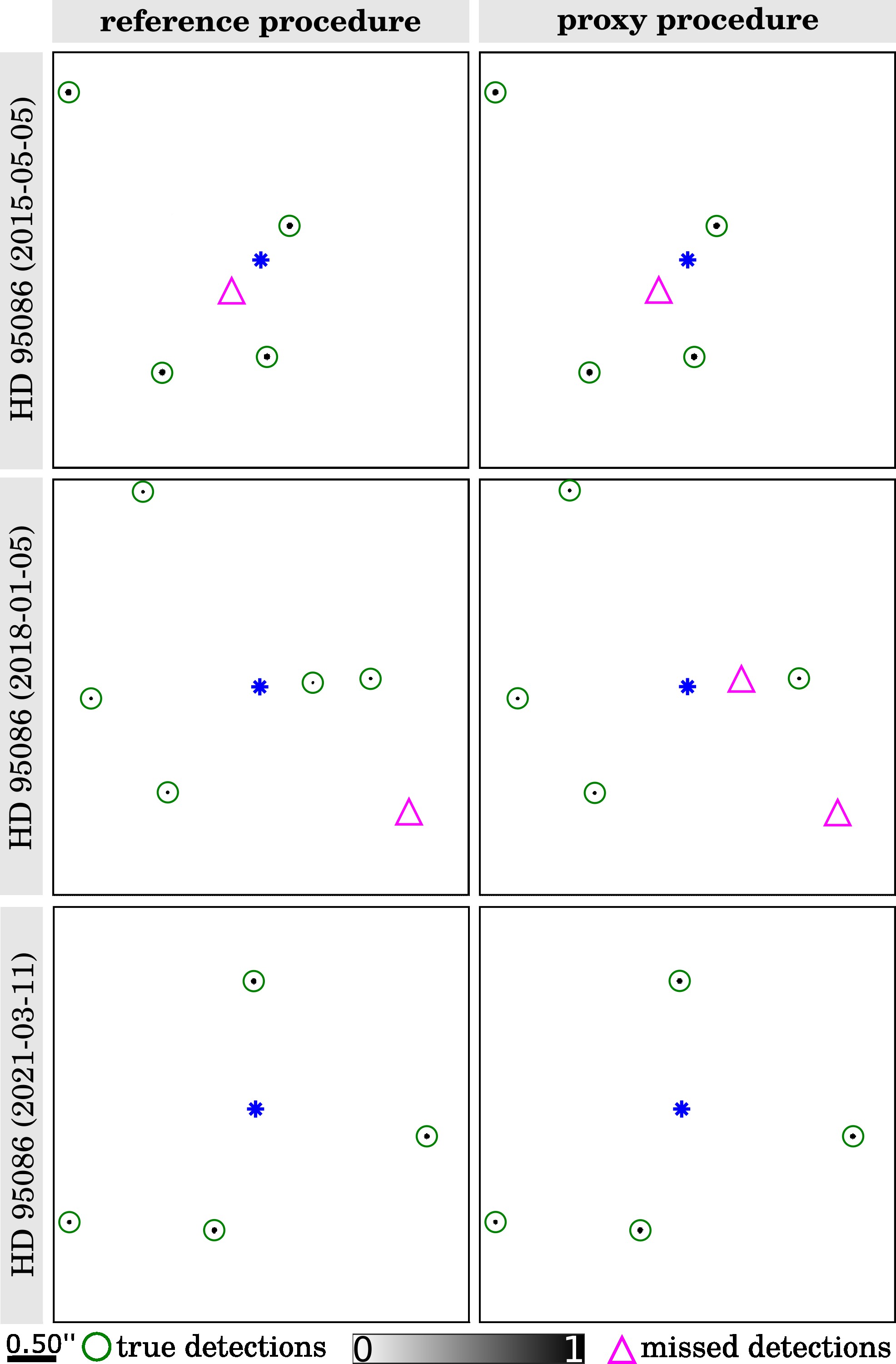}
		\caption{Comparison between the reference procedure and its proxy version for evaluation of the detection performance of the proposed algorithm. Synthetic sources are classified as true or missed detections using a detection threshold at $\tau=0.5$. Examples of detection maps obtained in the presence of injected synthetic sources are given on three datasets of HD 95086 (2015-05-05, 2018-02-23, 2021-03-11), see Sect. \ref{sec:results} for the observation logs.}
		\label{fig:HD95086_20150505_IRDIS_ADI_roc_injections_validation_method}
	\end{center}
\end{figure}

To ground in details the precision-recall trade-off of the proposed detection algorithm, it is necessary to build consistent ROCs and contrast curves by resorting to massive injections of synthetic sources --unknown at training time but that we aim to detect at inference time-- for various levels of contrast. 
For that purpose, the most realistic procedure, hereafter called \textit{reference procedure}, consists in (i) splitting the whole set of synthetic sources in small subsets, (ii) injecting  synthetic sources of one subset in the dataset of interest so that injected fake sources mimic the behaviour of real (possibly unknown) sources, (iii) training the detection model with the procedure described in Sect. \ref{sec:algorithm_detection}, and (iv) applying the trained model to the dataset containing the synthetic sources injected in step (ii). Steps (ii) to (iv) are repeated for all subsets of synthetic sources. This procedure simulates the real situation when we face a new dataset with real unknown exoplanets that we aim to detect at inference time. Due to the computational burden of step (iii), repeating this full procedure for all subsets of synthetic sources that we aim to detect at inference time is not realistic to build ROCs and contrast curves. To circumvent this issue, we resort to a \textit{proxy procedure}: instead of training a different model for each subset of injected sources, we train a unique model without synthetic sources mimicking the behaviour of real (possibly unknown) sources. Synthetic sources are injected \textit{a posteriori}, i.e., after completing the training step (iii), and the trained model is applied on the resulting dataset. Figure \ref{fig:reference_proxy_procedures} illustrates the principle of the reference and proxy procedures. The proxy procedure leads to an improvement in terms of algorithmic complexity by a factor equal to the number of subsets (typically 2,000, with in average 5 synthetic sources in each subset), that is determinant to be used in practice. At this stage, we still need to show that the proxy procedure we defined leads to reliable results so that our comparisons with state-of-the-art algorithms are fair. In particular, we aim to show that the model resulting from the proxy procedure is not prone to over-fitting induced by an imperfect separation between training and testing data. In the proxy procedure, over-fitting could possibly occur if the model partially memorizes the nuisance component that is seen by the network without the injected sources we aim to detect at inference time. Figure \ref{fig:HD95086_20150505_IRDIS_ADI_roc_injections_validation_method} shows examples of detection maps obtained with the procedure of reference described previously and its proxy version on the three datasets of HD 95086 (2015-05-05, 2018-02-23, 2021-03-11) considered in this work. In these experiments, we considered more than 700 synthetic sources spread over the whole field of view. Table \ref{tab:proxy_vs_reference_detection} compares quantitatively the two approaches in terms of detection performance. These results show that the proxy procedure leads to reliable and conservative estimations of the overall performance of \dPACO. In addition, since synthetic sources mimicking the behaviour of real unknown sources are recovered with comparable rates between the two procedures, it emphasizes that the custom data-augmentation strategy of the training step, including a random permutation of the images of the data series, is efficient to circumvent the absence of ground-truth about real sources. In the following, we safely use the proxy procedure to compare the detection capability of the proposed \dPACO algorithm with other post-processing methods.

\begin{table}
	\caption{Comparison between the reference procedure and its proxy version for evaluation of the detection performance of the proposed algorithm. Synthetic sources are classified as true, missed, or false detections using a detection threshold at $\tau=0.5$. Mean detection results are averaged for the three datasets of HD 95086 (2015-05-05, 2018-02-23, 2021-03-11), see Sect. \ref{sec:results} for the observation logs.}
	\label{tab:proxy_vs_reference_detection}
	\centering
	\begin{tabular}{ccc}
		\hline
		& \textbf{reference procedure} & \textbf{proxy procedure}\\
		\hline
		\textbf{true detections} & 583/728 (80.0\%) & 564/728 (77.5\%)\\
		\textbf{missed detections} & 145/728 (20.0\%) & 162/728 (22.5\%)\\
		\textbf{false detections} & 0 & 0\\
		\hline	
	\end{tabular}
\end{table} 


Following the previously defined proxy procedure, we present detailed results obtained on HIP 88399 (2018-04-11), which is a SPHERE-IRDIS dataset representative of the mean results we obtained over the eleven ones we consider in this work. Results for the ten other datasets are reported in the supplementary material. Figure  \ref{fig:HIP88399_20180411_IRDIS_ADI_roc_population} shows detection results on a sample of 10,000 synthetic sources in a diagram contrast \textit{versus} angular separation for \PACO and the proposed method. Each synthetic source is classified as missed, true, or false detection using the detection thresholds defined in Sect. \ref{subsec:detection_results_known_real_sources}. For \PACO, setting the detection threshold at $\tau=5$ corresponds, in average, to a realistic control \citep{flasseur2018unsupervised,flasseur2018exoplanet,flasseur2018SPIE,flasseur2020robust} of the probability of false alarms (PFA) at $5\sigma$ (i.e., PFA $\simeq 3 \times 10^{-7}$). While the PFA should theoretically be controlled by the other algorithms producing a S/N map (cADI and PCA), we have shown in previous works \citep{flasseur2018unsupervised,flasseur2018exoplanet,flasseur2018SPIE,flasseur2020robust} that the contrast curves are over-optimistic for these algorithms (i.e., there are significantly more false alarms than expected) due to a miss-modeling of the nuisance component. This claim is also supported by the detection maps given in Figs. 1 to 20 of the  supplementary material for which the number of experienced false alarms is significantly higher than expected at $\text{S/N}=5$. 
For the proposed method, converting pseudo-probabilities into S/N scores is not feasible given that the pseudo-probabilities are very close either to 0 or 1 due to the underlying binary pixel-wise classification task considered at training time. For this reason, we can only check empirically that the targeted false alarm rate at $5\sigma$ is satisfied. To do so, we capitalize on our experiments with synthetic sources by counting the number of false alarms, i.e. detection blobs above the threshold $\tau=0.5$, that do not correspond to the location of an injected synthetic source. The number of false alarms is then converted into an empirical PFA by dividing the number of counts by the total number of possible detection blobs (each with a radius of one resolution element, i.e. four pixels radius for SPHERE-IRDIS observations) within the detection maps. By applying this procedure on several dozens of detection maps obtained with the proposed algorithm, we experienced in average a lower or equal empirical PFA than statistically expected at $5\sigma$. As a conclusion of this study, the contrast estimates we will derive for the proposed approach can be fairly compared with the \PACO results since they correspond to similar probabilities of false alarms and of detections. Figure \ref{fig:HIP88399_20180411_IRDIS_ADI_roc_population} illustrates the capability of the proposed method to detect fainter sources than \PACO. 
From the large number of synthetic sources classified as true, missed and false detections in Fig. \ref{fig:HIP88399_20180411_IRDIS_ADI_roc_population}, we now derive the contrast curve of each algorithm. Concerning the proposed approach, given that we showed empirically that the probability of false alarms is controlled at $5\sigma$, it simply remains to compute the contrast level for which and equal amount of true and of missed detections is experienced. This procedure is repeated for the full range of angular separations with a sliding window of 0.05 arcsec wide. The same procedure is applied for the other algorithms. Figure \ref{fig:HIP88399_20180411_IRDIS_ADI_roc_graph_contrast_vs_separation_summary} summarizes the resulting contrast curves obtained with the five considered algorithms. The proposed method achieves the best detection sensitivity with an improvement in contrast up to a factor four with respect to the \PACO algorithm. We also compare the detection sensitivity of \dPACO with the fundamental detection limit driven by the photon noise. The procedure to compute the photon noise limit is based on a careful evaluation of the contribution of the different sources of noise (i.e., photon, thermal background, and detector readout noise) combined with a statistical evaluation of the underlying S/N. This procedure will be described in details in a paper currently in preparation. We observe that \dPACO can reach for some datasets (see e.g. HIP 88399 (2018-04-11) in Fig. \ref{fig:HIP88399_20180411_IRDIS_ADI_roc_graph_contrast_vs_separation_summary}, and HIP 88399 (2015-05-10) as well as HIP 88399 (2016-04-16) in the supplementary material) at large separations the best achievable detection limit driven by the photon noise, which corresponds to an optimal unmixing between the signal of the sources of interest and the nuisance component. Near the star, an important gap remains (by a factor 5 to 10) between the actual performance and the theoretical lower limit, which a sign of a lack of angular diversity in this area. In practice, the gap between the actual contrast and the lower bound limit depends on several characteristics such as the quality and the stability of the observing conditions, the total amount of parallactic rotation, the number $T$ of temporal frames, etc., as illustrated in Figs. 21 and 22 of the supplementary materials. Reducing this gap at close separation could be addressed by investigating ways to perform the training step of our model from various datasets for which the presence of sources at similar locations is very unlikely. This adaptation requires specific developments that are left for future work.

\begin{figure}
	\begin{center}
		\includegraphics[width=0.48\textwidth]{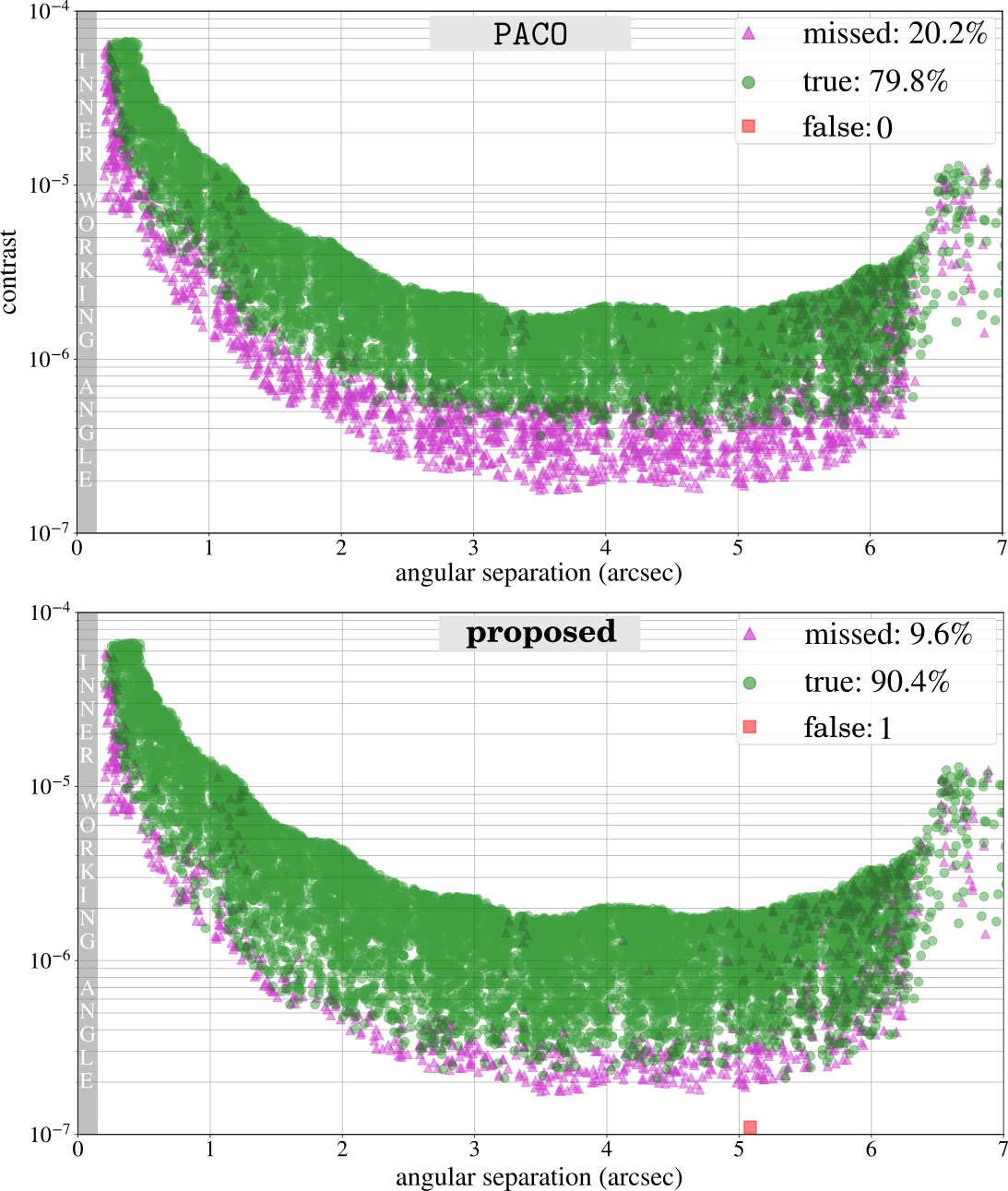}
		\caption{Detection results for 10,000 synthetic sources in a diagram plotting contrast \textit{versus} angular separation. Each synthetic source is classified as missed, true or false detection. The angular separation of false alarms is reported on the $x$-axis, and they are not associated to contrast value since they do not correspond to injected synthetic sources. Dataset: HIP 88399 (2018-04-11), see Sect. \ref{sec:results} for the observation logs.}
		\label{fig:HIP88399_20180411_IRDIS_ADI_roc_population}
	\end{center}
\end{figure}

\begin{figure}
	\begin{center}
		\includegraphics[width=0.48\textwidth]{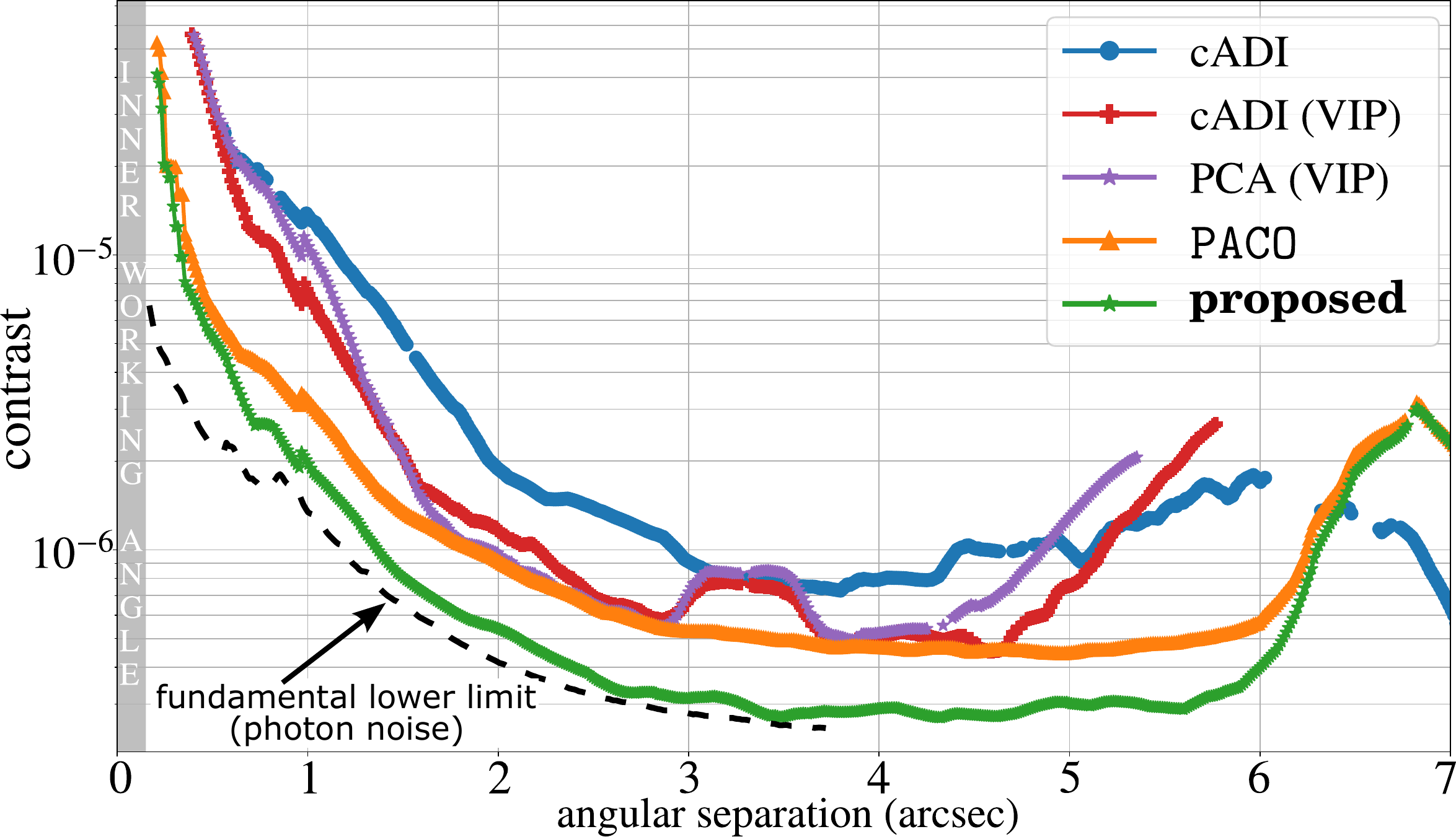}
		\caption{Contrast as a function of the angular separation. These results are based on the classification between true and missed detections of massively injected synthetic sources at various levels of contrast, see Fig. \ref{fig:HIP88399_20180411_IRDIS_ADI_roc_population}. It can be noted that the contrast curves of cADI and PCA are over-optimistic (see text) since the targeted PFA is not reached. For that reason, the angular separations leading to a PFA locally higher than ten times the targeted PFA at $5\sigma$ are not reported. The black dashed line represents the ultimate detection limit driven by the photon noise. Dataset: HIP 88399 (2018-04-11), see Sect. \ref{sec:results} for the observation logs.}
\label{fig:HIP88399_20180411_IRDIS_ADI_roc_graph_contrast_vs_separation_summary}
	\end{center}				
\end{figure}

\begin{figure}
	\begin{center}
		\includegraphics[width=0.48\textwidth]{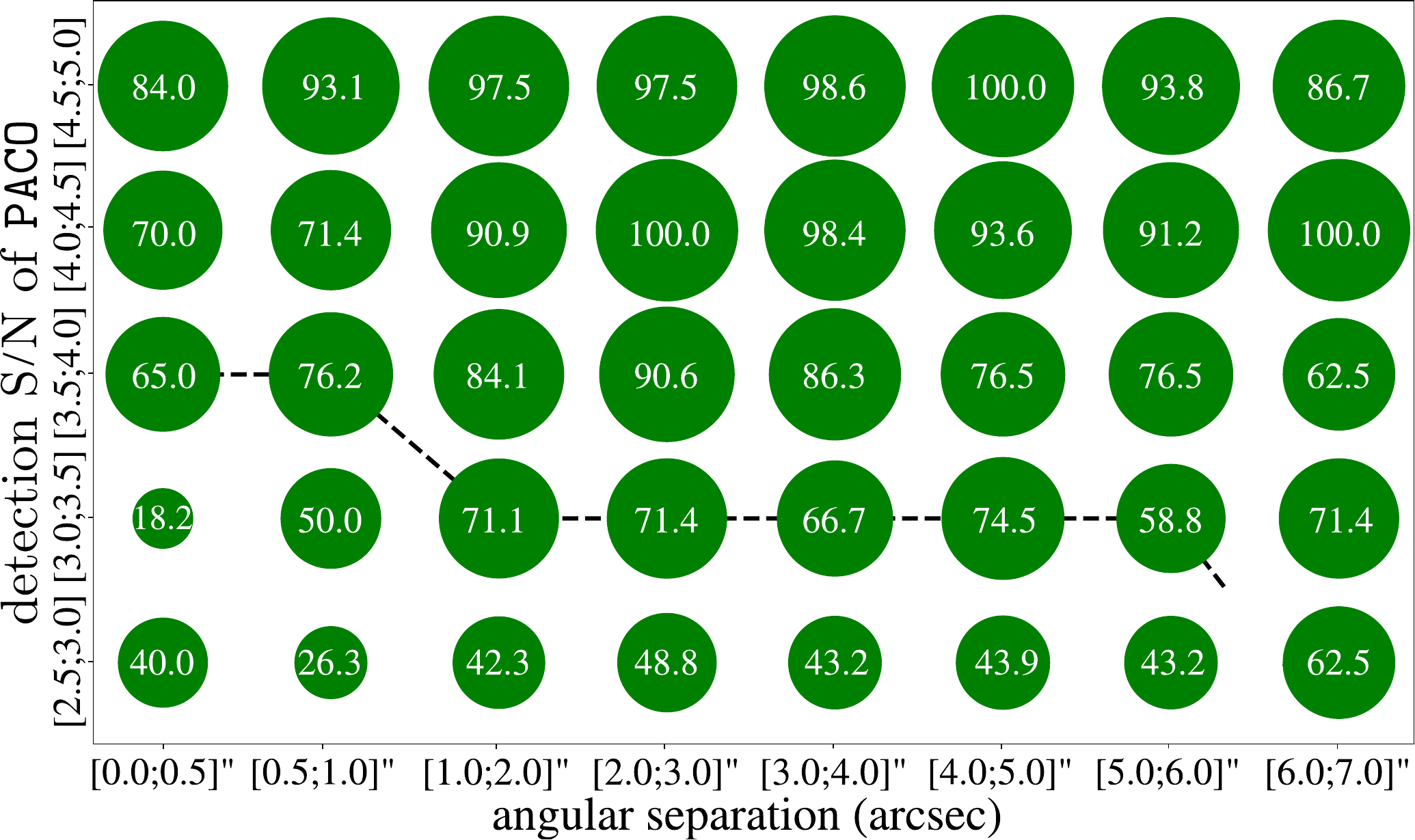}
		\caption{TPR (in percent) of \dPACO for synthetic sources missed by \PACO (i.e., for $\text{S/N} \le 5$) as a function of the angular separation (on the $x$-axis) and of the \PACO's S/N of detection (on the $y$-axis). The black dashed line represents the equivalent \PACO's detection threshold to reach $\text{TPR} = 50\%$ with \dPACO (we recall that the classical detection threshold at $5\sigma$ with \PACO also corresponds to $\text{TPR} = 50\%$). Dataset: HIP 88399 (2018-04-11), see Sect. \ref{sec:results} for the observation logs.}
	\label{fig:HIP88399_20180411_IRDIS_ADI_roc_scatter_deep_paco}
	\end{center}				
\end{figure}

\begin{figure*}	
	\begin{center}
		\includegraphics[width=\textwidth]{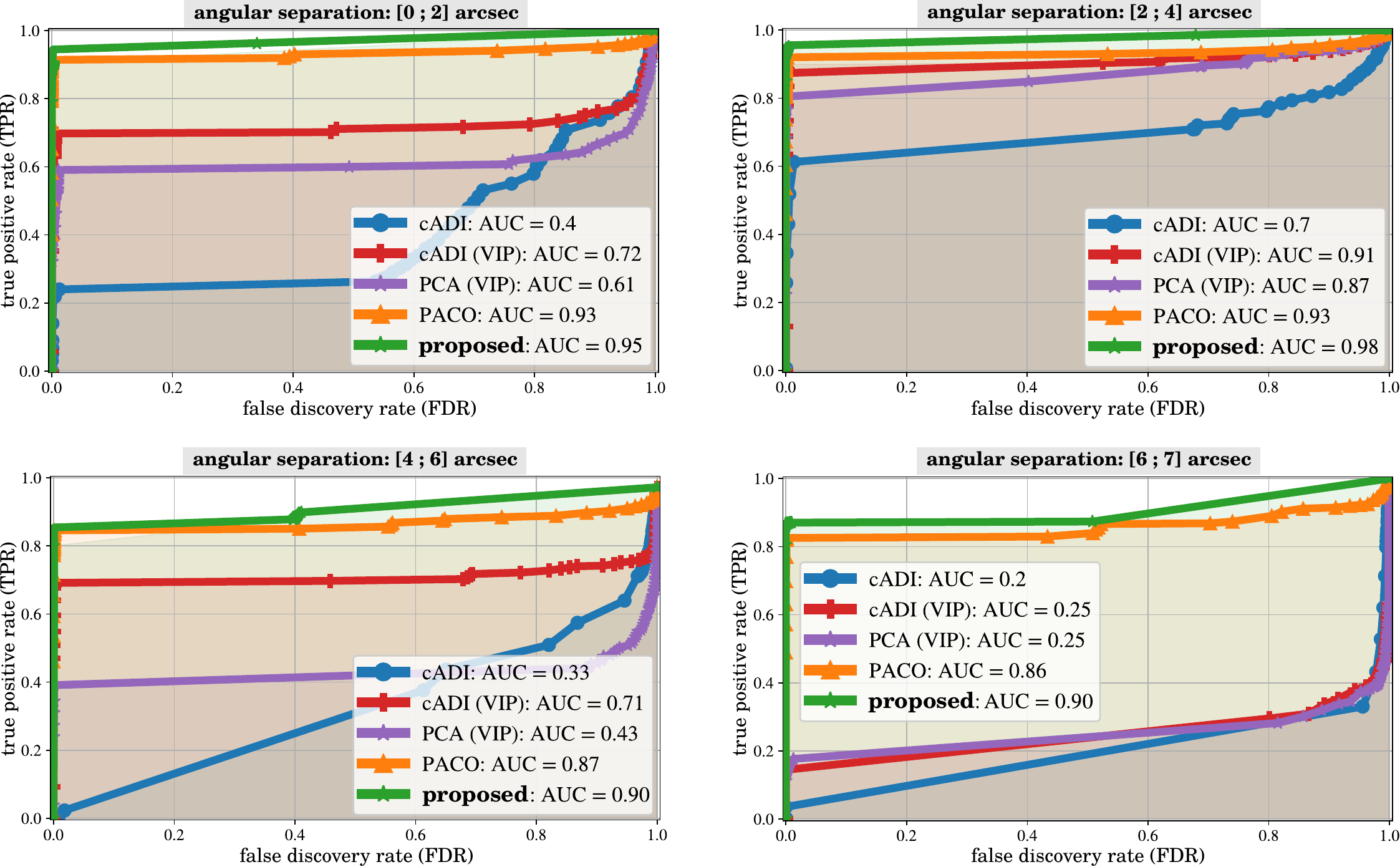}
		\caption{ROCs showing the TPR as a function of the FDR for injected synthetic sources. Dataset: HIP 88399 (2018-04-11), see Sect. \ref{sec:results} for the observation logs.}
		\label{fig:HIP88399_20180411_IRDIS_ADI_TPR_f(FPR_FDR)_synthetic_sources_only_detailed_ang_sep}
	\end{center}
\end{figure*}

\begin{table}
	\caption{Mean results of AUC for ROCs giving the TPR as a function of the FDR. The scores are averaged over the eleven SPHERE-IRDIS datasets considered in this paper, see Sect. \ref{sec:results} for the observation logs. Only massively injected sources (10,000 per dataset) were considered. Figures 8 of \citep{flasseur2022exoplanet} displays the corresponding ROC from which these mean results were aggregated.}
	\label{tab:roc_mean_results}
	\centering
	\begin{tabular}{cccccc}
		\toprule
		\textbf{sep. ('')} & \textbf{cADI} & \textbf{cADI (VIP)} & \textbf{PCA (VIP)} & \textbf{PACO} & \textbf{proposed}\\
		\midrule
		$\left[ 0 ; 2 \right]$ & 0.53 & 0.60 & 0.62 & 0.81 & \textbf{0.91}\\
		$\left[ 2 ; 4 \right]$ & 0.60 & 0.65 & 0.67 & 0.83 & \textbf{0.92}\\
		$\left[ 4 ; 6 \right]$ & 0.11 & 0.59 & 0.55 & 0.71 & \textbf{0.88}\\
		$\left[ 6 ; 7 \right]$ & 0.15 & 0.38 & 0.33 & 0.77 & \textbf{0.90}\\
		\bottomrule
	\end{tabular}
\end{table} 

Figure \ref{fig:HIP88399_20180411_IRDIS_ADI_roc_scatter_deep_paco} focuses on the comparison between \PACO and the proposed \dPACO algorithm. It represents the TPR of \dPACO for synthetic sources missed by \PACO (i.e., for $\text{S/N} \le 5$) as a function of the angular separation and of the \PACO's S/N of detection. For instance, on this dataset about 65\% of sources below 0.5 arcsec with a S/N of detection between 3.5 and 4.0 with \PACO are detected with \dPACO (i.e., above the detection threshold $\tau=0.5$). Similarly, more than 86\% of sources between 2.0 and 4.0 arcsec with a S/N of detection between 3.5 and 4.0 with \PACO are detected with \dPACO (above the detection threshold $\tau=0.5$). For these two examples, achieving the same TPR with \PACO would require to decrease the detection threshold at the price to an increase of the FPR up to a mean factor of 800. Figures 23 and 24 of the supplementary material give similar type of representation than Fig. \ref{fig:HIP88399_20180411_IRDIS_ADI_roc_scatter_deep_paco} for the ten other SPHERE-IRDIS datasets analyzed in this work.

Figure \ref{fig:HIP88399_20180411_IRDIS_ADI_TPR_f(FPR_FDR)_synthetic_sources_only_detailed_ang_sep} gives ROCs representing the TPR as a function of the FDR for the HIP 88399 (2018-04-11) dataset. There results are  obtained with the 10,000 synthetic sources considered for results presented in Fig. \ref{fig:HIP88399_20180411_IRDIS_ADI_roc_population}. The results are split in four different angular separation ranges: $\left[ 0 ; 2 \right]$, $\left[ 2 ; 4 \right]$, $\left[ 4 ; 6 \right]$, and $\left[ 6 ; 7 \right]$ arcsec. Table \ref{tab:roc_mean_results} complements this study by presenting averaged results over the eleven SPHERE-IRDIS datasets we consider in this study, and the detailed scores for each dataset are given in Table 2 of the supplementary material. These results illustrate again the benefits of the proposed method in terms of precision-recall trade-off: the AUC under ROC is improved by 9 to 17$\%$ with respect to the best comparative algorithm for the four angular separation ranges we consider.

\subsubsection{Identification of candidate background point-like sources}
\label{subsubsec:confirmation_bckgd_sources}

\begin{figure}
	\begin{center}
		\includegraphics[width=0.5\textwidth]{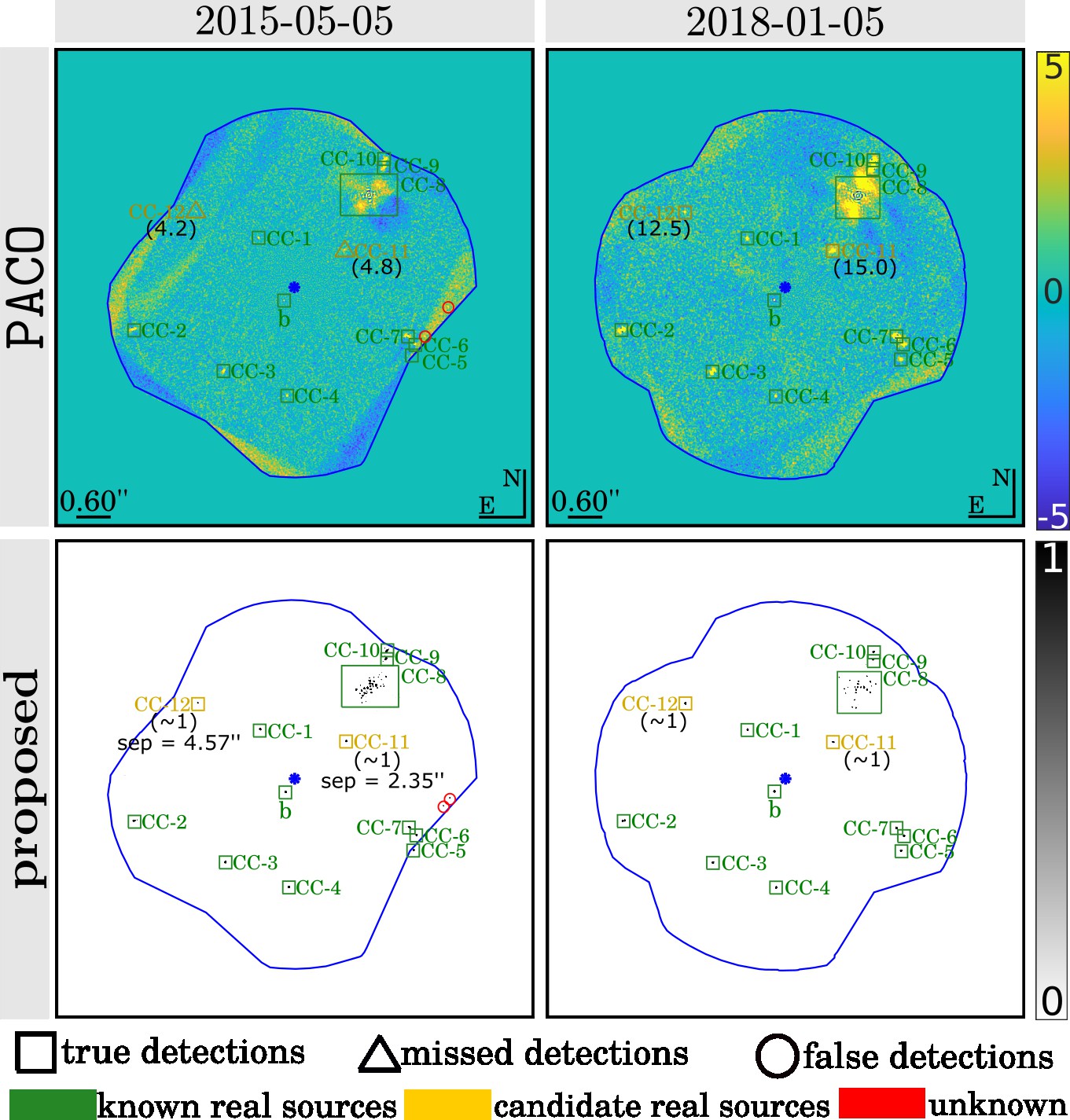}
		\caption{Detection maps obtained with \PACO and \dPACO on two epochs of the HD 95086 star observed with SPHERE-IRDIS. Symbols classify sources as true, missed, and false detections based on our analysis. The detection threshold was set to $\tau=5$ for \PACO producing a S/N map, and to $\tau=0.5$ for \dPACO producing a pseudo-probability map. Colors classify sources as known real sources, candidate (background) sources, and detections with unknown status. Datasets: HD 95086 (2015-05-05 and 2018-01-05), see Sect. \ref{sec:results} for the observation logs.}
	\label{fig:hd95086_2_sources}
	\end{center}
\end{figure}

In this section, we take the example of a joint analysis of datasets of HD 95086 considered in this work to illustrate the ability of the proposed approach to detect candidate faint point-like sources. Figure  \ref{fig:hd95086_2_sources} shows the detection maps obtained with \PACO and the proposed \dPACO approach on the 2015 and 2018 observations of HD 95086. It emphasizes that two candidate point-like sources (respectively denoted CC-11 and CC-12) are detected with \PACO at a S/N above 5 only for one of the two epochs (in 2018, which is also one of the best SPHERE-IRDIS observations of this star) while they remain just below the classical detection threshold at $5\sigma$ on the 2015 epoch. These two point-like sources are detected with \dPACO on the same two epochs. Based on their astrometry, these two faint point sources are co-moving with the other background stars seen in the projected field of view of the instrument, so that they could be background stars too\footnote{The goal of this paper is not to study in details these CCs, but rather to present a new post-processing algorithm. These CCs should be taken with caution, and a detailed analysis is needed, in particular to exclude systematic sources of errors.}. These candidate point-like sources are not reported in the last detailed study of the HD 95086 architecture based on the analysis of ten SPHERE-IRDIS datasets (including the 2015 epoch but not the 2018 epoch) with classical post-processing algorithms (i.e., cADI, PCA, TLOCI), see Fig. 2 of \cite{chauvin2018investigating}. The analysis of more recent and better datasets (including the 2018 one) in the SHINE survey \citep{langlois2021sphere} of the SPHERE instrument allows to identify by visual inspection (i.e., not based on a strict measure of S/N above the classical detection threshold at $5\sigma$) CC-11 while CC-12 has not been identified yet. The detection of the two CCs in the worst epoch data (2015-05-05) emphasizes again the benefits of the proposed \dPACO algorithm. This example also illustrates the complementarity of \PACO and \dPACO. Even if \dPACO does not provide tight confidence scores (i.e., its outputs can not be directly interpreted in terms of a S/N), it allows to identify candidate point-like sources. Given the locations found by \dPACO, an estimation of the PFA can be derived from the S/N extracted on the \PACO detection maps at the same locations. Using this procedure, the theoretical PFA (not including systematic sources of errors) for the co-localization of the candidate point sources CC-11 and CC-12 in the two epochs is small. None of CC-11 and CC-12 are detected in the 2021 epoch of HD 95086, as shown by Figs. 3 and 4 of the supplementary material. This observation is consistent with the fact that the achievable contrast is worse for this dataset than for the 2015 and 2018 observations, likely because the observing conditions were quite average for the 2021 observations and that only half of the total integration time was spent on the target star (the 2021's epoch being recorded with the star-hopping technique, and the resulting reference dataset remaining not exploited in this paper). 

\PACO and \dPACO detection maps from the 2015 and 2018 epochs also display a few blobs, circled in red, above the detection threshold that are not consistently detected on multiple epochs. These detections are located very near the borders of the (extended) field of view. They are likely artifacts due to the lack of temporal samples in this area to estimate the model parameters. This hypothesis is also supported by the fact that \PACO detection maps are not stationary and does not follow a centered Gaussian distribution with unit variance in this area. \dPACO detection map     from the 2021 epoch display a blob at 0.25'', that we likely identified as a false alarm since we have experienced a few false alarms in the same area during training time due to a strong stellar leakage (see Figs. 3, 4, and 22(b) of the supplementary material). 

\subsection{Characterization results}
\label{subsec:characterization_results}

In Sect. \ref{subsubsec:results_synthetic_sources_characterization}, we ground the performance of the proposed photometry estimation module on the same datasets than the considered ones for the detection stage by resorting to massive injections of synthetic sources. For that purpose, we use the ARE metric (Eq. (\ref{eq:loss_characterization})) averaged per angular separation. We also put in perspective the results of the detection and of the characterization stages to ground the global gain brought by the proposed algorithm. 
In Sect. \ref{subsubsec:faster_procedure}, we propose a fast (and approximate) evaluation procedure suited when large amounts of synthetic sources are considered. Finally, in Sect. \ref{subsubsec:results_real_sources_characterization} we briefly discuss and compare photometry estimations provided by our method with estimations coming from the literature on some known sources (either exoplanets, brown dwarfs or background sources) present in the eleven SPHERE-IRDIS datasets considered in this paper. 

\subsubsection{Characterization of synthetic sources}
\label{subsubsec:results_synthetic_sources_characterization}

\begin{figure*}
	\begin{center}
		\includegraphics[width=\textwidth]{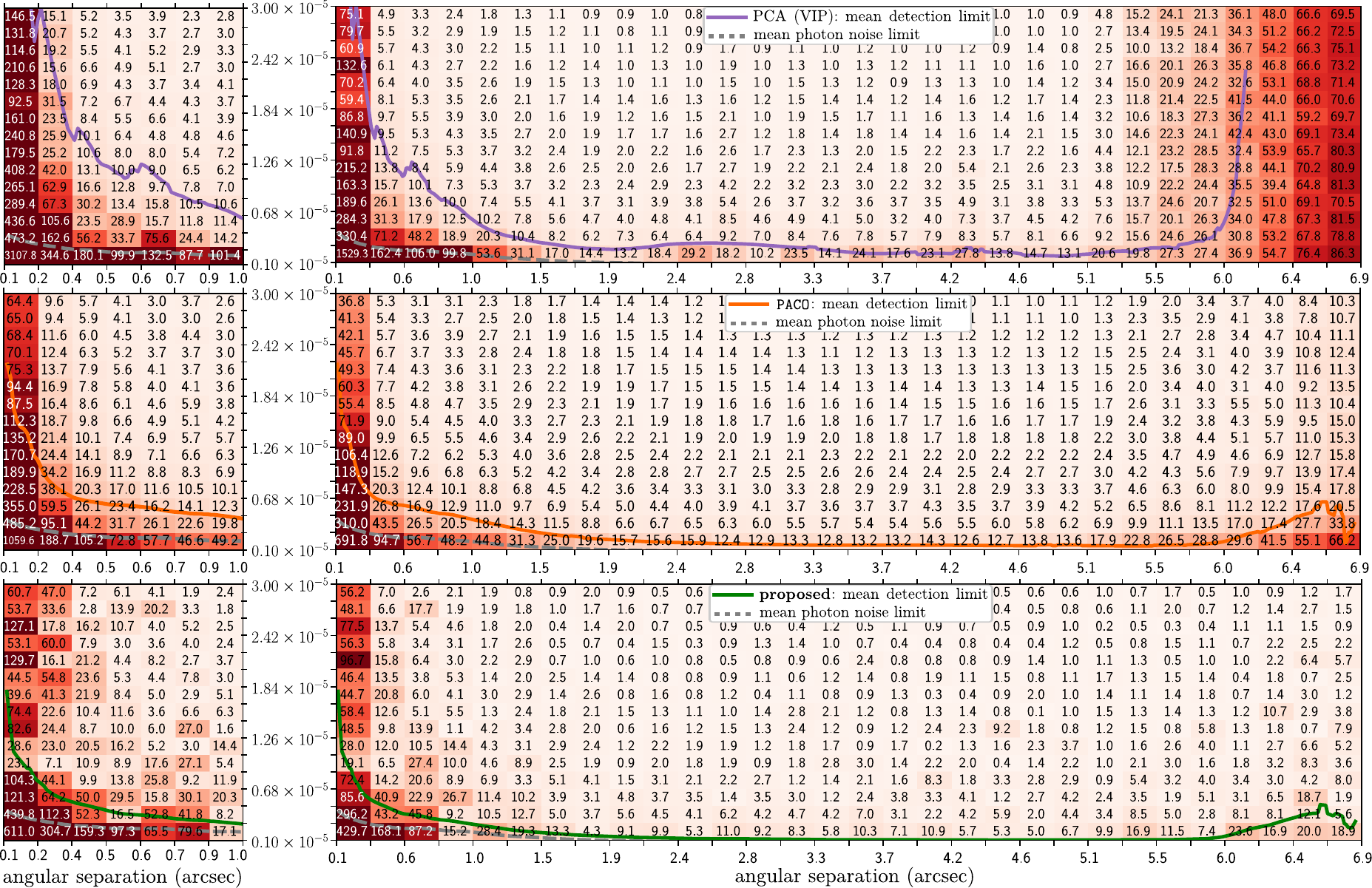}
		\caption{Mean ARE score on the estimated photometry of injected synthetic sources as a function of the their contrast and of their angular separation. From top to bottom, the panels corresponds respectively to PCA (VIP), \PACO, and the proposed algorithm. For each panel, the mean detection limit (straight line) and the mean photon noise limit (dashed line) are superimposed. The results are averaged azimuthally for 40,000 sources of flux drawn uniformly between $1 \times 10^{-6}$ and  $3 \times 10^{-5}$. Datasets: the eleven SPHERE-IRDIS datasets considered in this work, see Sect. \ref{sec:results} for the recording logs.}
	\label{fig:contrast_curve_cells_proposed}
	\end{center}
\end{figure*}

\begin{figure}
	\begin{center}
		\includegraphics[width=0.48\textwidth]{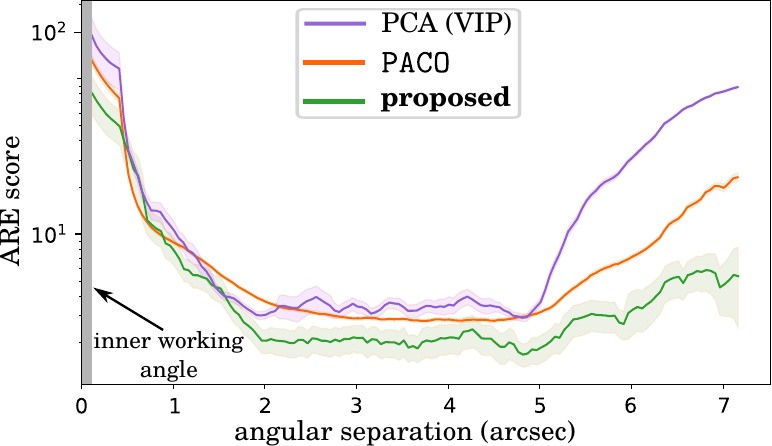}
		\caption{Mean ARE score on the estimated photometry of injected synthetic sources as a function of their angular separation. The results are averaged azimuthally for sources, of flux drawn uniformly between $1 \times 10^{-6}$ and  $3 \times 10^{-5}$, that have been considered in Fig. \ref{fig:contrast_curve_cells_proposed}. Note that because of the azimuthal average, some sources belonging below the detection limit contribute to the display results, especially at short angular separations, see Fig. \ref{fig:contrast_curve_cells_proposed}. Datasets: the eleven SPHERE-IRDIS datasets considered in this work, see Sect. \ref{sec:results} for the recording logs.}
	\label{fig:curves_aggregated_photometry}
	\end{center}
\end{figure}

To assess the performance of our method, we reproduce a \textit{real} setting, hereafter called \textit{reference procedure}, in which we estimate the flux of injected synthetic sources representative of (possibly unknown) exoplanets. 
To do so, we first inject synthetic test sources in the considered dataset, and then extract the patches used for training the model. It is crucial to perform these steps in that order otherwise the model could have the opportunity to learn the residual nuisance component below the test sources, and a bias could be introduced.

Following this strategy, Fig. \ref{fig:contrast_curve_cells_proposed} shows the ARE score on the estimated photometry of synthetic sources massively injected with the reference procedure as a function of the their contrast and of their angular separation. Figure 26 of the supplementary material gives the same type of plots for the throughput (i.e., the ratio $\widehat{\alpha} / \alpha$ between the retrieved and ground-truth source's contrast). In both cases, results are averaged over the eleven SPHERE-IRDIS datasets of this study. Figure \ref{fig:curves_aggregated_photometry} complements this study by aggregating the ARE score over the source contrast. These results show that the proposed algorithm leads, in average, to better characterization performance than PCA (VIP) and \PACO for angular separations larger than 0.8'', with a reduction of the ARE by a factor between 1.10 and 10 with respect to PCA (VIP), and by a factor between 1.10 and 5 with respect to \PACO. Closer to the star, the advantage is on average to \PACO and to the proposed approach, the best of the two algorithms depending on the dataset and of the source location. The fact that \PACO can perform better than the proposed algorithm is due to the complexity to train a deep model, without leak between the train and the test sets, from a unique dataset of interest. This effect occurs only at short angular separations since this is the region of the field of view where generating multiple non-redundant training samples is the most tricky. As illustrated in Sect. \ref{subsec:detection_synthetic_sources}, this side effect does not occur in the detection stage of the proposed approach thanks to the included whitening procedure, which removes most of the quasi-static speckles (i.e., only residual structures non-captured by the statistical model remain). Except these residual structures that we aim to capture by deep learning, each new training set thus contain, before injection of synthetic sources, a quasi-random realization of uncorrelated Gaussian noise. This key property prevents a leak between the train and the test sets as well as the memorization of the nuisance structures by the network during its training.

\subsubsection{Efficient (and approximate) evaluation procedure}
\label{subsubsec:faster_procedure}

\begin{figure}
	\begin{center}
		\includegraphics[width=0.48\textwidth]{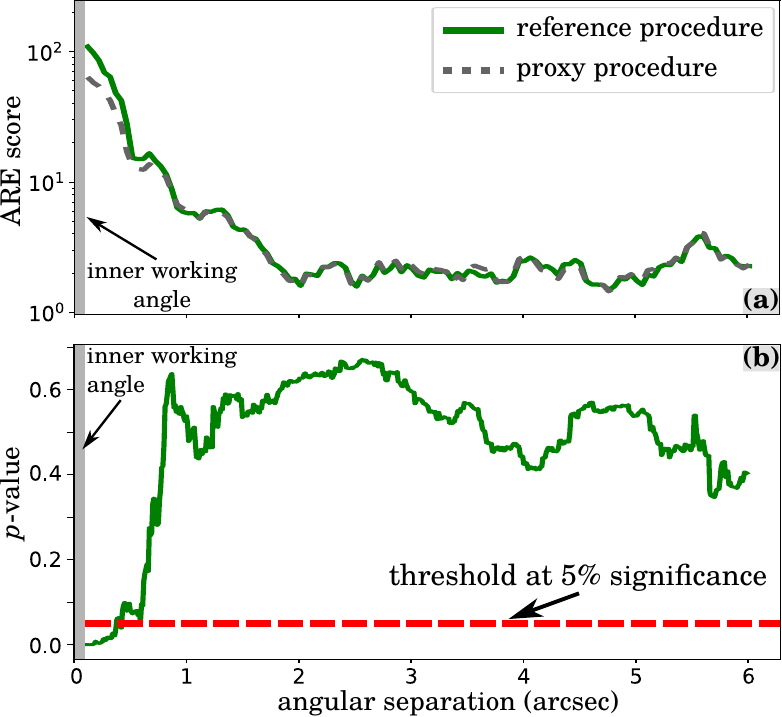}
		\caption{Comparison between the reference procedure and its proxy version for evaluation of the characterization performance of the proposed algorithm. (a) Mean ARE scores (see definition in Eq. (\ref{eq:loss_characterization})) are reported as a function of the angular separation. (b) $p$-values are reported as a function of the angular separation, and can be compared with the 5\% significance threshold given by the dashed red curve. Datasets: the eleven SPHERE-IRDIS datasets considered in this work, see Sect. \ref{sec:results} for the recording logs.}
	\label{fig:bias_and_statistical_test_characterization}
	\end{center}
\end{figure}

\noindent When the massive injection of synthetic sources withing multiple datasets is needed to ground the performance of a built detector, the reference procedure described and applied in Sect. \ref{subsubsec:results_synthetic_sources_characterization} is computationally expensive. The computational bottleneck is related to the need to train a new model for every test dataset containing a dozen of synthetic test sources. As an illustration, 30 different models must be trained to get a test set of only 300 samples. In this context and in a similar fashion to  the detection stage (see Sect. \ref{subsec:detection_results_known_real_sources} and Fig. \ref{fig:reference_proxy_procedures}), we also propose a fast and approximate version of the reference procedure, that is referred as the \textit{proxy procedure} in the following. Instead of training a different model for each subset of injected sources, we train a unique model without additional synthetic sources mimicking the behaviour of real sources. Synthetic sources are injected \textit{a posteriori}, i.e., after training. Finally, the trained model is applied on the resulting dataset. The gain in terms of algorithmic complexity (about a factor 2,000 with in average 5 synthetic sources in each subset) is similar to the one brought by the proxy procedure of the detection step. 

At this stage, we have to measure the ability of the proxy procedure to provide a fair estimate of the real performance that would be achieved with the reference procedure. Figure \ref{fig:bias_and_statistical_test_characterization}(a) compares the performance of the two procedures in terms of mean ARE on the eleven SPHERE-IRDIS datasets considered in this paper. In these experiments, we considered around 3,300 synthetic sources spread over the whole field of view. We observe that the proxy procedure leads to results very close to the ones provided by the reference procedure, excepted near the star where the proxy procedure is over-optimistic, i.e. a positive bias is present (up to a factor five, at worst). This bias at short angular separations can be attributed to \textit{data leakages} between the train and the test sets in the absence of whitening procedure. Basically, as for the detection part, the model is data-dependent and the absence of whitening procedure as well as of the associated temporal shuffling of the frames induces that some parts of the nuisance component are seen and memorized by the network using the proxy procedure. This effect occurs only at short angular separations since this is the area of the field of view where training patches contain most likely some similar parts of the nuisance component. This hypothesis is also supported by the absence of bias between the proxy and the reference procedures of the detection stage. The latter encompasses a whitening procedure and a temporal shuffling of the frames that prevents data leakages between the train and test sets. 

So, it remains to decide for a suited cut-off in terms of angular separation to switch from the reference to the proxy procedure. For that purpose, we resort to a binary hypothesis test. It performs a paired $t$-test \citep{kendall1948advanced}, where the $\mathcal{H}_0$ (null) hypothesis represents the equality between the reference and the proxy procedures while the $\mathcal{H}_1$ (alternative) hypothesis represents better results with the proxy procedure. From the results presented in Fig. \ref{fig:bias_and_statistical_test_characterization}(a), we conduct in Fig. \ref{fig:bias_and_statistical_test_characterization}(b) this statistical test on the same datasets. It displays the resulting $p$-values as a function of the angular separation, and compares them with a 5\% significance threshold. For the prescribed significance level, the two procedures can be considered as equivalent for angular separations larger than 0.5 arcsec. 

As a conclusion of this study, when the number of synthetic sources for which the photometry should be evaluated at inference time constitutes a computational bottleneck, the efficient and approximate proxy procedure can be safely used in the main part of the field of view and the more expensive procedure of reference should be used near the star.

\subsubsection{Characterization of real sources}
\label{subsubsec:results_real_sources_characterization}

\begin{small}
\begin{table*}
	\caption{Estimated photometry on some known real sources present in the considered datasets. For HD 95086, the (candidate) point-like sources are denoted as in Fig. \ref{fig:hd95086_2_sources}. For HIP 88399 (respectively, HIP 72192), point-like sources are denoted by CC1 to CC5 (respectively, CC1 and CC2) through increasing angular separations. The astrometry (angular separation and true-North angle) has been estimated with \PACO. Photometry estimations extracted from the literature has been obtained with TLOCI by \citep{langlois2021sphere} for (a), TLOCI by \citep{desgrange2022depth} for (b), ANDROMEDA by \citep{desgrange2022depth} for (c), PCA by \citep{chauvin2017discovery} for (e), and TLOCI by \citep{cheetham2019spectral} for (f). Reported uncertainties correspond to three times the estimated standard-deviation.}
	\label{tab:real_sources_photometry}
	\centering
	\hspace*{-4mm}
	\begin{tabular}{ccccc|cccc}
		\toprule
		\multicolumn{5}{c}{\textbf{identification}} & \multicolumn{4}{c}{\textbf{estimated photometry $\widehat{\V \alpha}$ (contrast)}}\\
		\text{target} & \text{obs. date} & \text{source} & ang. sep. ('') & angle (°) & \text{literature} & \text{PCA (VIP)} & \PACO & \textbf{proposed}\\
		\midrule
		HD 95086 & 2015-05-05 & b & 0.619 $\pm$ 0.003 & 148.4 $\pm$ 0.2 & $^\text{(a)}$(1.13 $\pm$ 0.46)$\times 10^{-5}$ & (1.06 $\pm$ 0.06)$\times 10^{-5}$ & (1.14 $\pm$ 0.17)$\times 10^{-5}$ & 1.11$\times 10^{-5}$\\
		& & & & & $^\text{(b)}$(1.10 $\pm$ 0.55)$\times 10^{-5}$ & & \\
		& & & & & $^\text{(c)}$(1.58 $\pm$ 1.75)$\times 10^{-5}$ & \\
		HD 95086 & 2015-05-05 & CC1 & 2.207 $\pm$ 0.003 & 35.1 $\pm$ 0.1 & $^\text{(a)}$(6.19 $\pm$ 1.94)$\times 10^{-6}$ & (6.33 $\pm$ 0.57)$\times 10^{-6}$ & (7.05 $\pm$ 0.72)$\times 10^{-6}$ & 6.41$\times 10^{-6}$\\
		HD 95086 & 2015-05-05 & CC11 & 2.346 $\pm$ 0.006 & 305.4 $\pm$ 0.2 & -- & (2.06 $\pm$ 4.61)$\times 10^{-6}$ & (3.14 $\pm$ 0.70)$\times 10^{-6}$ & 1.80$\times 10^{-6}$\\
		HD 95086 & 2015-05-05 & CC12 & 4.546 $\pm$ 0.010 & 52.3 $\pm$ 0.1 & -- & (2.53 $\pm$ 1.31)$\times 10^{-6}$ & (2.70 $\pm$ 0.90)$\times 10^{-6}$ & 1.94$\times 10^{-6}$\\
		\midrule
		HD 95086 & 2018-01-05 & b & 0.617 $\pm$ 0.002 & 145.7 $\pm$ 0.2 & $^\text{(b)}$(1.20 $\pm$ 0.32)$\times 10^{-5}$ & (1.23 $\pm$ 0.23)$\times 10^{-5}$ & (1.20 $\pm$ 0.06)$\times 10^{-5}$ & 1.18$\times 10^{-5}$\\
		& & & & & $^\text{(c)}$(1.32 $\pm$ 0.35)$\times 10^{-5}$ & \\
		HD 95086 & 2018-01-05 & CC1 & 2.241 $\pm$ 0.002 & 37.4 $\pm$ 0.1 & -- & (6.25 $\pm$ 1.88)$\times 10^{-6}$ & (6.59 $\pm$ 0.37)$\times 10^{-6}$ & 6.39$\times 10^{-6}$\\
		HD 95086 & 2018-01-05 & CC11 & 2.262 $\pm$ 0.004 & 305.9 $\pm$ 0.1 & -- & (2.56 $\pm$ 0.28)$\times 10^{-6}$ & (2.75 $\pm$ 0.35)$\times 10^{-6}$ & 1.90$\times 10^{-6}$\\
		HD 95086 & 2018-01-05 & CC12 & 4.606 $\pm$ 0.006 & 53.0 $\pm$ 0.1 & -- & (2.73 $\pm$ 3.00)$\times 10^{-6}$ & (2.39 $\pm$ 0.40)$\times 10^{-6}$ & 2.40$\times 10^{-6}$\\
		\midrule
		HD 95086 & 2021-03-11 & b & 0.630 $\pm$ 0.003 & 142.4 $\pm$ 0.2 & -- & (1.21 $\pm$ 0.23)$\times 10^{-5}$ & (1.25 $\pm$ 0.18)$\times 10^{-5}$ & 1.20$\times 10^{-5}$\\
		HD 95086 & 2021-03-11 & CC1 & 2.293 $\pm$ 0.005 & 40.8 $\pm$ 0.1 & -- & (5.66 $\pm$ 0.73)$\times 10^{-6}$ & (6.30 $\pm$ 1.56)$\times 10^{-6}$ & 5.48$\times 10^{-6}$\\
		HD 95086 & 2021-03-11 & CC11 & \textit{S/N too low} & \textit{S/N too low} & \textit{S/N too low} &  \textit{S/N too low} & \textit{S/N too low} & \textit{S/N too low}\\
		HD 95086 & 2021-03-11 & CC12 & \textit{S/N too low} & \textit{S/N too low} & \textit{S/N too low} &  \textit{S/N too low} & \textit{S/N too low} & \textit{S/N too low}\\
		\midrule
		HIP 88399 & 2015-05-10 & CC1 & 1.302 $\pm$ 0.002 & 210.4 $\pm$ 0.1 & -- & (7.09 $\pm$ 0.72)$\times 10^{-6}$ & (7.39 $\pm$ 0.66)$\times 10^{-6}$ & 6.67$\times 10^{-6}$\\
		HIP 88399 & 2015-05-10 & CC2 & 3.488 $\pm$ 0.006 & 142.4 $\pm$ 0.1 & $^\text{(a)}$(7.94 $\pm$ 2.10)$\times 10^{-6}$ & (7.10 $\pm$ 0.36)$\times 10^{-6}$ & (7.56 $\pm$ 0.27)$\times 10^{-6}$ & 7.61$\times 10^{-6}$\\
		HIP 88399 & 2015-05-10 & CC3 & 4.281 $\pm$ 0.006 & 227.7 $\pm$ 0.1 & -- & (2.36 $\pm$ 0.17)$\times 10^{-7}$ & (3.42 $\pm$ 1.03)$\times 10^{-7}$ & 2.52$\times 10^{-7}$\\
		HIP 88399 & 2015-05-10 & CC4 & 5.208 $\pm$ 0.003 & 322.1 $\pm$ 0.1 & $^\text{(a)}$(5.11 $\pm$ 6.00)$\times 10^{-7}$ & (5.57 $\pm$ 5.06)$\times 10^{-7}$ & (4.89 $\pm$ 0.95)$\times 10^{-7}$ & 4.94$\times 10^{-7}$\\
		HIP 88399 & 2015-05-10 & CC5 &\textit{S/N too low} & \textit{S/N too low} & \textit{S/N too low} &  \textit{S/N too low} & \textit{S/N too low} & \textit{S/N too low}\\
		\midrule
		HIP 88399 & 2016-04-16 & CC1 & 1.237 $\pm$ 0.003 & 212.8 $\pm$ 0.1 & $^\text{(a)}$(9.38 $\pm$ 6.99)$\times 10^{-6}$ & (1.04 $\pm$ 0.13)$\times 10^{-5}$ & (1.22 $\pm$ 0.18)$\times 10^{-5}$ & 1.11$\times 10^{-5}$\\
		HIP 88399 & 2016-04-16 & CC2 & 3.422 $\pm$ 0.001 & 141.7 $\pm$ 0.1 & $^\text{(a)}$(1.32 $\pm$ 2.97)$\times 10^{-5}$ & (1.47 $\pm$ 0.05)$\times 10^{-5}$ & (1.64 $\pm$ 0.05)$\times 10^{-5}$ & 1.57$\times 10^{-5}$\\
		HIP 88399 & 2016-04-16 & CC3 & 4.240 $\pm$ 0.003 & 228.6 $\pm$ 0.1 & $^\text{(a)}$(5.35 $\pm$ 4.20)$\times 10^{-7}$ & (5.10 $\pm$ 4.65)$\times 10^{-7}$ & (7.64 $\pm$ 1.36)$\times 10^{-7}$ & 5.78$\times 10^{-7}$\\
		HIP 88399 & 2016-04-16 & CC4 & 5.282 $\pm$ 0.003 & 322.6 $\pm$ 0.1 & $^\text{(a)}$(5.11 $\pm$ 6.00)$\times 10^{-7}$ & (7.21 $\pm$ 1.14)$\times 10^{-7}$ & (8.02 $\pm$ 1.25)$\times 10^{-7}$ & 7.58$\times 10^{-7}$\\
		HIP 88399 & 2016-04-16 & CC5 & 6.561 $\pm$ 0.003 & 201.8 $\pm$ 0.1 & -- & (1.10 $\pm$ 2.19)$\times 10^{-6}$ & (1.22 $\pm$ 0.94)$\times 10^{-6}$ & 1.18$\times 10^{-6}$\\	
		\midrule
		HIP 88399 & 2018-04-11 & CC1 & 1.106 $\pm$ 0.001 & 217.9 $\pm$ 0.1 & $^\text{(a)}$(1.61 $\pm$ 1.27)$\times 10^{-5}$ & (2.01 $\pm$ 0.25)$\times 10^{-5}$ & (2.22 $\pm$ 0.15)$\times 10^{-5}$ & 2.22$\times 10^{-5}$\\
		HIP 88399 & 2018-04-11 & CC2 & 3.289 $\pm$ 0.001 & 140.1 $\pm$ 0.1 & $^\text{(a)}$(2.21 $\pm$ 1.74)$\times 10^{-5}$ & (2.65 $\pm$ 2.19)$\times 10^{-5}$ & (2.82 $\pm$ 0.06)$\times 10^{-5}$ & 2.88$\times 10^{-5}$\\
		HIP 88399 & 2018-04-11 & CC3 & 4.140 $\pm$ 0.003 & 230.3 $\pm$ 0.1 & $^\text{(a)}$(8.79 $\pm$ 6.90)$\times 10^{-7}$ & (1.10 $\pm$ 0.46)$\times 10^{-6}$ & (9.70 $\pm$ 1.31)$\times 10^{-7}$ & 1.03$\times 10^{-6}$\\
		HIP 88399 & 2018-04-11 & CC4 & 5.411 $\pm$ 0.002 & 323.6 $\pm$ 0.1 & $^\text{(a)}$(1.21 $\pm$ 0.95)$\times 10^{-6}$ & (1.31 $\pm$ 0.25)$\times 10^{-6}$ & (1.40 $\pm$ 0.12)$\times 10^{-6}$ & 1.38$\times 10^{-6}$\\
		HIP 88399 & 2018-04-11 & CC5 & 6.429 $\pm$ 0.002 & 202.5 $\pm$ 0.1 & -- & (4.68 $\pm$ 1.85)$\times 10^{-7}$ & (2.83 $\pm$ 0.33)$\times 10^{-6}$ & 2.81$\times 10^{-6}$\\
		\midrule
		HD 131399 & 2015-06-12 & Ab & 0.845 $\pm$ 0.004 & 195.6 $\pm$ 0.2 & $^\text{(d)}$(7.94 $\pm$ 2.10)$\times 10^{-6}$ & (9.36 $\pm$ 8.01)$\times 10^{-6}$ & (1.05 $\pm$ 0.17)$\times 10^{-5}$ & 8.23$\times 10^{-6}$\\
		\midrule
		HD 131399 & 2016-05-07 & Ab & 0.825 $\pm$ 0.003 & 194.2 $\pm$ 0.2 & $^\text{(d)}$(7.66 $\pm$ 2.21)$\times 10^{-6}$ & (7.58 $\pm$ 2.10)$\times 10^{-6}$ & (7.92 $\pm$ 1.37)$\times 10^{-6}$ & 7.86$\times 10^{-6}$\\
		\midrule
		HIP 65426 & 2017-02-09 & b & 0.834 $\pm$ 0.001 & 149.9 $\pm$ 0.1 & $^\text{(e)}$(9.91 $\pm$ 7.38)$\times 10^{-5}$ & (6.20 $\pm$ 0.46)$\times 10^{-5}$ & (5.87 $\pm$ 0.17)$\times 10^{-5}$ & 6.17$\times 10^{-5}$\\
		\midrule
		HIP 65426 & 2018-05-13 & b & 0.825 $\pm$ 0.001 & 149.9 $\pm$ 0.1 & $^\text{(a)}$(8.47 $\pm$ 2.24)$\times 10^{-5}$ & (9.08 $\pm$ 2.44)$\times 10^{-5}$ & (8.98 $\pm$ 0.09)$\times 10^{-5}$ & 9.06$\times 10^{-5}$\\
		& & & & & $^\text{(f)}$(8.39 $\pm$ 2.22)$\times 10^{-5}$ &\\
		\midrule
		HIP 72192 & 2015-06-11 & CC1 & 4.632 $\pm$ 0.012 & 166.7 $\pm$ 0.1 & -- & (2.94 $\pm$ 2.31)$\times 10^{-6}$ & (3.24 $\pm$ 1.07)$\times 10^{-6}$ & 2.88$\times 10^{-6}$\\
		HIP 72192 & 2015-06-11 & CC2 & 4.891 $\pm$ 0.006 & 342.8 $\pm$ 0.1 & -- & (5.44 $\pm$ 1.85)$\times 10^{-6}$ & (6.07 $\pm$ 1.10)$\times 10^{-6}$ & 5.95$\times 10^{-6}$\\
		\bottomrule
	\end{tabular}
\end{table*} 
\end{small}

In this section, we compare the photometry estimations obtained with the tested algorithms on some real known sources present in the considered datasets. When available, we also compare them with measurements published in the literature that have been obtained either with the SPHERE Data Center implementations of TLOCI and PCA or with ANDROMEDA (for HD 95086 b only). For PCA (VIP), the reported uncertainties are computed from the residual combined image (i.e., after subtraction of the on-axis PSF, derotation, and stacking) in an annulus at the same angular separation than the source of interest. For \PACO, the photometry is estimated by accounting for the spatial correlations of the nuisance component. As concerns the proposed algorithm, and as for its detection stage, the control of the uncertainties is an intricate task given that the core of the deep model works as a \textit{black-box}. For that reason, we did not produce estimations of the standard-deviation, which is left for future work. For the experiments we conduct with PCA (VIP), \PACO and the proposed method, we use the same pre-reduction of the raw observations and the same off-axis PSF template that has been measured prior the science observations. It is not ensured that these specificity also hold for photometry estimates extracted from the literature, which could induce  (unknown) systematic variations that are not taken into account. Similarly, given the variability of the observing conditions between different observations, absolute comparisons between multiple epochs of the same target should be done with caution.

\medskip

Table \ref{tab:real_sources_photometry} reports the photometry estimations on known real sources. Even in the absence of absolute ground-truth, we can make some relative comparisons for a given dataset and for a given source. In that view, photometry estimates produced by the proposed method are compatible with published and/or obtained results with PCA (VIP) and \PACO for almost all sources. The largest discrepancy is for CC5 of HIP 88399 (2018-04-11), where a factor six lies between the estimates from PCA (VIP) and the proposed algorithm. Very likely, the PCA (VIP) estimate is not reliable as the source is located at large angular separation, i.e. in an area of the field of view where PCA (VIP) is prone to large errors (see Fig. \ref{fig:contrast_curve_cells_proposed}). Besides, \PACO and the proposed method lead to quite close estimations, which also supports the previous claim. 

Photometry estimates of CC-12 that we identified from HD 95086 datasets in Sect. \ref{subsubsec:confirmation_bckgd_sources} are consistent among the different algorithms. They are also consistent between the two epochs where CC-12 has been detected. For CC-11 that we identified from the same datasets, we note a discrepancy by about 40\% between the estimates obtained (i) by the proposed method, and (ii) by the other tested algorithms. However, this two groups of estimations are consistent between the 2015 and 2018 epochs. The discrepancy could be attributed to the presence of a bright background source (denoted by CC-8 in Fig. \ref{fig:hd95086_2_sources}) that could lead to an overestimation of the photometry with algorithms that do not rely on a training step with synthetic sources. In any case, these two candidate point-like sources should be considered with caution. 

\section{Conclusion}
\label{sec:conclusion}

\noindent We have described the key principles of a new algorithm for detecting and characterizing point-like sources at high contrast from ADI observations. The detection stage combines the statistics-based model of \PACO with deep learning in a three step procedure: (i) the data are centered and whitened using the \PACO framework, (ii) a CNN is trained to detect synthetic sources from the pre-processed images, and (iii) a detection map is inferred. While the CNN itself works as a black-box approach, the proposed method encompasses prior domain knowledge such as the apparent motion of sources and the expected shape of the exoplanetary signal inside the ADI datasets. More importantly, the proposed detection approach capitalizes on the statistical model of the nuisance component embedded in \PACO to improve the stationarity and the contrast during a pre-processing step. Once a candidate source has been identified, its photometry can be estimated using a dedicated deep learning module, also trained in a supervised fashion. 

Tested on eleven SPHERE-IRDIS datasets, the proposed detection method performs better than standard algorithms of the field like PCA as well as \PACO in terms of precision-recall trade-off. The detection sensitivity is improved by a factor between two and five in average with respect to \PACO for the whole range of angular separations. This gain in even more important compared to baseline methods as cADI or PCA. Interestingly, we showed experimentally that the proposed approach is able to (mostly) reach the fundamental detection sensitivity driven by the photon noise limit for angular separations above 0.5''. This corresponds to an optimal extraction of the sought signals in that regime of separations. The gain brought by the characterization stage of the proposed approach is more moderate but non-negligible. The absolute error of photometric estimation is reduced by a factor two with respect to \PACO for sources of contrast up to $10^{-6}$ located above 0.5''. Nearer the star, the advantage is on average either to \PACO or to the proposed approach, depending on the dataset and of the source location. The gain brought by the characterization module of the proposed approach could allow to further constrain the physical properties of detected exoplanets, usually performed by fitting atmospheric models to the extracted photometry. 

Based on the analysis of these results, several lessons can be drawn regarding the deployed methodology. As a major point, this work illustrates the feasibility to build a complex model (with millions of parameters) of the nuisance component through supervised deep learning from a single ADI dataset of observations. The underlying detection model is learned, without over-fitting, thanks to a custom data-augmentation strategy based on two key ingredients: (i) a random temporal shuffling of the individual images applied before injections of each set of synthetic training sources, (ii) a pre-processing step encompassing a whitening procedure that removes most of the spatial correlations of the nuisance component. Prior to the injection of synthetic training sources, each training set is thus formed by a quasi-random realization of (mostly) uncorrelated Gaussian noise. This property prevents a leak between the train and the test sets as well as a potential memorization of the nuisance structures by the network during its training. Beyond that aspects, the pre-processing step allows to improve the stationarity of the data, thus preventing the high number of false alarms that can occur with other approaches of the field based on supervised deep learning. Our results also emphasize that deriving a flux estimate is a more complex task than providing a qualitative result related to the presence or to the absence of a source with our hybrid modeling of the nuisance component. In particular, we illustrate numerically that the whitening process is detrimental for source characterization (hence, it is not applied) because it modifies both the shape and the amplitude of the exoplanetary signature. In the absence of the  whitening procedure, a shallower architecture should be used to avoid over-fitting.

We are currently working on the extension of the proposed algorithm for the joint processing of multi-spectral datasets such as the ones provided by the SPHERE-IFS instrument using the angular plus spectral differential imaging technique. Besides, we are currently investigating the three main limitations of the proposed approach: (i) the lack of control of the uncertainties, (ii) its task-dependence which is not adapted to reconstruct spatially resolved objects like circumstellar disks, and (iii) the data-dependence of the learning procedure which does not take benefits from multiple observations to build a more general and robust model of the nuisance component. Concerning the later point, building a model from multiple observations could be a promising step to reduce the remaining gap (by a factor 10 to 30) between the current detection performance and the theoretical ultimate detection sensitivity driven by the fundamental photon noise limit. Besides, we would like to incorporate within our deep models some meta-data (e.g., monitoring of the observing conditions, telemetry of the adaptive optics), with the aim to further improve their sensitivity and robustness. 

\section*{Acknowledgements}

We thank the anonymous referee for her/his careful reading of the manuscript as well as her/his insightful comments and suggestions.

This project is supported in part by the European Research Council (ERC) under the European Union's Horizon 2020 research and innovation programme (COBREX; grant agreement n° 885593). The work of TB  and JP was supported in part by the Inria/NYU collaboration, the Louis Vuitton/ENS chair on artificial intelligence and the French government under management of Agence Nationale de la Recherche as part of the \textit{Investissements d'avenir} program, reference ANR19-P3IA0001 (PRAIRIE 3IA Institute). The work of JM was supported in part by the ERC grant number 714381 (SOLARIS project) and by ANR 3IA MIAI@Grenoble-Alpes (ANR-19-P3IA-0003).

This work was granted access to the HPC resources of IDRIS under the allocation 2022-AD011013643 made by GENCI.
 
OF, TB, JM, JP, ML, and AML conceived and designed the method as well as the analysis presented in this paper. OF and TB developed, tested, and implemented the algorithm. OF and ML selected the raw data. ML pre-reduced them through the SPHERE Data Centre. OF and TB performed the analysis of the data. OF, TB, JM, JP, ML, and AML wrote the manuscript.

\section*{Data Availability}
 
The raw data used in this article are freely available on the ESO archive facility at \href{http://archive.eso.org/eso/eso\_archive\_main.html}{http://archive.eso.org/eso/eso\_archive\_main.html}. They were pre-reduced with the SPHERE Data Centre, jointly operated by OSUG/IPAG (Grenoble), PYTHEAS/LAM/CESAM (Marseille), OCA/Lagrange (Nice), Observatoire de Paris/LESIA (Paris), and Observatoire de Lyon/CRAL (Lyon, France). The resulting pre-processed datasets will be shared based on reasonable request to the corresponding author.



\bibliographystyle{mnras}
\bibliography{main} 




\appendix

\section{Discussion about the statistical model and the whitening of the observations}
\label{app:discussion_statistical_model_whitening}

\subsection{Refinement of the statistical model}
\label{subapp:refinement_statistical_model}

\begin{table}
	\caption{Comparison between a multi-variate Gaussian model (Eqs. (\ref{eq:sample_estimators})) and a GSM model (Eqs. (\ref{eq:sample_estimators_robust})) for the statistical modeling of the nuisance component in the pre-processing step of the proposed detection approach. Reported scores are AUC of F1R score (best when close to 1)  as an overall measurement of the precision-recall trade-off of the underlying detector. Mean results and standard-deviation are obtained on the eleven SPHERE-IRDIS datasets considered in this work, see Sect. \ref{sec:results} for the recording logs.}
	\label{tab:comparison_gaussian_vs_gsm}
	\centering
	\begin{tabular}{ccc}
		\hline
		\textbf{Ang. Sep. ('')} & \textbf{Gaussian model \textit{(default)}} & \textbf{GSM model \textit{(variant)}}\\
		\hline
		\textbf{$\boldsymbol{\left[0 ; 2\right]}$''} & $0.88 \pm 0.03$ & $0.90 \pm 0.03$\\
		\textbf{$\boldsymbol{\left[2 ; 4\right]}$''} & $0.90 \pm 0.05$ & $0.91 \pm 0.04$\\
		\textbf{$\boldsymbol{\left[4 ; 6\right]}$''} & $0.89 \pm 0.04$ & $0.91 \pm 0.01$\\
		\textbf{$\boldsymbol{\left[6 ; 7\right]}$''} & $0.88 \pm 0.06$ & $0.90 \pm 0.03$\\
		\hline	
	\end{tabular}
\end{table}

Concerning the statistical model of the nuisance component, the multi-variate Gaussian assumption described in Sect. \ref{subsubsec:statistical_modeling} is a convenient approximation leading to closed-form expressions for the underlying estimators. However, this formulation neglects all temporal fluctuations of the data. In \cite{flasseur2020robust,flasseur2020pacoasdi}, we consider a refinement of this model using a multi-variate Gaussian scaled mixture (GSM; \cite{conte1995asymptotically,wainwright2000scale}). It amounts to model the distribution of patch $\V f_{n,t}$ centered around pixel $n$ at time $t$ by a temporally weighted multi-variate Gaussian $\mathcal{N}(\V m_n, \sigma_{n,t}^2 \, \M C_n)$. Under this model, the sample estimates $\lbrace \widehat{\V m}_n ; \widehat{\sigma}_{n,t}^2 ; \widehat{\M S}_n \rbrace$ of the local mean, of the temporal scaling factors, and of the covariance coming from the maximum likelihood are obtained with a fixed-point iterative method as follows \citep{flasseur2020robust}:
\begin{equation}
	\begin{cases}
		\widehat{\V m}_n = \frac{1}{ \sum\limits_{t=1}^T 1/ \hat{\sigma}_{n,t}^2  } \cdot \sum\limits_{t=1}^T \frac{1}{\hat{\sigma}_{n,t}^2} \M E_{n,t} \, \V r \, \in \mathbb{R}^K\,, \vspace{1mm} \\
		\widehat{\sigma}_{n,t}^2 = \frac{1}{K} (\M E_{n,t}\, \V r - \widehat{\V m}_n)\T \, \widehat{\M S}_n^{-1} \, (\M E_{n,t}\, \V r - \widehat{\V m}_n) \, \in \mathbb{R}_+ \,, \vspace{1mm} \\
		\widehat{\M S}_n = \frac{1}{T} \sum\limits_{t=1}^T \frac{1}{\hat{\sigma}_{n,t}^2} (\M E_{n,t}\, \V r - \widehat{\V m}_n) (\M E_{n,t}\, \V r - \widehat{\V m}_n)\T \, \in \mathbb{R}^{K \times K}\,.
	\end{cases}
	\label{eq:sample_estimators_robust}
\end{equation}
The regularized covariance $\widehat{\M C}_n$ is estimated on the fly by shrinkage of $\widehat{\M S}_n$ as defined in Eq. (\ref{eq:shrinkfactor}), with $Q= \left( \sum_{t=1}^T 1/\widehat{\sigma}_{n,t}^2 \right)^2 \big/ \left( \sum_{t=1}^T 1/\widehat{\sigma}_{n,t}^4 \right)$ the equivalent number of patches involved in the computation of $\widehat{\M S}_n$ in the presence of the weighting factors $\lbrace \widehat{\sigma}_{n,t}^2 \rbrace_{t=1:T}$, see \cite{flasseur2020robust}. The scaling factors $\lbrace \widehat{\sigma}_{n,t}^2 \rbrace_{t=1:T}$ can be interpreted as the local variance of the the residual data (i.e., after centering and whitening). This approach allows to identify and to neutralize outliers, taking the form both of spatially interpolated defective pixels and of local areas displaying fluctuations on some temporal frames larger than on other ones. These properties transfer to the estimators of the statistics of the nuisance component with an improved robustness against bad data, thus leading to a better detection sensitivity and characterization accuracy. 

We tested to integrate a GSM model in the pre-processing step of the proposed algorithm. Table \ref{tab:comparison_gaussian_vs_gsm} compares, with the same procedure as described in Sect. \ref{subsubsec:loss_metrics}, the detection performance in terms of precision and recall of the proposed algorithm using either a multi-variate Gaussian model (Eqs. (\ref{eq:sample_estimators})) or a GSM model (Eqs.  (\ref{eq:sample_estimators_robust})) in the pre-processing step. It illustrates that, as for \PACO, the GSM model leads to an improved detection sensitivity, with an improved stability over datasets (i.e., smaller error bars). However, the gain is significantly smaller than the typical gain, reaching 10 to 15\%, obtained when substituting the multi-variate Gaussian assumption with a GSM model in the \PACO algorithm \citep{flasseur2020robust}. This observation can be attributed to the presence of the additional learning stage of our proposed method that, hopefully, partially correcting for the approximate fidelity (with respect to the observations) of the statistical model embedded in the pre-processing step.

Given the increased computational burden of the proposed detection approach with a GSM model by a factor between 10 and 30 (which represents the typical number of iterations needed to reach convergence of the estimators (\ref{eq:sample_estimators_robust})), we recommend, for practical reasons, to use the standard version of \dPACO embedding a multi-variate Gaussian model, as defined in Sect. \ref{subsubsec:statistical_modeling}. This is also the choice we made for the presentation of the results in the main core of this paper. The alternative version of the algorithm embedding a GSM model can be reserved to refine, in a second step, the reduction of some datasets with ambiguous detections. When comparing with state-of-the-art detection algorithms, we use for \PACO the version embedding a GSM model to compare fairly the proposed method against the best setting of existing methods. 
 
\subsection{Refinement of the whitening of the observations}
\label{subapp:refinement_whitening_observations}

\begin{table}
	\caption{Comparison between non-normalized (Eqs. (\ref{eq:whitening_general_case})) and normalized (Eqs. (\ref{eq:whitening_variant_case})) outputs produced by the pre-processing step of the proposed detection approach. Reported scores are AUC of F1R score (best when close to 1)  as an overall measurement of the precision-recall trade-off of the underlying detector. Mean results and standard-deviation are obtained on the eleven SPHERE-IRDIS datasets considered in this work, see Sect. \ref{sec:results} for the recording logs.}
	\label{tab:comparison_nonnormalized_vs_normalized}
	\centering
	\begin{tabular}{ccc}
		\hline
		\textbf{Ang. Sep. ('')} & \textbf{Not normalized \textit{(default)}} & \textbf{Normalized \textit{(variant)}}\\
		\hline
		\textbf{$\boldsymbol{\left[0 ; 2\right]}$} & $0.88 \pm 0.03$ & $0.88 \pm 0.04$\\
		\textbf{$\boldsymbol{\left[2 ; 4\right]}$} & $0.90 \pm 0.05$ & $0.92 \pm 0.05$\\
		\textbf{$\boldsymbol{\left[4 ; 6\right]}$} & $0.89 \pm 0.04$ & $0.92 \pm 0.03$\\
		\textbf{$\boldsymbol{\left[6 ; 7\right]}$} & $0.88 \pm 0.06$ & $0.91 \pm 0.05$\\
		\hline	
	\end{tabular}
\end{table}

\begin{figure}
	\begin{center}
		\includegraphics[width=0.48\textwidth]{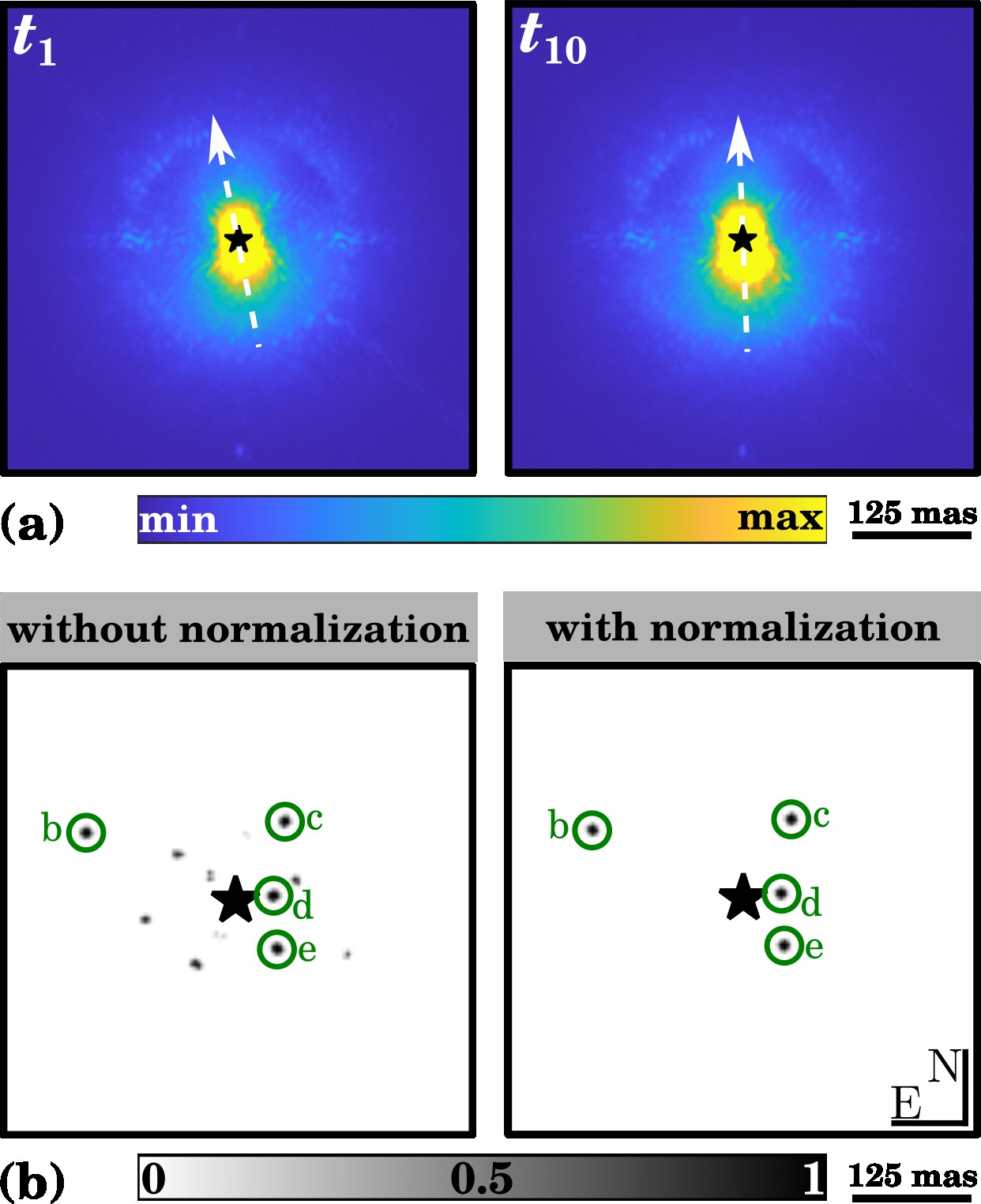}
		\caption{(a): Illustration of two data frames impacted by a wind-driven halo near the star. The direction of main elongation of the halo is symbolized by the white arrow. (b): Detection map  obtained on the corresponding dataset with the proposed detection algorithm embedding the whitening procedure either defined in Sect. \ref{subsubsec:preprocessing_centering_whitening} (i.e., without normalization by the whitened off-axis PSF) or defined in Appendix \ref{subapp:refinement_whitening_observations} (i.e., with normalization by the whitened off-axis PSF). The four known exoplanets (HR 8799 b, c, d, e) are circled in green. Dataset: HR 8799 (2015-07-04), see Sect. \ref{sec:results} for the observation logs.}
		\label{fig:hr8799_whitening}
	\end{center}
\end{figure}	

\noindent The pre-processing step (centering and local whitening) described in Sect. \ref{subsubsec:preprocessing_centering_whitening} is computationally quite efficient since it involves, for non-overlapping square patches, only $\lfloor N / K \rceil$ ($\simeq 10^4$ for the SPHERE-IRDIS instrument) matrix multiplications of size $K \times K$. However, it has two drawbacks: (i) it leads to some spatial discontinuities between adjacent patches, (ii) it does not account for the spatially non-stationary transformation induced by the whitening process on the off-axis PSF in terms of shape and of intensity. In practice, we observe that this approach sometimes lacks of robustness for observations recorded under medium to bad conditions, i.e., where the unmixing between the nuisance component and the sought objects is even more difficult. Limitation (i) can be addressed by considering overlapping patches (e.g., with a patch stride of one pixel). Limitation (ii) can be addressed by adding an output term explicitly accounting for the shape and intensity transformation induced by the whitening process. In this context, the pre-processed images $\widetilde{\V r}$ in $\mathbb{R}^{N \times 2}$ are now formed by the concatenation of two images $\widetilde{\V a}$ in $\mathbb{R}^N$ and $\widetilde{\V b}$ in $\mathbb{R}^N$ defined by:
\begin{equation}
	\begin{cases}
		\widetilde{b}_{n'} = \frac{\left[ \sum\limits_{n \in \mathbb{P}} \M E_n\T \, \V h\T \, \widehat{\M C}^{-1}_n \, (\V r_n - \widehat{\V m}_n) \right]_{n'}}{q_{n'}}\,, \forall n' \in \llbracket 1 ; N \rrbracket \,, \\
		\widetilde{a}_{n'} = \frac{\left[ \sum\limits_{n \in \mathbb{P}} \M E_n\T \, \V h\T \, \widehat{\M C}^{-1}_n \, \V h \right]_{n'}}{q_{n'}}\,, \forall n' \in \llbracket 1 ; N \rrbracket \,,
	\end{cases}
	\label{eq:whitening_variant_case}
\end{equation}
with $q_{n'}$ the number of patches averaged at each location $n'$ of the field of view, as defined in Eq. (\ref{eq:nb_patches_general_case}). For overlapping square patches\footnote{Given the typical circular shape of speckles, we also considered circular overlapping patches. Square and circular patches lead to very comparable detection performance.} with a unit patch stride, $q_{n'}$ is equal to $K$ almost everywhere, excepted on the borders of the field of view where it progressively tends to zero. The term $\V b$ can be interpreted as the correlation between the whitened off-axis PSF and the centered plus whitened observations, while $\V a$ is a normalization term representing the auto-correlation of the whitened off-axis PSF. It can be noted that for each pixel $n'$ of the field of view, the ratio $\widetilde{b}_{n'} / \sqrt{\widetilde{a}_{n'}}$ (respectively, the ratio $\widetilde{b}_{n'} / \widetilde{a}_{n'}$) corresponds to the S/N of detection (respectively, to the source flux) that would be estimated by \PACO at pixel $n'$ \citep{flasseur2018exoplanet}. To prevent the deep model (built from the outputs of the pre-processing step) to learn only the mapping $\widetilde{\V r} \rightarrow \widetilde{\V b} / \sqrt{\widetilde{\V a}}$, we resort to a residual learning procedure. It consists in evaluating the loss function (see Sect. \ref{subsubsec:loss_metrics}) jointly on the detection map produced by the proposed algorithm and also on the \PACO S/N map $\widetilde{\V b} / \sqrt{\widetilde{\V a}}$ computed on the fly. This strategy explicit rewards the deep model to perform better than \PACO. 

Table \ref{tab:comparison_nonnormalized_vs_normalized} compares, with the same procedure as described in Sect. \ref{subsubsec:loss_metrics}, the detection performance in terms of precision and recall of the proposed algorithm using either the whitening procedure described in Sect. \ref{subsubsec:preprocessing_centering_whitening}, and the whitening procedure described in this Appendix. It shows that for the eleven SPHERE-IRDIS datasets we study in details in this work, the variant approach accounting for the transformation induced by the whitening process on the off-axis PSF leads only to a slight improvement of the overall detection performance. This is due to the fact that none of the eleven considered datasets was recorded under bad observing conditions. Figure \ref{fig:hr8799_whitening} gives a qualitative comparison between the two approaches for a dataset of HR 8799 (see Sect. \ref{sec:results} for the recording logs) impacted by the wind-driven halo effect. The variant procedure described in this appendix allows to avoid numerous evident false alarms occurring when the normalization of the pre-processed frames is omitted.

Compared to the whitening procedure described in Sect. \ref{subsubsec:preprocessing_centering_whitening}, the variant described in this appendix has a computational burden increased by a factor $4\times K$, i.e. by typically a factor between 200 and 500 for the VLT/SPHERE instrument. For practical reasons, we recommend to use the standard whitening procedure by default, as defined in Sect. \ref{subsubsec:preprocessing_centering_whitening}. This is also the choice we made for the presentation of the results in the main core of this paper. The alternative version of the algorithm accounting for the whitening of the off-axis PSF can be reserved to refine, in a second step, the reduction of some datasets for which obvious and numerous false alarms are experienced at inference time.

\section{Main settings of the detection and characterization stages of the proposed algorithm}
\label{app:settings}

Table \ref{tab:summary_settings} gives a quick-look summary of the main settings for the proposed algorithm, both for the detection and for the characterization modules. Fields are classified in three categories: \textit{pre-processing}, \textit{generation of the training set}, and \textit{deep learning model}.

\bsp	
\label{lastpage}

\newpage
\pagestyle{empty}

\begin{landscape}
\begin{table*}
	\vspace*{-0.5cm}	
	\caption{Summary of the main settings used for the detection and characterization modules of the proposed algorithm. $^{\text{(a)}}$We recall that the training sets are generated on the fly for the detection stage, i.e., the notion of epochs is used only to schedule the learning rate. For the characterization stage, the term epoch is used in a classical meaning so that the generated patches are seen multiple times (corresponding to the number of epochs) by the network.}
	\label{tab:summary_settings}
	\hspace*{-7.0cm}
	\begin{tabular}{c|cc}
		\toprule
		\textbf{parameters} & \textbf{detection} & \textbf{characterization} \\
		\midrule
		\multicolumn{3}{c}{$\blacktriangleright$ \textit{pre-processing}}\\
		\midrule
		input & observations $\V r \in \mathbb{R}^{N \times T}$  & observations $\V r \in \mathbb{R}^{N \times T}$\\
		strategy for known sources (training only) & temporal shuffling & source masking + temporal shuffling\\ 		
		temporal centering & yes & yes\\
		spatial whitening & yes & no\\
		whitening patch shape & square & --\\
		whitening patch area $K$ & $\in \llbracket 7^2 ; 12^2 \rrbracket$ pixels (automatic) & --\\
		whitening patch stride & $\lfloor\sqrt{K} \, \rceil$ pixels (default) or 1 pixel (variants of App. \ref{app:discussion_statistical_model_whitening}) & --\\
		whitening statistical model & multi-variate Gaussian (default) or GSM (variant of App. \ref{subapp:refinement_statistical_model}) & --\\
		whitening output quantity & $\widehat{\M{L}}_n\T \, (\V r_n - \widehat{\V m}_n)$ s.t. $\widehat{\M{L}}_n \, \widehat{\M{L}}_n\T = \widehat{\M C}_n^{-1}\,,$ $\forall n \in \mathbb{P}$ (default) or & --\\
		& $\text{concat.} \left( \left[ \sum\limits_{n \in \mathbb{P}} \M E_n\T \, \V h\T \, \widehat{\M C}^{-1}_n \, (\V r_n - \widehat{\V m}_n) \right]_{n'} ; \left[ \sum\limits_{n \in \mathbb{P}} \M E_n\T \, \V h\T \, \widehat{\M C}^{-1}_n \, \V h \right]_{n'} \right) \,, \forall n' \in \llbracket 1 ; N \rrbracket$ (variant of Appendix \ref{subapp:refinement_whitening_observations}) & \\
		parallactic derotation & yes & yes\\
		temporal collapsing & no & yes (after injections if any)\\
		implantation & GPUs or CPUs (parallelized) & GPUs or CPUs (parallelized)\\
		output & pre-processed observations $\widetilde{\V r} \in \mathbb{R}^{N \times T}$ & pre-processed observations $\widetilde{\V r} \in \mathbb{R}^{N \times T}$\\
		\midrule 
		\multicolumn{3}{c}{$\blacktriangleright$ \textit{generation of the training set}}\\
		\midrule
		input & pre-processed observations $\widetilde{\V r} \in \mathbb{R}^{N \times T}$ & pre-processed observations $\widetilde{\V r} \in \mathbb{R}^{N \times T}$\\
		pre-processing update on injection arcs & yes & yes\\
		data augmentation & yes & yes\\
		total number $P$ of sources & $\in \llbracket 25,000 ; 50,000 \rrbracket$ & $\simeq 40,000$\\
		number $S$ of training sets & $\in \llbracket 500 ; 1,000 \rrbracket$ & $\simeq 8,000$\\
		number $P^{\left[ s \right]}$ of sources per set $s$ & $\in \llbracket 1, 10 \rrbracket$ & $\in \llbracket 1 ; 10 \rrbracket$\\
		location $\phi$ of sources & uniform in polar system & uniform in Cartesian system\\
		contrast $\alpha$ of sources & uniform in $\left[ 3\widehat{\sigma}_{\phi_p}^{\texttt{PACO}} ; 12\widehat{\sigma}_{\phi_p}^{\texttt{PACO}} \right]$  & uniform in $\left[ 1 \times 10^{-6} ; 3 \times 10^{-5}\right]$ (default)\vspace{0.5mm}\\
		cropping patch area $J$ & -- & $31^2$ pixels\\
		implantation & CPUs (parallelized) & CPUs (parallelized)\\
		output & sets of pre-processed observations with injections $\lbrace \widebreve{\V r}^{\left[ s \right]} \in \mathbb{R}^{N \times T} \rbrace_{s=1:S}$ & sets of pre-processed patches with injections $\lbrace \widebreve{\V p}^{[p]} \in \mathbb{R}^J \rbrace_{p=1:P}$\\
\midrule
		\multicolumn{3}{c}{$\blacktriangleright$ \textit{deep learning model}}\\
		\midrule
		input & a set of pre-processed observations with injections $\widebreve{\V r}^{\left[ s \right]} \in \mathbb{R}^{N \times T}$ & a pre-processed patch with injection $\widebreve{\V p}^{[p]} \in \mathbb{R}^J$\\
		architecture & U-Net (Res-Net 18 backbone) & custom VGG-like\\
		task & pixel-wise classification & regression\\
		number of weights & $\simeq 11$ millions & $\simeq 1.2$ millions\\
		pre-trained weights & no & no\\ 
		optimization loss & Dice2 (overlap measure) & absolute relative error\\
		optimizer & AMSGrad & Adam\\
		validation metric & F1R score (precision / recall trade-off) & absolute relative error\\
		batch size & 1 (set of pre-processed observations with injections) & 1024 (pre-processed patches with  injections)\\
		weight decay & $10^{-5}$ & 0\\
		initial learning rate & $10^{-3}$ & $10^{-3}$\\
		learning rate scheduling & yes (-10\% every 10 epochs) & yes (-70\% every 50 epochs)\\
		number of epochs$^{\text{(a)}}$ & $\in \llbracket 1; 100 \rrbracket$ (on the fly) & 300 (fixed)\\		
		implantation & GPUs & GPUs\\
		output & detection map $\widehat{\V y} \in \left[ 0; 1\right]^M$ & photometry estimates $\widehat{\alpha} \in \mathbb{R}_+$\\
		\bottomrule
	\end{tabular}
\end{table*} 
\end{landscape}

\includepdf[pages={1-}]{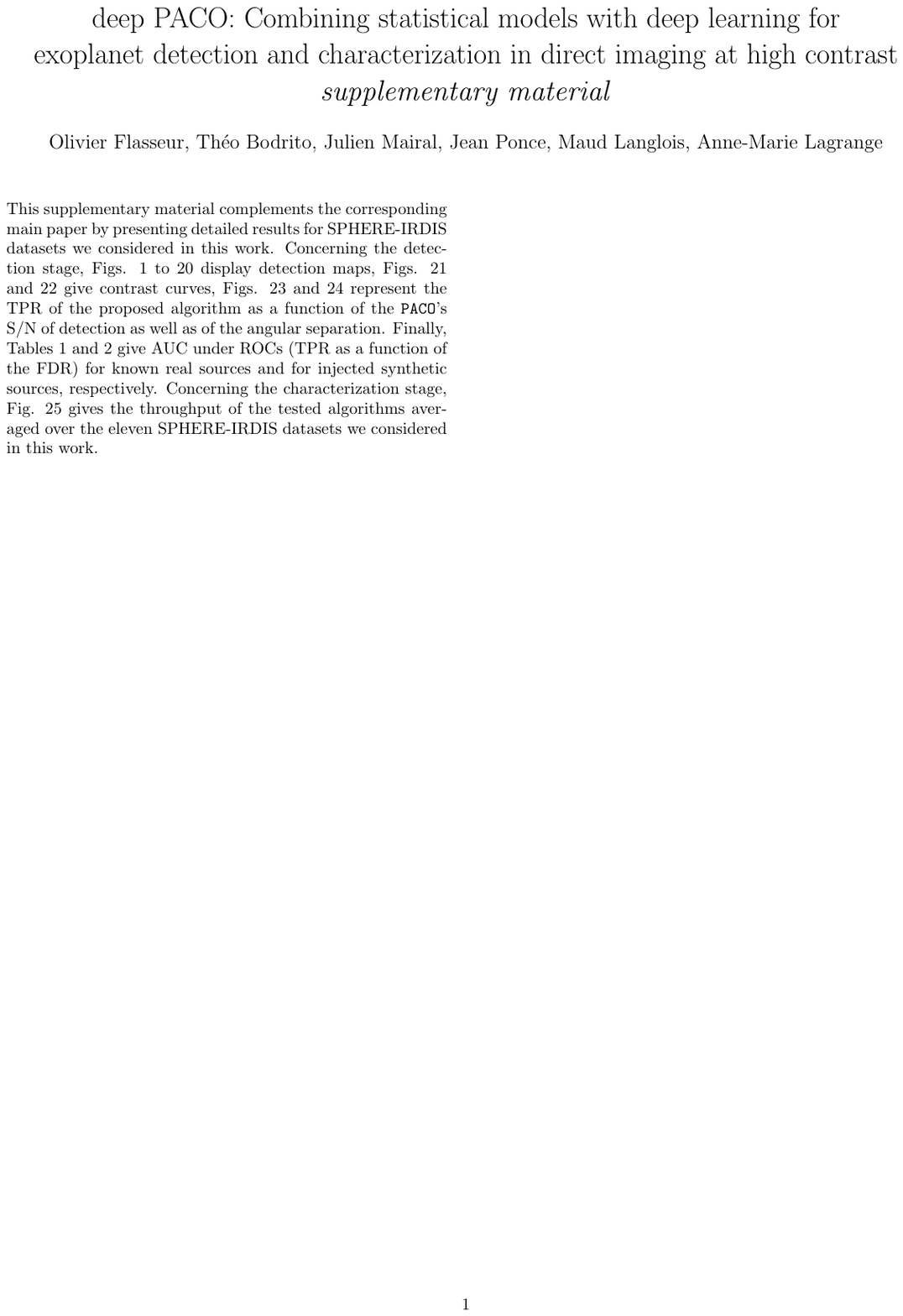}




\end{document}